\documentclass[fleqn,usenatbib,useAMS]{rasti}

\usepackage{newtxtext,newtxmath}

\usepackage[T1]{fontenc}

\usepackage{graphicx}	
\usepackage{amsmath}	
\usepackage{graphicx}
\usepackage{subcaption}
\usepackage{comment}
\usepackage{pdflscape}	

\title[Transfer learning for galaxy feature detection using Faster R-CNN]{Transfer learning for galaxy feature detection: Finding Giant Star-forming Clumps in low redshift galaxies using Faster R-CNN}

\author[J. J. Popp et al.]{
Jürgen J. Popp,$^{1}$\thanks{E-mail: jurgen.popp@open.ac.uk}
Hugh Dickinson,$^{1}$
Stephen Serjeant,$^{1}$
Mike Walmsley,$^{2}$
Dominic Adams,$^{3}$
Lucy Fortson,$^{3}$
\newauthor
Kameswara Mantha,$^{3}$
Vihang Mehta,$^{4}$
James M. Dawson,$^{5}$
Sandor Kruk$^{6}$
and Brooke Simmons$^{7}$
\\
$^{1}$School of Physical Sciences, The Open University, Milton Keynes, MK7 6AA, UK\\
$^{2}$Jodrell Bank Centre for Astrophysics, Department of Physics \& Astronomy, University of Manchester, Oxford Road, Manchester M13 9PL, UK \\
$^{3}$School of Physics and Astronomy, University of Minnesota, 116 Church Street SE, Minneapolis, MN 55455, USA\\
$^{4}$IPAC, Mail Code 314-6, California Institute of Technology, 1200 E. California Blvd., Pasadena, CA, 91125, USA\\
$^{5}$Centre for Radio Astronomy Techniques \& Technologies, Rhodes University, Artillery Road, Grahamstown, 6140, SA\\
$^{6}$ESAC/ESA, Camino Bajo del Castillo, s/n., Urb. Villafranca del Castillo, 28692 Villanueva de la Ca\~nada, Madrid, Spain\\
$^{7}$Physics Department, Lancaster University, Lancaster, LA1 4YB, UK
}

\date{Accepted XXX. Received YYY; in original form ZZZ}

\pubyear{2023}

\begin{document}
\label{firstpage}
\pagerange{\pageref{firstpage}--\pageref{lastpage}}
\maketitle

\begin{abstract}
Giant Star-forming Clumps (GSFCs) are areas of intensive star-formation that are commonly observed in high-redshift ($z \gtrsim 1$) galaxies but their formation and role in galaxy evolution remain unclear. Observations of low-redshift clumpy galaxy analogues are rare but the availability of wide-field galaxy survey data makes the detection of large clumpy galaxy samples much more feasible. Deep Learning (DL), and in particular Convolutional Neural Networks (CNNs), have been successfully applied to image classification tasks in astrophysical data analysis. However, one application of DL that remains relatively unexplored is that of automatically identifying and localising specific objects or features in astrophysical imaging data. In this paper we demonstrate the use of DL-based object detection models to localise GSFCs in astrophysical imaging data. We apply the Faster R-CNN object detection framework (FRCNN) to identify GSFCs in low-redshift ($z\lesssim0.3$) galaxies. Unlike other studies, we train different FRCNN models on observational data that was collected by the Sloan Digital Sky Survey (SDSS) and labelled by volunteers from the citizen science project `Galaxy Zoo: Clump Scout' (GZCS). The FRCNN model relies on a CNN component as a `backbone' feature extractor. We show that CNNs, that have been pre-trained for image classification using astrophysical images, outperform those that have been pre-trained on terrestrial images. In particular, we compare a domain-specific CNN -- `\textit{Zoobot}' -- with a generic classification backbone and find that \textit{Zoobot} achieves higher detection performance. Our final model is capable of producing GSFC detections with a completeness and purity of $\geq 0.8$ while only being trained on $\sim$ 5,000 galaxy images.
\end{abstract}





\section{Introduction}
Deep field observations of high-redshift star-forming galaxies (SFGs) with the Hubble Space Telescope (HST) showed galaxy morphologies which differ from low-redshift galaxies. The dominating spiral and elliptical shapes in the local Universe are replaced by more irregular and chaotic morphologies at higher redshifts \citep{Cowie1995, Bergh1996, Elmegreen2005a, Elmegreen2007a, Elmegreen2009, FoersterSchreiber2009, FoersterSchreiber2011, Guo2015, Guo2018}. While these early HST-based studies suggest that the formation of galaxies with disk morphologies happened late in the cosmological timeline, recent studies using data from the James Webb Space Telescope (JWST) Early Release observations \citep{Ferreira2022} and the JWST CEERS observations \citep{Ferreira2023} find a high number of regular disk galaxies already at early times. With the longer wavelength filters and higher spatial resolution from JWST more faint morphological features of galaxies could be resolved revealing different morphologies for previously peculiar galaxy types.

H$\alpha$ line emission and rest-UV/optical continuum emissions show that most galaxies at $z > 1$ are dominated by several giant star-forming knots or `clumps' (GSFC, or clumps for short) which appear much more luminous and larger in extent than H~II regions of local galaxies. Unlensed observations report clump sizes of $\sim 1$ kpc \citep{Elmegreen2007a, FoersterSchreiber2011} and stellar masses ranging from $10^7 - 10^9 M_{\odot}$ \citep{Elmegreen2007a, Guo2012, Guo2018, Zanella2019, Mehta2021}. For these extended regions of star-formation, highly elevated specific star-formation rates (sSFR) have been observed \citep{Guo2012, Guo2018, Fisher2016}. 

However, clumps that are observed in high-redshift galaxies are likely to be unresolved. Observations by HST from lensed galaxies found clump sizes of $\sim 30-100$ pc \citep{Livermore2012, Adamo2013, Cava2017} and from recent JWST Early Release observations of lensed galaxies at $z=1-8.5$ clumps with sizes of $<10$ to $100$s of pc have been detected \citep{Claeyssens2023}. These JWST observations also revealed clump masses as low as $10^5\,M_\odot$. Other reports indicate that kpc-scale clumps observed at redshifts of $z \sim$ 0.1 to 0.3 consist of smaller coalesced clumps which are not resolved with existing instruments \citep{Overzier2009, Fisher2014, Messa2019}.

The formation and evolution of GSFCs are still debated in the literature. There are believed to be two principal modes of GSFC formation, (1) formation by gravitational instabilities in a gas-rich disk \citep{Elmegreen2005, Bournaud2007, Bournaud2013, Mandelker2014, Romeo2014, Fisher2016} and (2) formation due to galaxy interactions and mergers \citep{Conselice2009, Mandelker2016, Zanella2019}. However, the ways in which clumps do contribute to the evolution of the host galaxy towards modern elliptical and spiral types is not yet fully understood.

The fraction of clumpy galaxies appears to peak at $\sim 55-65\%$ around $z \sim 2$ \citep{Guo2015, Shibuya2016} but this fraction decreases with decreasing redshift \citep[e.g.][]{Adams2022}. Due to the scarcity of clumpy galaxies in the local Universe most surveys of clumpy galaxies have focused on intermediate- and high-redshift galaxies. Therefore, their evolution and properties have not been fully studied for a continuous redshift range between $0 \leq z \lesssim 0.3$. Comparable studies for local galaxies are faced with the challenge of identifying enough clumps in galaxies to base population statistics on a reasonable sample size. Extensive surveys like the Sloan Digital Sky Survey \citep[SDSS, ][]{York2000}, the Dark Energy Camera Legacy Survey \citep[DECaLS,][]{Dey2019} and the Hyper Suprime-Cam Subaru Strategic Program \citep[HSC SSP,][]{Aihara2017} are providing wide field imaging data that make systematic searches for large numbers of low-redshift clumpy galaxies possible but are limited by the resolution constraints of ground-based telescopes.

With forthcoming instruments like the \textit{Euclid} space telescope and wide-field surveys like the Vera Rubin Observatory Legacy Survey of Space and Time (LSST), vast amounts of high-resolution imaging data of local galaxies will become available. 

Such huge data volumes require automatic analysis. Deep Learning (DL), and in particular Convolutional Neural Networks \citep[CNNs, e.g.][]{LeCun2015}, have been successfully applied to image classification tasks in astrophysical data analysis \citep[for an overview see][]{HuertasCompany2023}. However, one application of DL, that of automatically identifying and localising specific objects or features in astrophysical imaging data either through object detection \citep[e.g.][]{HuertasCompany2020} or image segmentation \citep[e.g.][]{AragonCalvo2019, Burke2019, Merz2023, Zavagno2023}, has only been recently used in astrophysical data analysis.

Modern object detection algorithms like the Faster Region-based Convolutional Neural Network framework \citep[Faster R-CNN or FRCNN for short,][]{ren2015} are widely used in `terrestrial' applications, e.g. self-driving cars or face-recognition software. Those networks usually incorporate a pre-trained classification backbone which can be used for transfer learning and so the whole object detection model only needs fine-tuning for a specific use case to be able to produce a high detection performance.

To quantify the benefits of transfer learning for astrophysical image data this paper tests `\textit{Zoobot}' \citep{Walmsley2023} as a feature extraction backbone for object detection in galaxy images. \textit{Zoobot} is a classification-CNN which has been already pre-trained on morphological features from $>$1,000,000 galaxies and will be benchmarked against FRCNN models with feature extraction backbones that have been trained using terrestrial (i.e. not related to galaxies) imaging data.

Even fine-tuning a classifier or an object detection model still requires \textit{some} training data. A major challenge for astronomy is the lack of sufficiently large labelled data sets to train supervised DL models. Previous studies have used simulated training images with known labels \citep[e.g.][]{Burke2019,HuertasCompany2020,Ginzburg2021} or classifications from publicly available catalogues \citep[e.g.][]{Chan2019}. In contrast, the object detection models that we describe in this paper were trained using  observational data labelled by volunteers from the citizen science project `Galaxy Zoo: Clump Scout' \citep[GZCS,][]{Adams2022, Dickinson2022}. 

To assess the sample size required to obtain a scientifically useful GSFC detection performance, we train the different FRCNN models using training data sets with different sizes. The results of these tests can be used to estimate the required effort if labels are needed to fine-tune the FRCNN model for a new data set.

This paper is organised as follows. Section \ref{sec:theory_object_detection} provides a brief introduction to the techniques of object detection with Deep Learning, followed by a section describing the data sources and the necessary pre-processing steps (Section \ref{sec:gzcs}). Section \ref{sec:frcnn_development} explains the details of our model design and training process together with an evaluation of the achieved detection performance. We describe applications of the object detection models on different sets of imaging data in Section \ref{sec:results} and discuss the implications of our findings in Section \ref{sec:discussion}. The paper concludes with a summary of our results in Section \ref{sec:conclusion}.

\section{Deep Learning for object detection}
\label{sec:theory_object_detection}
Object detection is one of many technologies used in computer vision and image processing. Its main application is in detecting and recognising instances of \textit{semantic} objects, i.e. objects of meaningful physical origin, in images or videos \citep[e.g.][]{Dasiopoulou2005}. Object detection algorithms generally make use of Machine Learning (ML) or Deep Learning (DL) to produce automatic detections, localisations and classifications on large data sets like video-feeds or image catalogues. It is commonly used in computer vision tasks like face recognition or traffic sign recognition in driver assistance systems \citep[see][for example]{Erhan2014,Pavel2022}.

Among computer vision tasks, object detection can be seen as a combination of image classification and object localisation \citep[e.g.][]{Szegedy2013}. Whereas image classification attempts to assign a label or class to an entire image, object localisation tries to locate a single instance of a specific object in an image and marks it with a tightly cropped bounding box centred on the instance. Object detection not only tries to locate all instances of multiple objects in an image but also assigns a label to each instance found. Dealing with a variable number of objects and instances of different sizes are the main challenges for object detection algorithms.

In this paper we train a version of the Faster R-CNN architecture proposed by \citet{ren2015}. We chose an object detection algorithm over object instance segmentation algorithms, like Mask R-CNN \citep{He2017}, as we are mainly interested in the object localisation and expect that the resolution of our imaging data is too low to extract useful segmentation masks of detected objects.

\begin{figure}
    \centering
    \includegraphics[width=1\columnwidth]{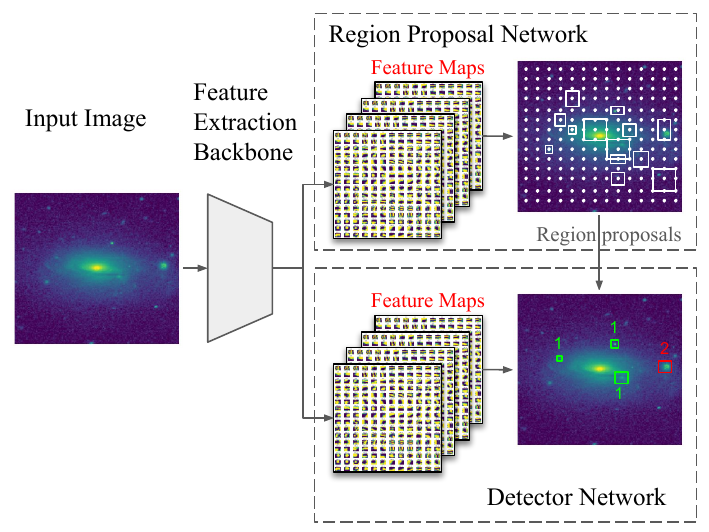}
    \caption[Faster R-CNN principle architecture.]{Schematic view of the Faster R-CNN architecture. The input image is fed into the backbone feature extractor and the resulting feature maps are used as input for the Region Proposal Network (RPN) and the Detector Network. Sample anchor points (white dots) and anchor boxes (white rectangles) are shown for the RPN. The detector network then uses the region proposals from the RPN and the feature maps from the feature extraction backbone to output the final, classified detections of class 1 objects (green) and class 2 objects (red), for example.}
    \label{fig:frcnn_schema}
\end{figure}

Briefly, FRCNN models comprise three components. First, a CNN is used as a `backbone' to extract spatial hierarchies of patterns or features from an input image. These features are then used as input to two separate sub-networks (Figure \ref{fig:frcnn_schema}).

The first is called the Region Proposal Network (RPN) and identifies (or proposes) regions in the image that are likely to contain objects. It sets anchor points at every pixel location of the output feature map of the feature extracting backbone and places at each anchor point position a set of $k$ anchor boxes with default sizes and aspect ratios. The RPN optimises these initial anchor boxes depending on the overlap with the ground-truth object boxes from the training set and generates a twofold output. The prediction scores (`objectness') for the two generic classes, `object' and `background', and for each of the $k$ anchor boxes are one output. The other output are regression coefficients for each of the four attributes: centre coordinates $x,y$, the width $w$ and the height $h$, of the $k$ anchor boxes.

The second sub-network, the detector network, is then used to classify the contents of the proposed regions of class `object' into one of the $n$ final object categories using the corresponding features for those parts of the image that were extracted by the backbone CNN. It also further refines the predicted bounding boxes.  

The final output of the FRCNN model is a collection of rectangular bounding boxes identifying groups of pixels in the image that contain objects and a classification identifying the type of object that each box contains.


\section{Data}
\label{sec:gzcs}
In this section we describe the criteria used to select the galaxy images that we use to train our models and the methods used to label them. Table \ref{tab:sample_sizes} lists the number of galaxy images that remain after each stage of our image selection and labelling pipeline.
\begin{table}
	\centering
	\caption[Reductions applied for the final galaxy sample.]{Reductions applied for the final galaxy sample.} 
        \label{tab:sample_sizes}
	\begin{tabular}{lr}
		\hline
		Selection & Galaxy count\\
		\hline
		Galaxy Zoo 2 (GZ2) & 304,122 \\
            With spectroscopic redshift & 243,500 \\
		With $0.02 \leq z \leq 0.25$ & 225,085 \\
		With $f_{\text{featured}} > 0.5$ (GZCS) & 53,613 \\
		After consensus aggregation & 20,683 \\ 
		With bulge markings removed & 20,646 \\
		After padding & 18,772 \\
		\hline
	\end{tabular}
\end{table}

Our starting point in this paper is the set of galaxy images that were used for the Galaxy Zoo: Clump Scout (GZCS) citizen science project \citep[][see also Appendix \ref{sec:galaxy_images_sdss} for a brief description of the image creation process]{Adams2022}. GZCS ran on the Zooniverse platform (\url{www.zooniverse.org/}) from the 19th September 2019 to the 11th February 2021. For the GZCS project, the participating volunteers were asked to annotate visible clumps on image cutouts of 53,613 SDSS galaxies. These were selected from over 300,000 galaxies that were classified by volunteers who contributed to the `Galaxy Zoo 2' citizen science project \citep[GZ2,][]{Willett2013}. For GZCS, galaxies were selected for which the majority of GZ2-volunteers answered with `No' to the question: ‘Is the galaxy simply smooth and rounded, with no sign of a disk?’, since it seemed unlikely that galaxies containing prominent GSFCs would match this description. The sample was further reduced to only contain galaxies with a documented spectroscopic redshift between $0.02 \leq z \leq 0.25$. The redshift constraint was applied to ensure that most of the clumps, which were anticipated to be of $\sim$kpc size, appear as point-like sources throughout all sample images (see also Appendix \ref{sec:galaxy_images_sdss}).

Volunteers who participated in the GZCS project were asked to identify the locations of the clumps within the selected galaxies. The annotation process is described in detail by \citet{Adams2022}, but a brief summary is given here. First the volunteers were asked to mark the central bulge of the galaxy to help them recognise that this should not be interpreted as a clump even though it has a similar appearance. The volunteers were also equipped with a `normal clump marker' and an `unusual clump marker'. The latter allowed volunteers to mark foreground stars which might overlap with the centre galaxy's spatial extent and could look similar to clumps in terms of colours and being a comparable point source in the SDSS images. We used the volunteers' markings for normal and unusual clumps as the classification label in our training set and further refer to them as `normal' and `odd' clumps, respectively.

Each galaxy image was inspected and annotated by at least 20 independent volunteers and their markings were then aggregated to derive consensus clump locations. \citet{Dickinson2022} developed a framework to aggregate two-dimensional image annotations into a consensus label which further reduced the sample to 20,683 clumpy galaxies (Table \ref{tab:sample_sizes}).

\begin{figure}
    \centering
    \includegraphics[width=0.8\columnwidth]{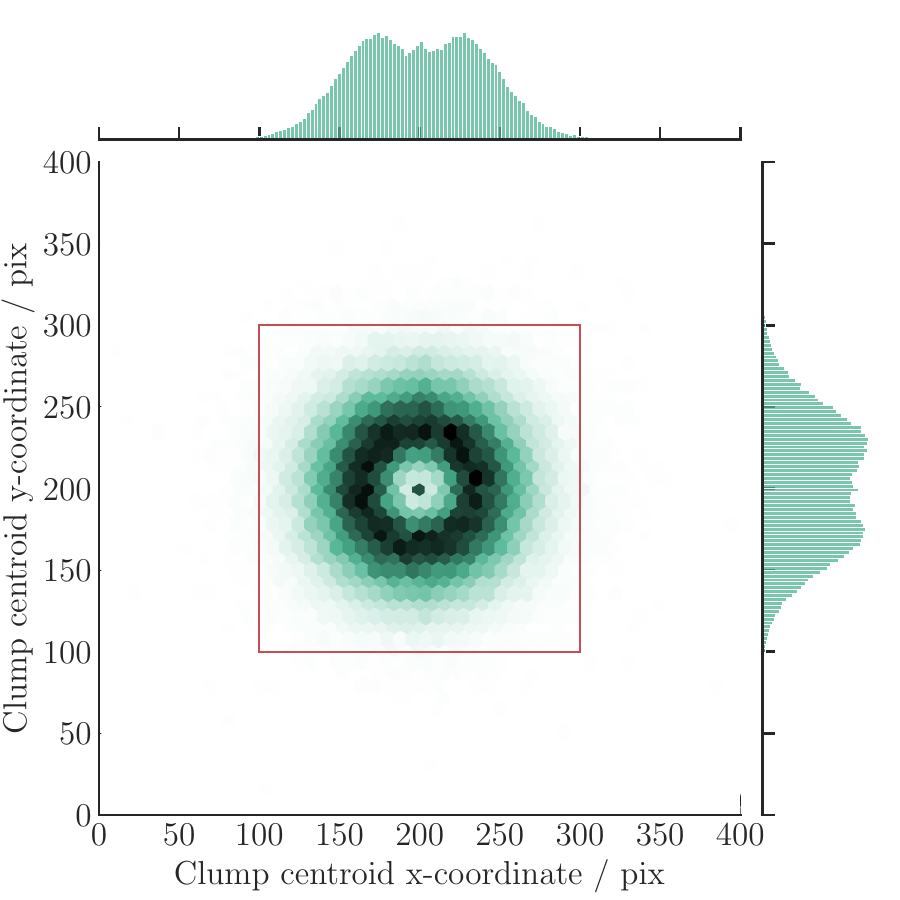}
    \caption[Spatial distribution of the clump centroids within the image dimensions.]{Spatial distribution of the clump centroids within the image dimensions from the final set of 18,772 galaxies containing 39,745 annotated clumps, before central bulge markings and clumps too close to the cropped image dimensions have been removed. The cropped imaged dimensions are marked by the red square and contains $\sim 99\%$ of the annotated clumps. After the image creation process, the median corresponding value for 1 pixel in the RGB-composite images is $\sim 0.2''$ (see Appendix \ref{sec:galaxy_images_sdss}).}
    \label{fig:clump_centroids}
\end{figure}

Figure \ref{fig:clump_centroids} shows that the vast majority of the aggregated clumps ($\sim 99\%$) fall into a central square of half the original side length of each image with the target galaxy at its centre. To reduce the computing time needed to train the network and to make the algorithm focus on the central galaxy only, the outer area was later cropped to the size of the central square during the image augmentation step for model training. Furthermore, any remaining bulge markings and clumps located within a 10\% pixel margin from the borders of each image were removed. This resulted in a final set of 18,772 galaxies containing 39,745 aggregated clump annotations (Table \ref{tab:sample_sizes}).

\begin{figure}
    \centering
    \includegraphics[width=1\columnwidth]{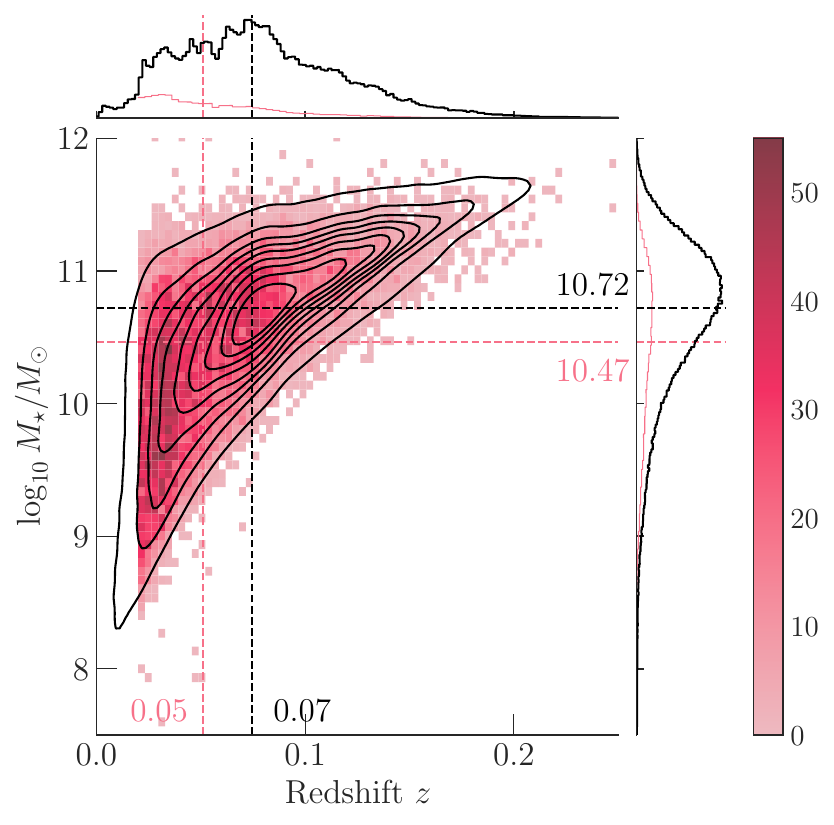}
    \caption[The galaxy stellar mass as a function of redshift.]{The galaxy stellar mass as a function of redshift for the galaxies used for developing the object detection models (red). Overlaid with black contours are the galaxies from the parent Galaxy Zoo 2 sample with spectroscopic redshifts. The dashed lines mark the median of each distribution.}
    \label{fig:galaxies_mass_redshift}
\end{figure}

Figure \ref{fig:galaxies_mass_redshift} illustrates the mass-redshift distribution of our final sample of galaxies with at least one off-centre clump we used for developing the object detection models. We did not apply further limits to our selection of host galaxies as these are used primarily to train the object detection models. Stellar mass estimates for galaxies in our final sample were taken from the SDSS DR7 MPA-JHU value-added catalogue \citep{Kauffmann2003, Brinchmann2004} and range between $10^{7}\,M_\odot \gtrsim M_\star \gtrsim 10^{12}\,M_\odot$.

\section{Developing the object detection model}
\label{sec:frcnn_development}
In the following sections we describe the specific FRCNN-implementation with the different feature extraction backbones that we compare with each other (Section \ref{sec:backbone}). The training setup and execution is covered in Section \ref{sec:training} and in Section \ref{sec:post_processing} we explain the post-processing steps that we perform. Finally, in Section \ref{sec:performance} the detection performance is evaluated.

\subsection{Feature extraction backbone}
\label{sec:backbone}
The CNN as a feature extraction backbone plays an important role in the FRCNN object detection framework. With its ability to extract features from the input images it provides crucial inputs for the RPN and the region classifier or detector network. In this paper we investigate the performance of five backbone CNNs that use different initial weight configurations and training strategies (Table \ref{tab:frcnn_classifiers}).

The \textit{Zoobot} model is a CNN developed to classify galaxies based on their morphological features by \citet{Walmsley2023}. We use a version of \textit{Zoobot} based on the ResNet50 architecture \citep{He2016}, which has been trained to morphologically classify galaxies in  SDSS, HST and DECaLS imaging data. In total, our \textit{Zoobot} version has been trained using more than 1,000,000 classified galaxy images, and training on more images is still on-going. 

The domain-specificity of the \textit{Zoobot} model, combined with the diversity of instruments that provided its training data, suggests that:
\begin{enumerate}
    \item its weights may extract features that are very well suited for the task of identifying clumps in galaxies and,
    \item using it as an FRCNN feature extraction backbone may allow the FRCNN model to be more easily adapted to novel imaging data sets using fewer labelled training examples.
\end{enumerate} 

However, neither of these hypotheses are necessarily true. CNNs used in computer vision applications have been very often pre-trained on massive `terrestrial' data sets. The high-level abstraction of such a CNN might have reached a high enough generalisation level after being pre-trained on a sample like the ImageNet data set, which consists of 1.2 million images belonging to more than 1,000 classes and has become a standard challenge for benchmarking DL models in computer vision \citep[ImageNet Large Scale Visual Recognition Challenge, ILSVRC,][]{ILSVRC15}.

To test whether a FRCNN model with \textit{Zoobot} as a feature extracting backbone (henceforth named \textit{Zoobot-backbone}) does indeed provide better performance and flexibility, we tested the performance versus a ResNet50 feature extraction backbone that has been pre-trained using the ImageNet data set. We refer to the backbone trained using these terrestrial images as \textit{Imagenet-backbone}. 

In addition to the unmodified \textit{Zoobot} model, we also tested a version of \textit{Zoobot} that we specifically train as a classifier to distinguish clumpy and non-clumpy galaxies (\textit{Zoobot-clumps-backbone}). With this approach we address the possibility that specific objects might have been underrepresented in the data set used for training \textit{Zoobot} and the feature extraction backbone has not learned to extract features resembling GSFCs in galaxies well enough. We outline the fine-tuning process in Appendix \ref{sec:zoobot_fine_tune}.

We kept the weights of the ResNet50 architecture with 48 convolutional layers in four blocks \citep[see][for details of the Resnet architecture]{He2016} fixed for all three models described above and only allowed the additional layers of the RPN and the detector network to adjust during training. In this mode, we tested the transfer learning ability of the three backbone feature extractors. 

For testing a fine-tuning approach, we added two more variants of the \textit{Imagenet-backbone} and \textit{Zoobot-backbone} models, in which some weights were allowed to vary during the FRCNN model training. Specifically, we allowed the upper ResNet50 blocks (2, 3 and 4) to vary their weights. We refer to these partially trainable backbones as \textit{Imagenet-backbone-finetuned} and \textit{Zoobot-backbone-finetuned}.

\begin{table*}
	\centering
	\caption[Backbone classifiers used for Faster R-CNN.]{Backbone classifiers used during the Faster R-CNN model development.}
        \label{tab:frcnn_classifiers} 
	\begin{tabular}{llllll}
		\hline
		   & Model name & Feature & Weight & Learning mode & Trainable \\
              & & extractor & initialisation & & blocks \\
              & & architecture & & & \\
		\hline
		1 & \textit{Imagenet-backbone}          & ResNet50 & ImageNet & Transfer learning & -- \\ 
		2 & \textit{Imagenet-backbone-finetuned}& ResNet50 & ImageNet & Fine-tuning & 2,3 and 4 \\ 
		3 & \textit{Zoobot-clumps-backbone}     & ResNet50 & \textit{Zoobot Clumps} & Transfer learning & -- \\ 
		4 & \textit{Zoobot-backbone}            & ResNet50 & \textit{Zoobot} & Transfer learning & -- \\ 
		5 & \textit{Zoobot-backbone-finetuned}  & ResNet50 & \textit{Zoobot} & Fine-tuning & 2,3 and 4 \\
		\hline
	\end{tabular}
\end{table*}

All five FRCNN models (Table \ref{tab:frcnn_classifiers}) were parameterised in the same way. The models expect the same RGB-composite images used for GZCS (see Appendix \ref{sec:galaxy_images_sdss}) as input together with a list of bounding box corner pixel-coordinates and class-labels as derived from the consensus aggregation process \citep{Dickinson2022}. As the galaxy cutouts were scaled to a size of $400 \times 400$ pixels for GZCS, they vary in pixel scale between $0.1''\,\mathrm{pixel}^{-1}$ and $1.3''\,\mathrm{pixel}^{-1}$ with a median of $0.2''\,\mathrm{pixel}^{-1}$, depending on the angular size of the central galaxy. The RPN was initialised with default anchor box sizes of $32 \times 32$, $64 \times 64$, $128 \times 128$, $256 \times 256$ and $512 \times 512$ pixels and aspect ratios of $0.5$, $1.0$ and $2.0$. 

\subsection{Model training}\label{sec:training}
We trained the models over 20 runs each with increasing training sample sizes, which we divided into random train/validation/test splits of size 70\%/20\%/10\% (see Appendix \ref{sec:training_runs} for details). The images were augmented by random horizontal and vertical flips and cropped to a size of $200 \times 200$ pixels, keeping the central galaxy but removing the parts where very few clumps have been marked (see Section \ref{sec:gzcs} and Figure \ref{fig:clump_centroids}). This helped to improve training time and made the FRCNN model focus on the central galaxy while removing the parts which would only produce unnecessary anchor boxes not containing any clumps. 

For all models and run-groups we used the `adaptive moment estimation’ optimiser \citep[\textit{Adam},][]{Kingma2014} with an initial learning rate of $10^{-4}$. We trained every configuration over 120 epochs each using PyTorch's `distributed data parallel' configuration on a multi-GPU environment of eight NVIDIA A100 GPUs and used batch sizes of 32.

\begin{figure*}
    \centering
    \includegraphics[width=1.\textwidth]{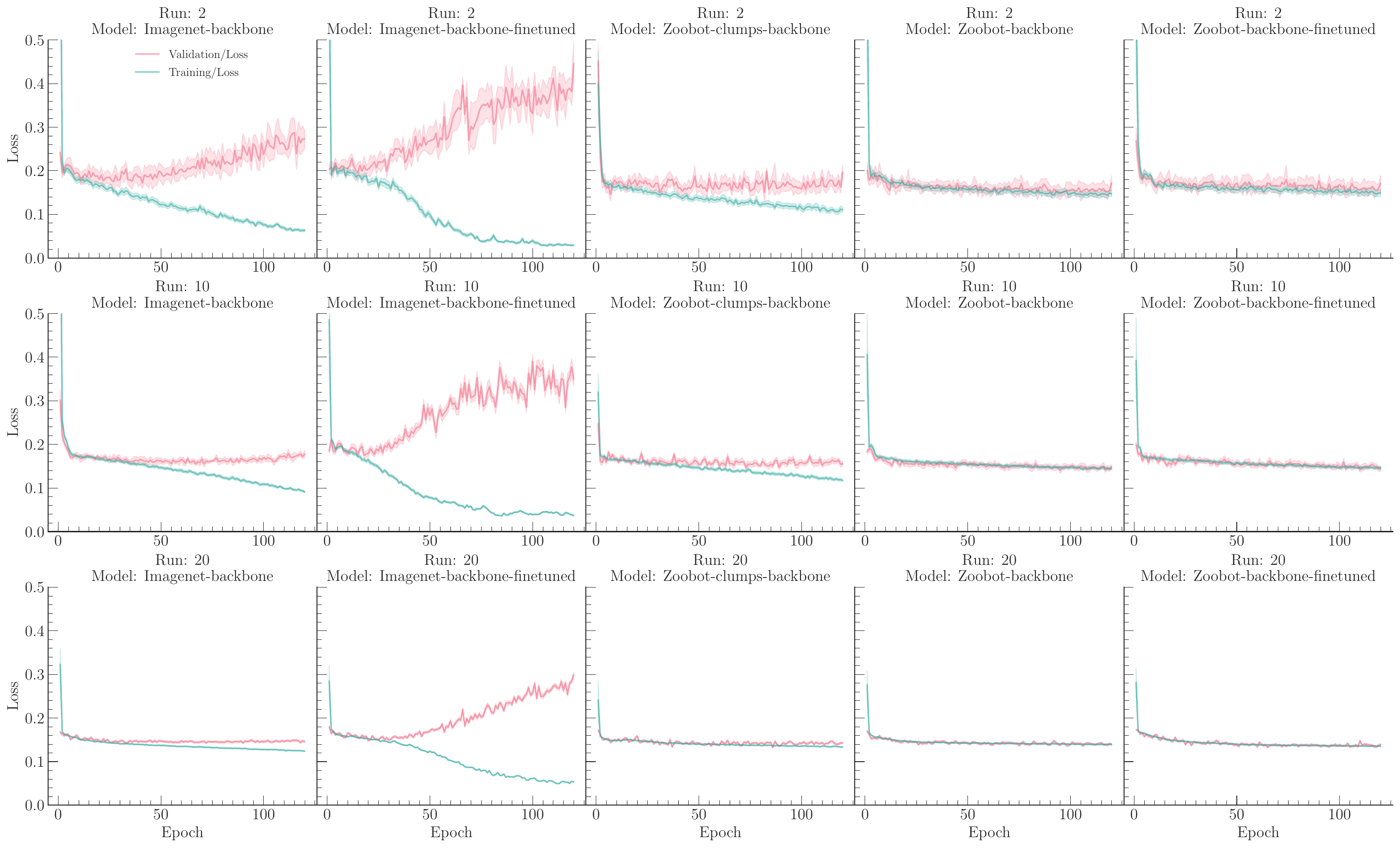}
    \caption[Training and validation loss]{The mean training and validation loss of the Faster R-CNN for run 2 (423 samples), run 10 (1,949 samples) and run 20 (13,140 samples). The training loss is plotted in blue-green and the validation loss in red, where the shaded areas show the corresponding $1\sigma$ standard error of the loss.}
    \label{fig:loss}
\end{figure*}

We monitored the training and validation loss at each epoch. Figure \ref{fig:loss} shows examples for run 2 (423 training samples), run 10 (1,949 training samples) and run 20 (13,140 training samples). Training and validation loss converge at the beginning of the training runs for all models. We observed over-fitting for the models \textit{Imagenet-backbone}, \textit{Imagenet-backbone-finetuned} and \textit{Zoobot-clumps-backbone} where the validation loss diverges from the training loss again. Over-fitting reduces with increasing size of the training data set for the models \textit{Imagenet-backbone} and \textit{Zoobot-clumps-backbone} (e.g. run 2 vs. run 20, Figure \ref{fig:loss}), but remains strong for the FRCNN model \textit{Imagenet-backbone-finetuned}. In contrast, both unchanged Zoobot models, \textit{Zoobot-backbone} and \textit{Zoobot-backbone-finetuned} are stable at all sample sizes tested and express a robust behaviour over all 120 epochs.

\subsection{Post-processing}\label{sec:post_processing}

\subsubsection{Non-maximum suppression}
The output of an object detection model consists of many overlapping bounding boxes with different objectness scores attached. We applied a process called non-maximum suppression (NMS) which uses the Jaccard distance  $J(A,B)$ \citep{Jaccard1912} to determine the Intersection over Union (IoU) of the areas $A$ and $B$,
\begin{equation}\label{eq:jaccard}
    J(A, B) = \frac{A \cap B}{A \cup B} \,
\end{equation}
and keeps only the bounding box with the highest score from the overlapping bounding boxes.

\begin{figure}
    \centering
    \subfloat[\centering Before non-maximum suppression.]{{\includegraphics[width=0.49\columnwidth]{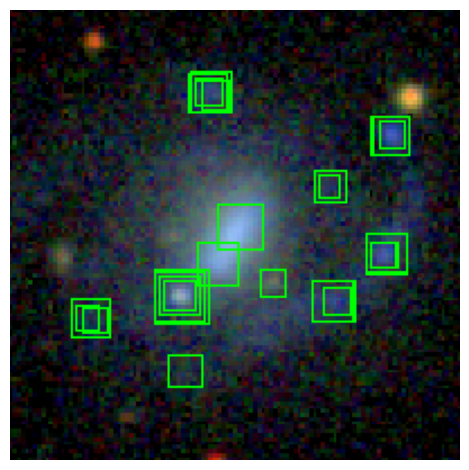} }}
    \subfloat[\centering After non-maximum suppression.]{{\includegraphics[width=0.49\columnwidth]{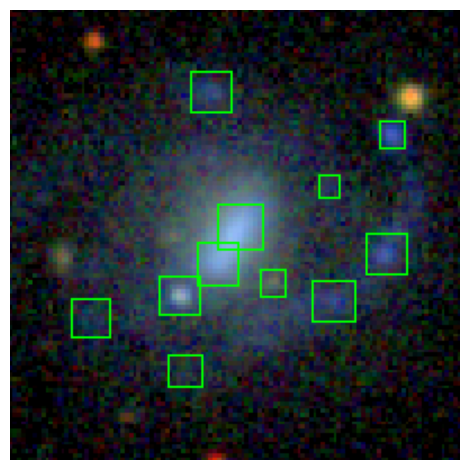} }}
    \caption[Non-maximum suppression.]{Comparison of bounding boxes for a sample galaxy before and after non-maximum suppression.}
    \label{fig:nms}
\end{figure}

Figure \ref{fig:nms} illustrates the result of NMS after being applied to a sample galaxy image. In most cases the raw model proposals consisted of multiple small bounding boxes which were fully contained within larger boxes with a lower objectness score. A threshold of $\text{IoU} \geq 0.2$, which we applied in this paper, proved to be suitable to discard most of those larger bounding box proposals while keeping partially overlapping, adjacent clump proposals. 

The output of this step is a set of clump candidates with either the `normal' or `odd' classification.

\subsubsection{Spatial exclusion and aggregation of clump candidates}
\begin{figure}
    \centering
    \subfloat[\centering Before merging.]{{\includegraphics[width=0.49\columnwidth]{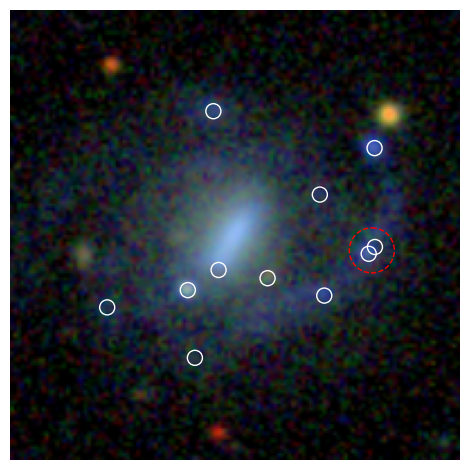} }}
    \subfloat[\centering After merging.]{{\includegraphics[width=0.49\columnwidth]{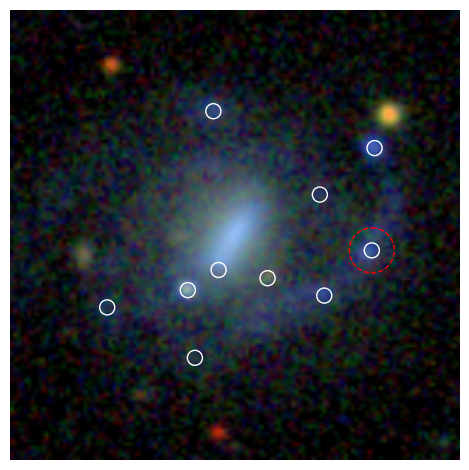} }}
    \caption[Merging adjacent clumps.]{If the distance between clump centroids is less than the $r$-band PSF-FWHM, the clumps are merged into one with a new location at the midpoint between the clump centroids. If at least one of the clumps is classified as an `odd' clump, that label is assigned to the new aggregated clump.}
    \label{fig:clump_merge}
\end{figure}

With the seeing of SDSS, clumps are assumed to appear as point-like sources (see also Section \ref{sec:gzcs}) with a light profile equal to the instrumental point-spread function (PSF). Therefore, we merged adjacent clump candidates not further apart than one $r$-band PSF-FWHM in each subject image into one single detection. We measured the distance between the clump centroids, i.e. the midpoint of the surrounding bounding boxes, and set the new location of the merged clump to the midpoint between the clump centroids. A new label was assigned so, that if at least one of the clumps is classified as an `odd' clump, that label is assigned to the new aggregated clump (Figure \ref{fig:clump_merge}).

\begin{figure}
    \centering
    \subfloat[\centering $r$-band FITS with outline of the galaxy segmentation map.]{{\includegraphics[width=0.49\columnwidth]{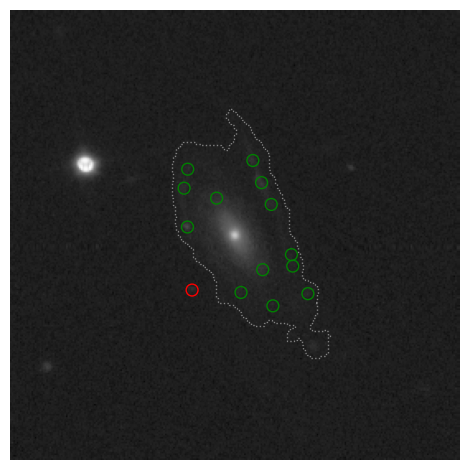} }}
    \subfloat[\centering RGB image with clumps marked.]{{\includegraphics[width=0.49\columnwidth]{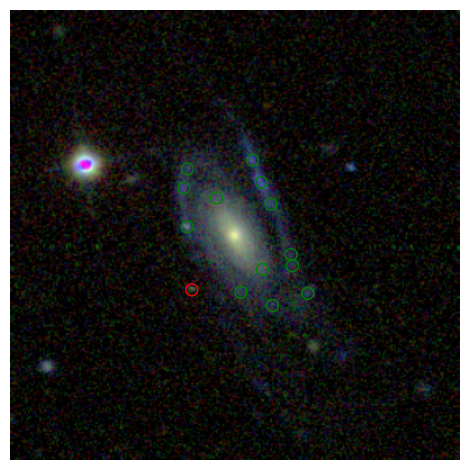} }}
    \caption[Applying a galaxy segmentation map.]{Applying the galaxy segmentation map to exclude detections located outside the host galaxy. The red circle marks a clump to be excluded, green circles mark clumps which will be kept. }
    \label{fig:galaxy_mask}
\end{figure}

Next, we removed all clump candidates located outside the extent of the host galaxy. For each galaxy, we smoothed the $r$-band image to generate a segmentation map with the \textit{photutils} Python package \citep{Bradley2023}, that outlines the extent of the host galaxy. The $r$-band images were smoothed using a Gaussian kernel with the size of the corresponding \textit{r}-band PSF-FHWM. Accounting for the low surface brightness of the galaxy outskirts, we applied a threshold of $1\sigma$ per pixel above the background noise to outline the central galaxy. Clumps located outside the galaxy outline were discarded (Figure \ref{fig:galaxy_mask}).

\subsection{Detection performance}\label{sec:performance}
We used a two-step approach to evaluate detection performance. In the first step, during the model training phase, performance was evaluated on the validation sample set of each run-group ($20\%$, see Section \ref{sec:training}) after each epoch. Based on the general detection performance and taking into account when the model first showed signs of over-fitting (e.g. Figure \ref{fig:loss}), we determined a best model version for each run. In the second step, we compared the trained model versions in an astrophysical context using the test sample set of each run-group ($10\%$, Section \ref{sec:training}).

\subsubsection{Determining the best model using the COCO metrics}
We evaluated the detection performance simultaneously on the validation set during the training process for all models using the metrics from the COCO Object Detection Challenge \citep{Lin2014}. A short description of the metrics can be found in Appendix \ref{sec:coco}.

We exported a current model version after each epoch during the training process and calculated average precision (AP) and average recall (AR) and F1-scores for detection score thresholds ranging from 0.1 to 0.9 after post-processing (see Section \ref{sec:post_processing}) the results. Based on the F1-score the best models were chosen from this pool of model versions, separately for each training run. The model versions we chose for training run 20, which used all of the 18,772 galaxy images for training, validation and testing, are listed in Table \ref{tab:best_models}.
\begin{table}
	\centering
	\caption[Best model versions for training run 20.]{Best model versions chosen for training run 20, based on a prediction score threshold of $\geq 0.3$ and IoU threshold of $\geq 0.5$.} 
        \label{tab:best_models}
	\begin{tabular}{lllll}
		\hline
		Model name & Epoch & AP & AR & $f_1$ \\
		\hline
		\textit{Imagenet-backbone}          &  20 & 0.40 & 0.39 & 0.40 \\ 
		\textit{Imagenet-backbone-finetuned}&  20 & 0.42 & 0.82 & 0.56 \\ 
		\textit{Zoobot-clumps-backbone}     &  60 & 0.07 & 0.73 & 0.13 \\ 
		\textit{Zoobot-backbone}            & 120 & 0.44 & 0.68 & 0.54 \\ 
		\textit{Zoobot-backbone-finetuned}  & 120 & 0.44 & 0.67 & 0.53 \\ 
		\hline
	\end{tabular}
\end{table}

\subsubsection{Completeness and purity for model detections}\label{sec:compl_purity}
In astrophysical applications, domain-specific post-processing steps are very often necessary and detections need to be reassessed with the help of additional morphological and physical parameters. Moreover, the appropriate object detection score threshold (or objectness threshold, see Section \ref{sec:theory_object_detection}) very often depends on the scientific questions asked. A complete sample can be a more appropriate outcome than a pure sample and vice versa.

As our clumps are based on assumed point-like sources not resolved by SDSS, we applied a method which is different to the IoU approach used by the COCO metrics (see Appendix \ref{sec:coco}) to compute the overlap between the ground-truth (i.e. the volunteers' labels from GZCS) and our model detections. A successful clump candidate is counted, if the distance between the centroid of a predicted clump candidate and the centroid of the ground-truth clump is less than 0.75 of the image-specific PSF-FWHM. 

\begin{figure}
    \centering
    \includegraphics[width=1\columnwidth]{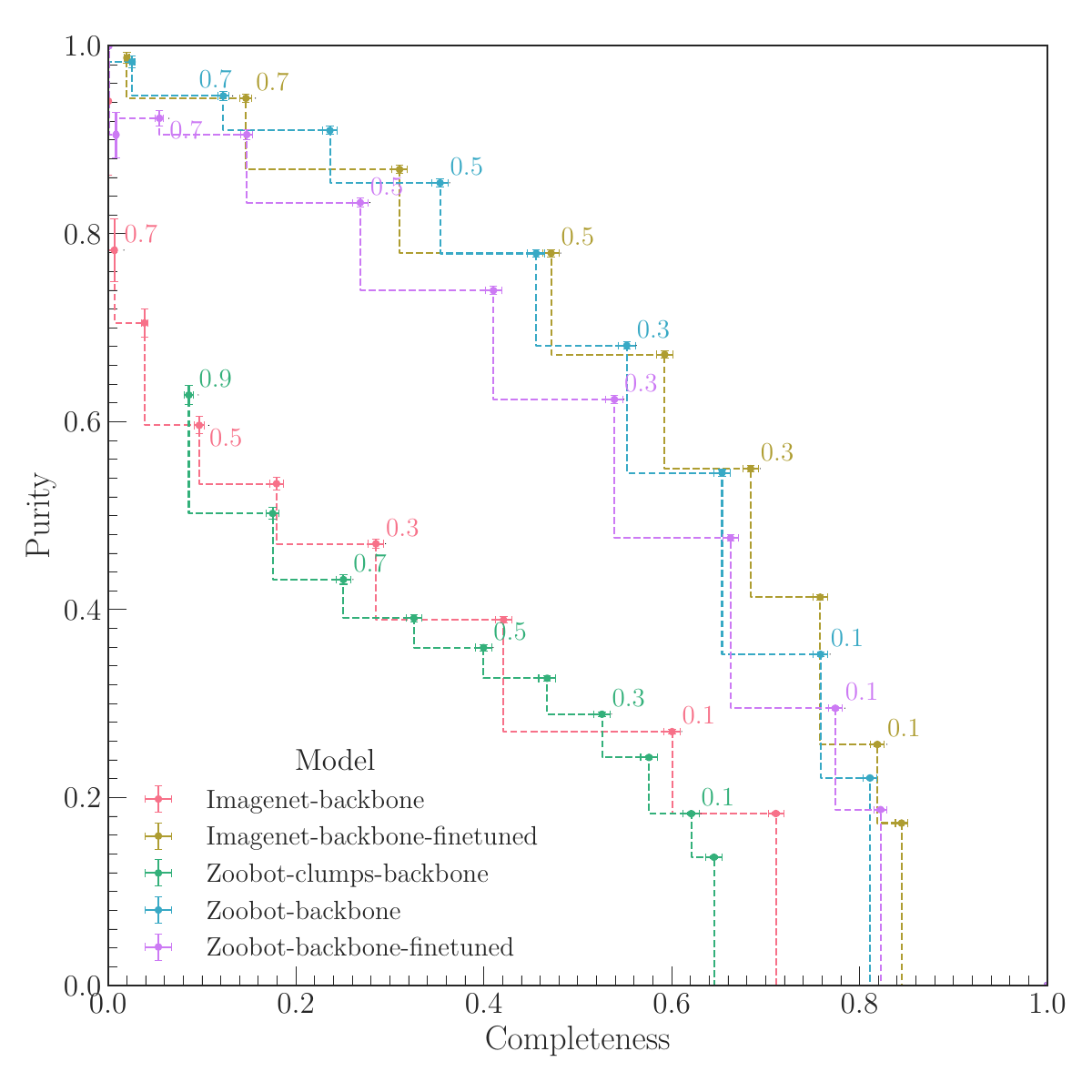}
    \caption[Purity and completeness for all models]{Purity and completeness for all five models. The detection score threshold $c_n$ is increasing from $0.0$ (right) to $0.9$ (left) as indicated by the annotations. A clump candidate is considered to be a true positive, if the distance between the centroid of a predicted clump candidate and the centroid of the ground-truth clump is less than 0.75 of the image-specific PSF-FWHM. All models have been trained on the full sample size (run 20).}
    \label{fig:recall_precision}
\end{figure}

Figure \ref{fig:recall_precision} shows completeness against purity for the final models from training run 20 and with an increasing detection score threshold $c_n$ from $0.0$ to $0.9$. The models using a feature extraction backbone initialised with the unmodified \textit{Zoobot} weights generally show a better detection performance compared to the \textit{Imagenet-backbone} model. Only if we allow the ImageNet-based model to adjust its weights for the last convolutional blocks of the backbone feature extractor (\textit{Imagenet-backbone-finetuned}), completeness and purity do reach similar levels. The object detection model which uses a backbone feature extractor pre-trained on detecting clumps (\textit{Zoobot-clumps-backbone}) appears to not benefit from the even more domain-specific initialisation.

\begin{figure}
    \centering
    \includegraphics[width=1\columnwidth]{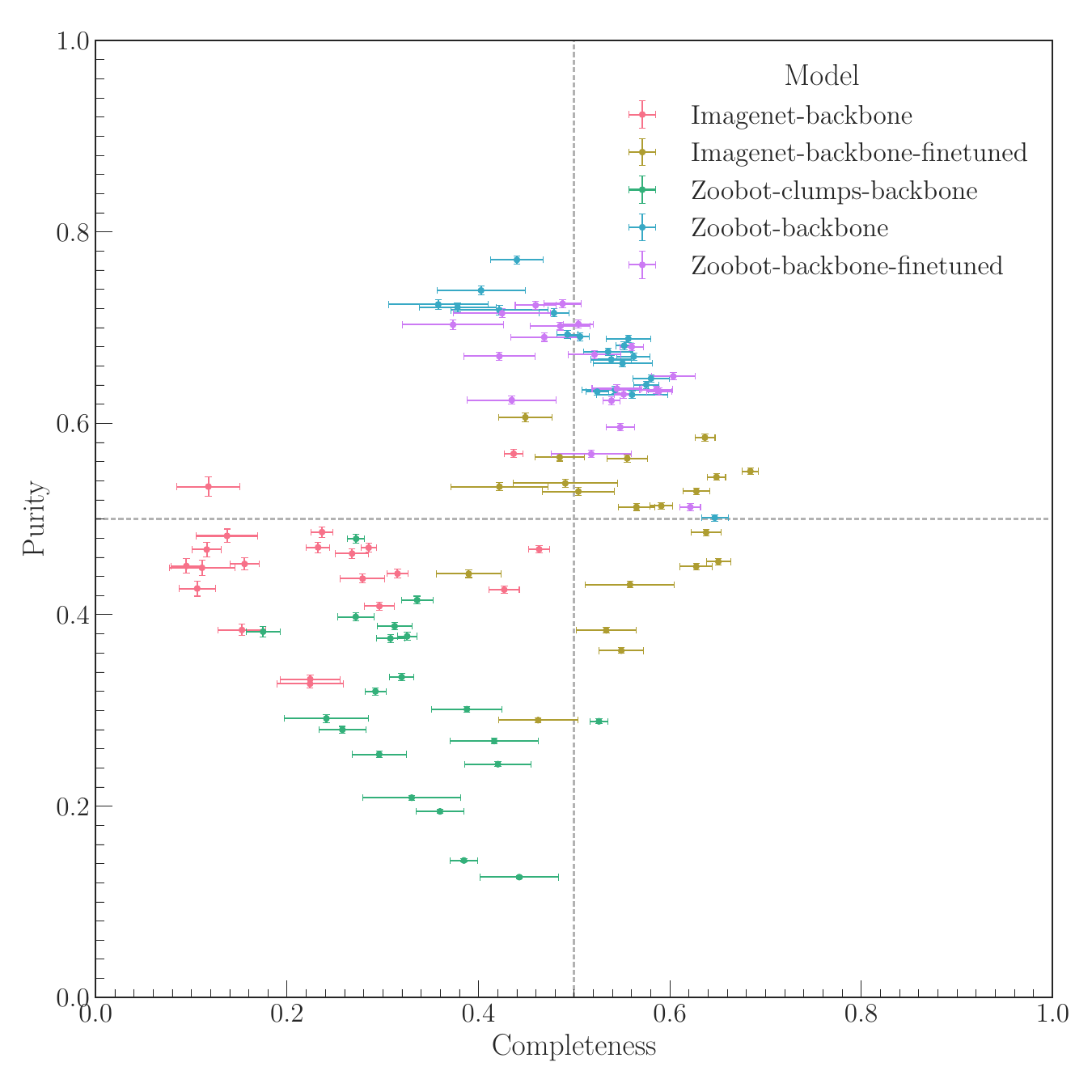}
    \caption[Model completeness and purity for a score threshold $\geq 0.3$.]{Model completeness and purity for a score threshold $\geq 0.3$. Error bars show the $95\%$ confidence interval. The different points for each model represent different training runs with different sample sizes but are not labelled for better visibility.}
    \label{fig:completeness_purity_single}
\end{figure}

We observe a similar performance ranking of the different FRCNN models if completeness is plotted against purity for all training runs (Figure \ref{fig:completeness_purity_single}, as an example for a score threshold of $\geq 0.3$ ).


\subsubsection{Performance per relative clump flux bin}\label{sec:performance_flux_ratio}
\citet{Guo2015} suggested a GSFC definition based on the ratio of clump to host galaxy UV-luminosity. A clump needs to exceed a ratio of $8\%$ to be classified as a GSFC and not as a normal star-forming region. We used this definition to determine the completeness and purity of our FRCNN models in terms of astrophysically relevant GSFC-detections and also applied a $3\%$ flux ratio threshold, similar to \citet{Adams2022}.

We measured the $u$-band flux from SDSS for each clump candidate (for details see Appendix \ref{sec:photometry}), as it is closest to UV, and retrieved the host galaxy $u$-band flux from the SDSS DR15 PhotoPrimary table \citep{Aguado2019}. The log of the ratio of clump to host galaxy flux was grouped into equally spaced bins of size $0.5$.

\begin{figure*}
    \centering
    \subfloat[\centering Completeness for the models trained on 5,061 galaxy images (run 15).\label{fig:completeness_run_15}]{{\includegraphics[width=0.49\textwidth]{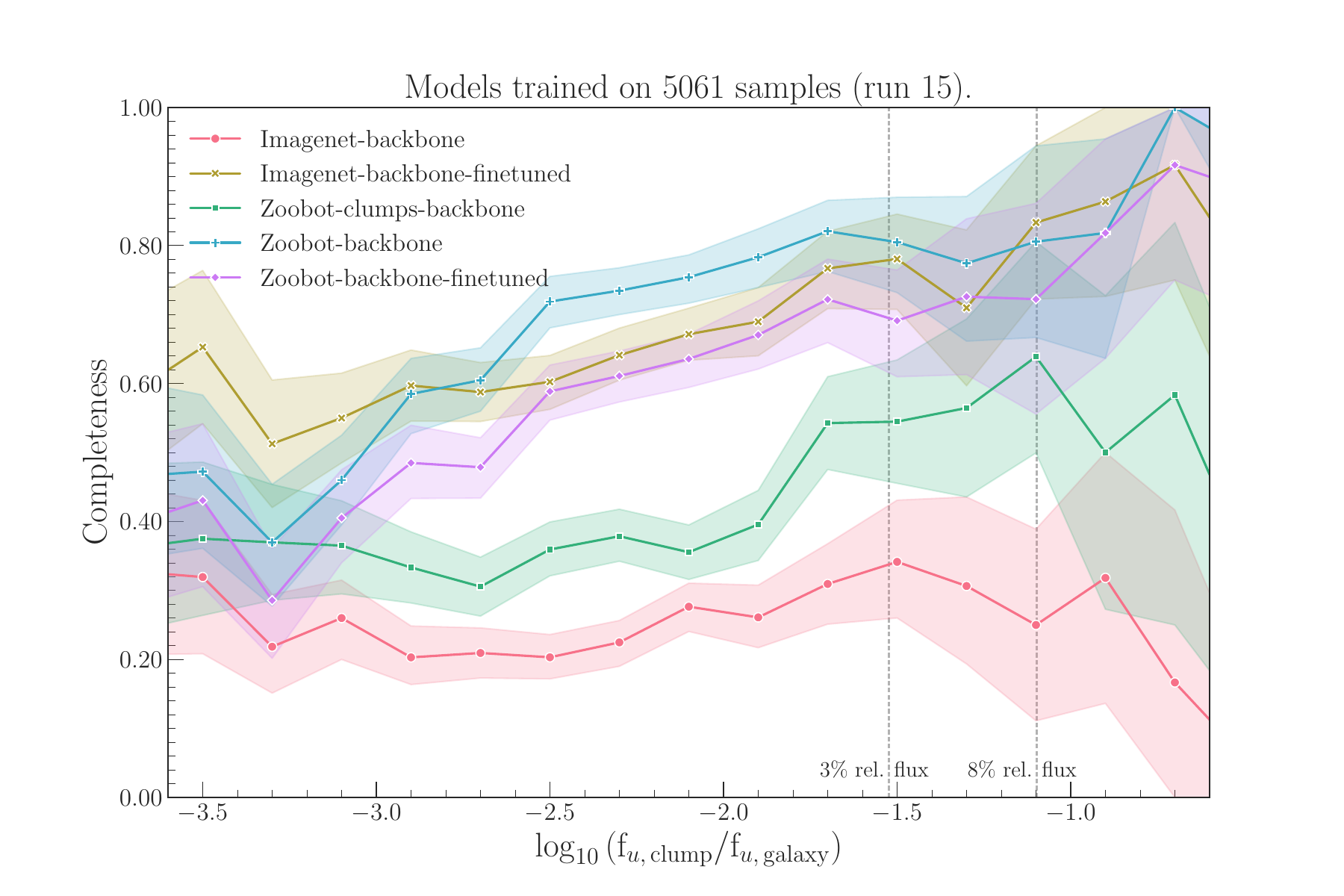} }}
    \subfloat[\centering Completeness for the models trained on 13,140 galaxy images (run 20).\label{fig:completeness_run_20}]{{\includegraphics[width=0.49\textwidth]{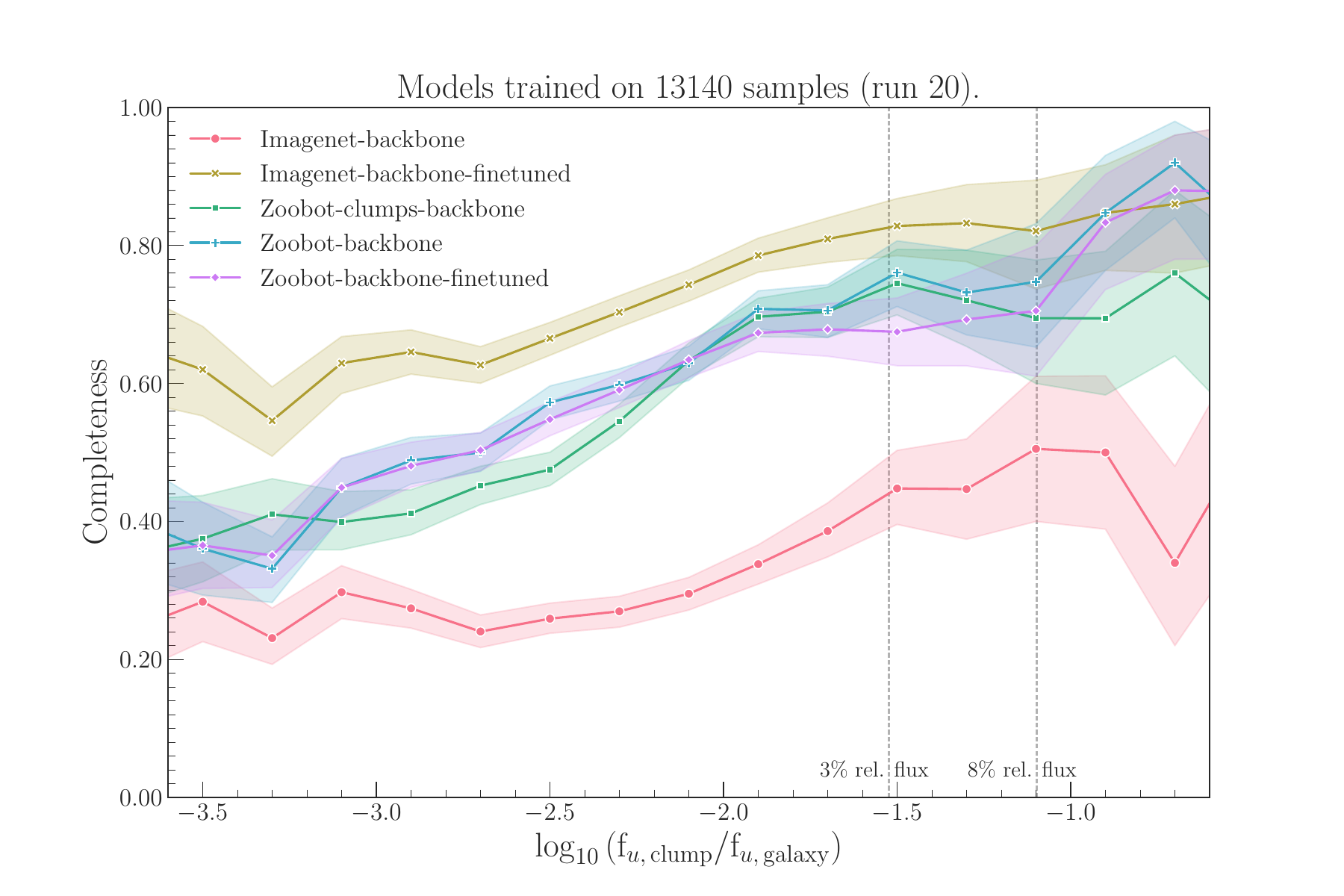} }}
    \caption[Model completeness per relative clump flux for training run 15 and run 20.]{Model completeness with respect to the volunteers' labels from GZCS per relative clump flux for training run 15 and run 20 (training sample size of 5,061 and 13,140, respectively, and for a score threshold of $\geq 0.3$). Shaded areas showing the $95\%$ confidence interval. The $3\%$ and $8\%$ threshold for the flux ratio are indicated with vertical dashed lines.}
    \label{fig:completeness_runs}
\end{figure*}

We then compared the completeness of the discovered clumps against the log of the relative flux for each model and each training run. As an example, Figure \ref{fig:completeness_run_15} plots completeness for the models trained on 5,061 galaxy images (run 15), which is considerably smaller than the full set of training samples. The figure shows the highest completeness for the model \textit{Zoobot-backbone} over most of the flux ratio range. Only for the faintest clumps does the completeness drop to $0.4$ and below the completeness values of the model \textit{Imagenet-backbone-finetuned}. \textit{Zoobot-backbone-finetuned} reaches similar completeness levels within the error margins but the remaining two models, \textit{Zoobot-clumps-backbone} and \textit{Imagenet-backbone}, are significantly lower in completeness.

Note, that the completeness is already high ($\gtrsim 0.8$) in the flux ratio ranges of $\geq 3\%$ and $\geq 8\%$ for the FRCNN model \textit{Zoobot-backbone} after we trained the model on only 5,061 galaxy images. For comparison, Figure \ref{fig:completeness_run_20} shows the completeness versus relative flux ratio for run 20, where we trained the models on the full set of training samples. If trained on the full set of 13,140 galaxy images, \textit{Imagenet-backbone-finetuned} now reaches the highest completeness values, although the completeness of the other models, apart from \textit{Imagenet-backbone}, are similar in the clump-specific flux ratio ranges.

The completeness shown by the two FRCNN models using a version of the unmodified \textit{Zoobot} as their feature extraction backbone is  similar after either being trained on a reduced train set (run 15) or on the full train set (run 20). This indicates that a reasonable detection performance can already be achieved through a relatively small labelled training sample. Figures \ref{fig:completeness_run_15} and \ref{fig:completeness_run_20} also show, that \textit{Imagenet-backbone-finetuned} and \textit{Zoobot-clumps-backbone} require a larger training set to achieve comparable completeness performance. Fine-tuning a terrestrial feature extraction backbone, in our case \textit{Imagenet-backbone-finetuned}, does result in a high completeness performance but only through using a large data set. In contrast, applying the same model in transfer learning mode (\textit{Imagenet-backbone}) does not result in comparable completeness levels as seen from the FRCNN models with domain-specific feature extracting backbones.

\begin{figure*}
    \centering
    \subfloat[\centering Purity for the models trained on 5,061 galaxy images (run 15).\label{fig:purity_run_15}]{{\includegraphics[width=0.49\textwidth]{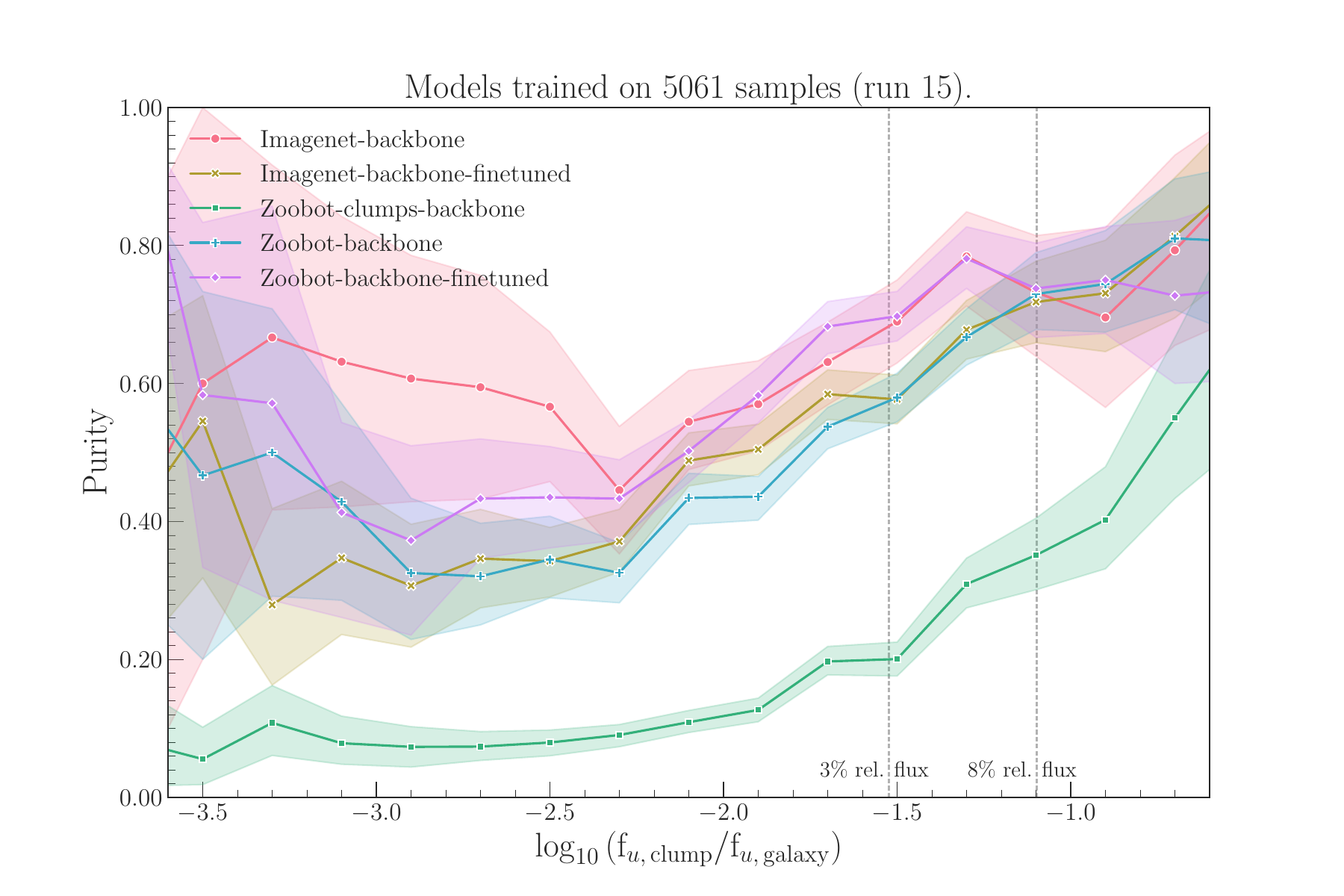} }}
    \subfloat[\centering Purity for the models trained on 13,140 galaxy images (run 20).\label{fig:purity_run_20}]{{\includegraphics[width=0.49\textwidth]{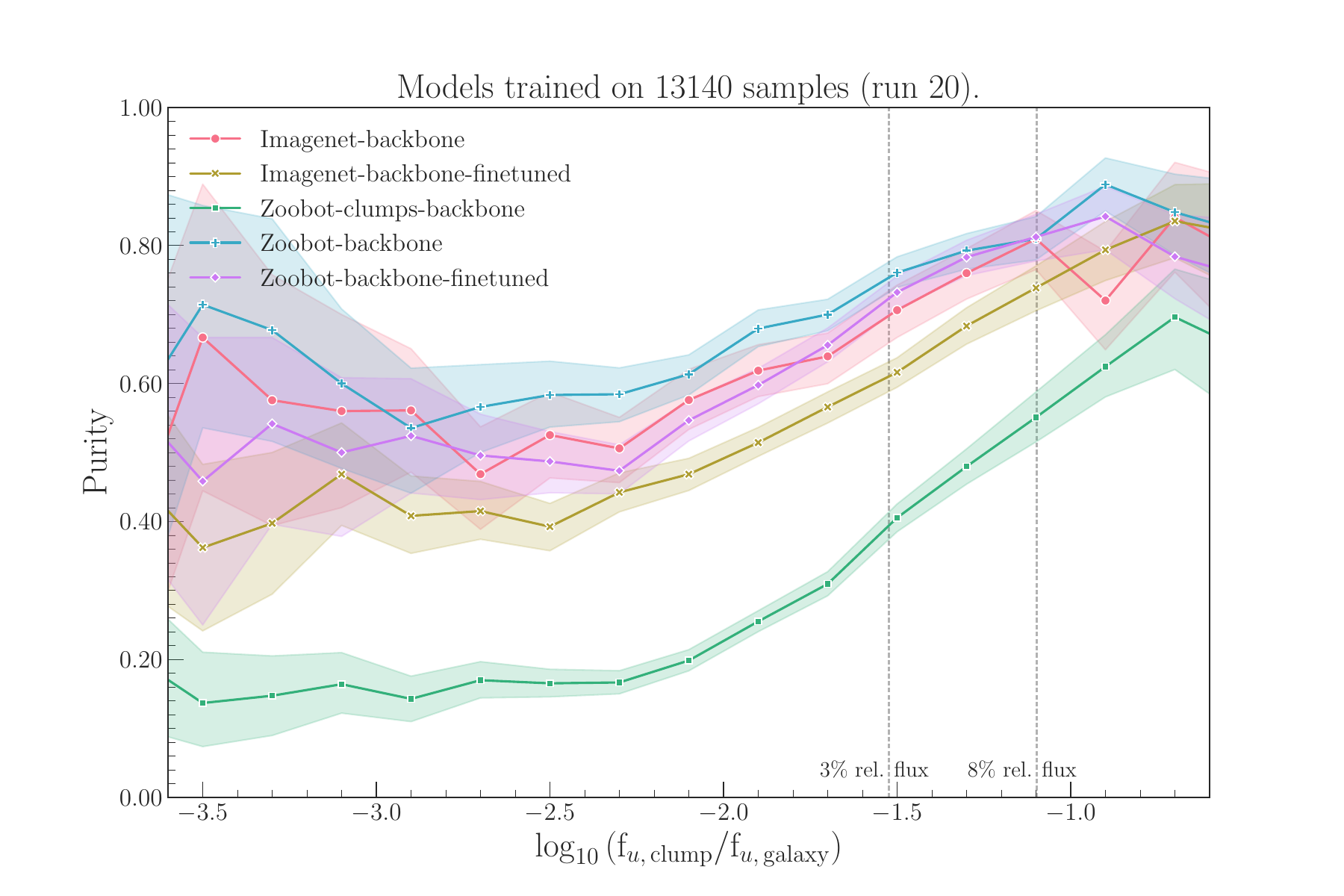} }}
    \caption[Model purity per relative clump flux for training run 15 and run 20.]{Model purity with respect to the volunteers' labels from GZCS per relative clump flux for training run 15 and run 20 (training sample size of 5,061 and 13,140, respectively, and for a score threshold of $\geq 0.3$). Shaded areas showing the $95\%$ confidence interval. The $3\%$ and $8\%$ threshold for the flux ratio are indicated with vertical dashed lines.}
    \label{fig:purity_runs}
\end{figure*}

We also calculated purity for the same flux ratio bins for all models and training runs. Figures \ref{fig:purity_run_15} and \ref{fig:purity_run_20} show the two resulting plots after training run 15 and 20. The achieved purity levels are highest for relatively bright clumps (flux ratios $>3\%$). The purity of the model detections tend to be closer together for all models except the \textit{Zoobot-clumps-backbone} model. Unlike before, where we compared the completeness, there does not appear to be a clear difference between models with terrestrial and domain-specific feature extraction backbones.

\begin{figure*}
    \centering
    \subfloat[\centering Purity and completeness for $u$-band flux ratio $\geq 3\%$.\label{fig:recall_precision_3pct}]{{\includegraphics[width=0.49\textwidth]{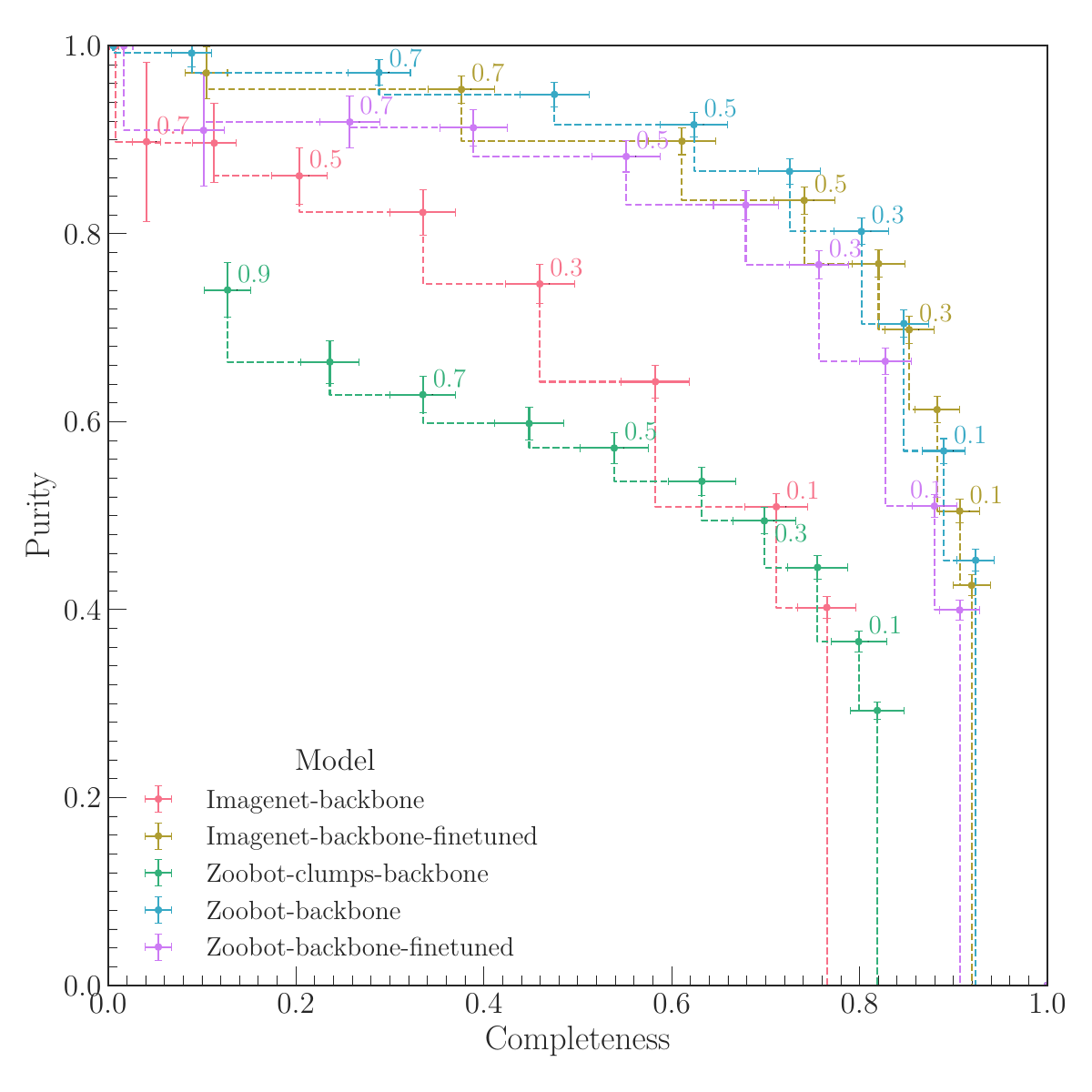} }}
    \subfloat[\centering Purity and completeness for $u$-band flux ratio $\geq 8\%$.\label{fig:recall_precision_8pct}]{{\includegraphics[width=0.49\textwidth]{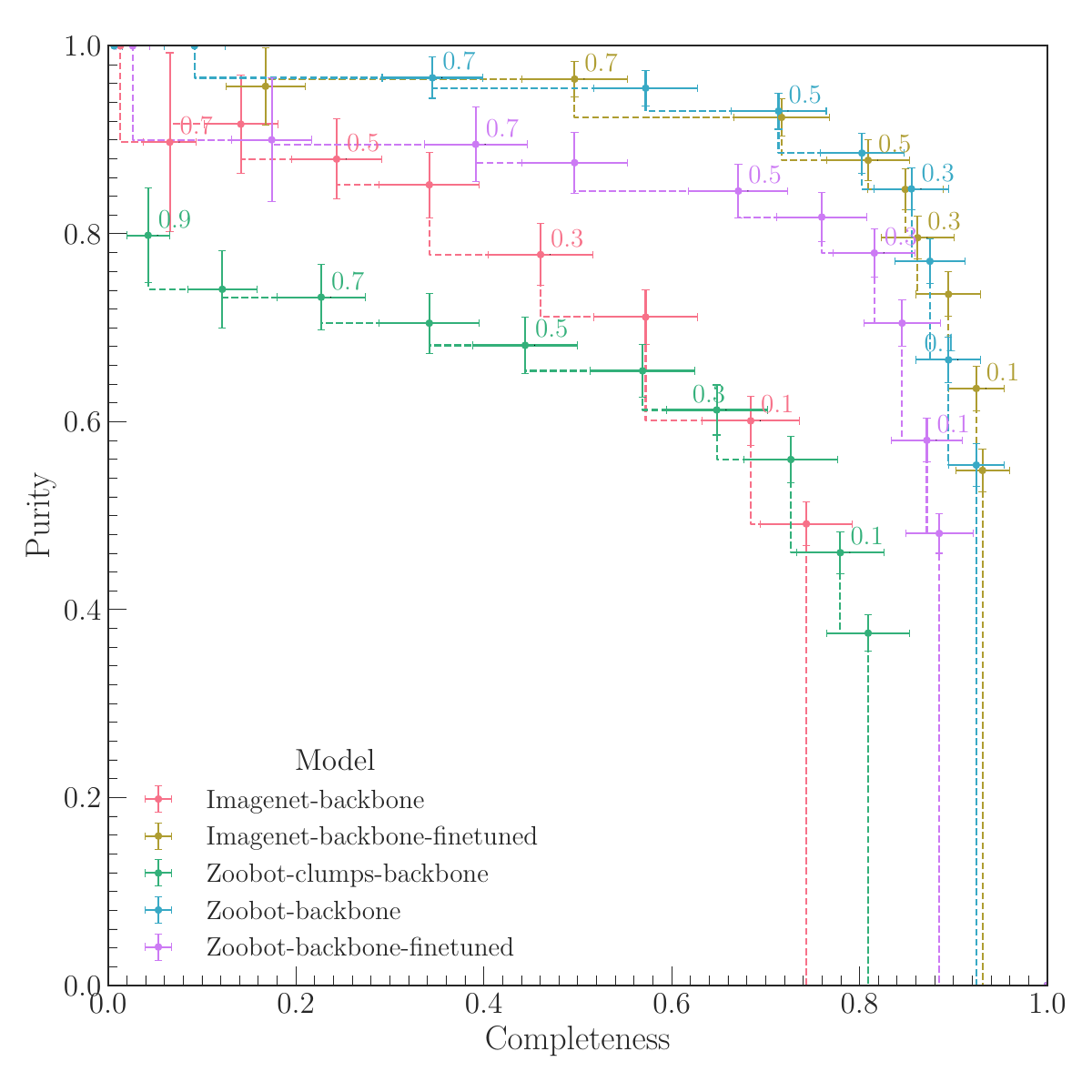} }}
    \caption[Purity and completeness for all models ($u$-band flux ratio $\geq 3\%$ and $\geq 8\%$).]{Purity and completeness for all five models for clumps and clump candidates with a measured $u$-band flux ratio compared to the host galaxy of $\geq 3\%$ and $\geq 8\%$. The detection score threshold $c_n$ is increasing from $0.0$ (right) to $0.9$ (left) as indicated by the annotations. Error bars show the $95\%$ confidence interval. All models have been trained on the full sample size (run 20).}
    \label{fig:recall_precision_flux_ratio}
\end{figure*}

To further illustrate the differences between the five feature extraction backbones and the increase in purity and completeness, if the clump candidates are limited to relevant flux ratios, we show completeness against purity plots for clump candidates with a $u$-band flux ratio of $\geq 3\%$ in Figure \ref{fig:recall_precision_3pct} and for a ratio $\geq 8\%$ in Figure \ref{fig:recall_precision_8pct}.

\section{Detection results}\label{sec:results}
We applied the five different FRCNN models to two different data sets, (1) Galaxy Zoo: Clump Scout (SDSS images) and (2) Hyper Suprime-Cam (HSC) images. 

In Section \ref{sec:detections_gzcs} we compare the detections from each model on SDSS imaging data to the ground-truth data set, which consists of the GZCS galaxies with annotated clumps from the volunteers. 
In Section \ref{sec:detections_hsc} we describe a first test of transferability of the object detection models after we applied the models `out-of-the-box', i.e. without any additional training, on the HSC imaging data. 

Unless otherwise indicated, all clump candidate detections were made with models developed on the full training set (run 20) of the GZCS sample and for detection scores $\geq 0.3$.

\subsection{Clump candidate detection - GZCS images}\label{sec:detections_gzcs}
\subsubsection{Visual comparison}
\begin{figure*}
    \centering
    \includegraphics[width=1\textwidth]{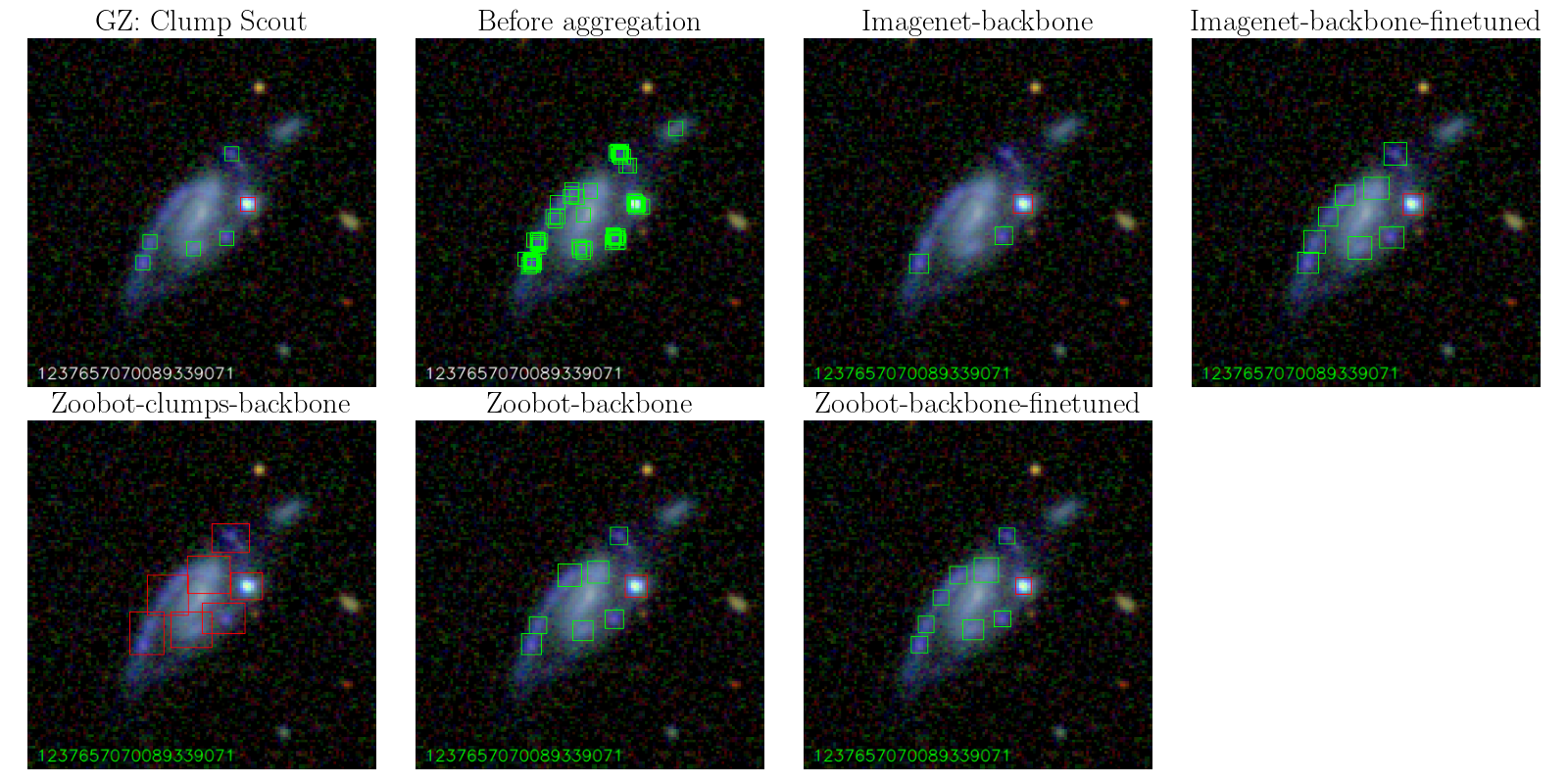}
    \caption[Clump candidates for a SDSS galaxy for all models.]{Clump candidates for a SDSS galaxy for all models. From left to right, top row: GZCS volunteers' labels, original GZCS volunteers' labels before the aggregation process, \textit{Imagenet-backbone} and \textit{Imagenet-backbone-finetuned}. From left to right, bottom row: \textit{Zoobot-clumps-backbone}, \textit{Zoobot-backbone} and \textit{Zoobot-backbone-finetuned}. Normal clumps are marked with green boxes, odd or unusual clumps with red boxes. The images are labelled with the SDSS-DR7 object number of the host galaxy. Detection score threshold is $\geq 0.3$.}
    \label{fig:sdss_example}
\end{figure*}

We first compared the detections by the five FRCNN models visually to each other and to the GZCS volunteers' labels. Figure \ref{fig:sdss_example} shows the differences in detection performance (Section \ref{sec:performance}) for one example galaxy.

This example illustrates the general observations we made while the resulting detections went through a visual vetting process. \textit{Zoobot-clumps-backbone} tends to produce far bigger bounding boxes whereas the model \textit{Imagenet-backbone} generally detects fewer clump instances than the other models, which explains the relatively low completeness for \textit{Imagenet-backbone}.

We observed that the models often predict additional clump instances which were not marked by the volunteers. These can be non-blue objects like foreground stars but in many cases these detections show typical visual clump characteristics and were not marked by a high enough number of volunteers for a consensus label. This can be seen from the second image in in Figure \ref{fig:sdss_example}, which shows the original GZCS volunteers' annotations before the aggregation process \citep[see Section \ref{sec:gzcs} and][]{Dickinson2022}.

Figures \ref{fig:sdss_examples1}, \ref{fig:sdss_examples2} and \ref{fig:sdss_examples3} in the Appendix provide additional comparisons of SDSS galaxy examples.

\subsubsection{Comparison to the GZCS sample}
We determined the number of detected normal and odd clumps predicted by the FRCNN models and the clump per galaxy ratios (Table \ref{tab:clumps_per_galaxy}). 

The clump predictions vary considerably between the models. For example, the models \textit{Zoobot-clumps-backbone} and \textit{Imagenet-backbone-finetuned} predict far more clumps than the other models and what has been originally annotated by the volunteers. These models detect clumps in almost all galaxies from the GZCS set, whereas the models \textit{Zoobot-backbone} and \textit{Zoobot-backbone-finetuned} both detect $\sim 15\%$ fewer clumpy galaxies. The model \textit{Imagenet-backbone} predicts the lowest number of clumps in only $\sim 60\%$ of the GZCS galaxies.

These differences result not only from normal clump candidates but also from the number of detected odd clump candidates. While both models based on the unmodified \textit{Zoobot} feature extractor, detect a similar number of normal clumps, they detect fewer odd clumps compared to the number marked by the GZCS volunteers. We also observed, that \textit{Imagenet-backbone-finetuned} detects $\sim 60\%$ more normal clumps, whereas a model like \textit{Zoobot-clumps-backbone} predicts more than five times the number of odd clumps annotated in the GZCS sample.
\begin{table*}
	\centering
	\caption[Clumpy galaxies, clumps and clumps per galaxy ratios for the final models.]{Clumpy galaxies, clumps and clumps per galaxy ratios for the final models per $u$-band clump/galaxy flux ratio for a detection threshold of $\geq 0.3$.}
        \label{tab:clumps_per_galaxy}
        \begin{tabular}{llrrrrrrr}
		\hline
		Model & $u$-band flux ratio & Clumpy galaxies & \multicolumn{2}{c}{Clumps, all} & \multicolumn{2}{c}{Clumps, normal} & \multicolumn{2}{c}{Clumps, odd} \\
		\hline
		    & $f_\text{u, clump}/f_\text{u, galaxy}$ & count & count & average & count & average & count & average \\
        \hline
                 GZ: Clump Scout                 & all         & 18,772 & 39,745 & 2.12 & 29,619 & 1.89 & 10,126 & 1.20 \\ 
                                                 & $\geq 0.03$ &  7,329 & 10,106 & 1.38 &  7,464 & 1.34 &  2,642 & 1.07 \\
                                                 & $\geq 0.08$ &  2,810 &  3,406 & 1.21 &  2,007 & 1.22 &  1,399 & 1.05 \\ \hline
        \textit{Imagenet-backbone}                & all         & 11,191 & 17,804 & 1.59 & 16,223 & 1.57 &  1,581 & 1.05 \\
                                                 & $\geq 0.03$ &  4,427 &  5,359 & 1.21 &  4,382 & 1.22 &    977 & 1.03 \\
                                                 & $\geq 0.08$ &  1,796 &  2,015 & 1.12 &  1,277 & 1.14 &    738 & 1.03 \\  \hline
        \textit{Imagenet-backbone-finetuned}     & all         & 17,836 & 48,659 & 2.73 & 46,271 & 2.66 &  2,388 & 1.06 \\ 
                                                 & $\geq 0.03$ &  8,300 & 12,698 & 1.53 & 10,995 & 1.54 &  1,703 & 1.04 \\
                                                 & $\geq 0.08$ &  3,085 &  3,882 & 1.26 &  2,592 & 1.33 &  1,290 & 1.04 \\ \hline
        \textit{Zoobot-clumps-backbone}             & all         & 17,491 & 69,966 & 4.00 & 14,625 & 1.61 & 55,341 & 3.95 \\ 
                                                 & $\geq 0.03$ &  8,472 & 15,149 & 1.79 &  2,802 & 1.23 & 12,347 & 1.76 \\
                                                 & $\geq 0.08$ &  2,951 &  4,027 & 1.36 &    605 & 1.20 &  3,422 & 1.32 \\ \hline
        \textit{Zoobot-backbone}            & all         & 15,858 & 31,199 & 1.97 & 29,018 & 1.93 &  2,181 & 1.05 \\ 
                                                 & $\geq 0.03$ &  7,210 &  9,712 & 1.35 &  7,913 & 1.35 &  1,799 & 1.05 \\
                                                 & $\geq 0.08$ &  2,752 &  3,272 & 1.19 &  1,879 & 1.23 &  1,393 & 1.04 \\ \hline
        \textit{Zoobot-backbone-finetuned} & all         & 15,937 & 32,923 & 2.07 & 30,301 & 2.03 &  2,622 & 1.07 \\ 
                                                 & $\geq 0.03$ &  7,122 &  9,684 & 1.36 &  7,800 & 1.36 &  1,884 & 1.06 \\
                                                 & $\geq 0.08$ &  2,821 &  3,330 & 1.18 &  1,871 & 1.22 &  1,459 & 1.05 \\
        \hline
	\end{tabular}
\end{table*}

Comparing the number of clumps per galaxy (Table \ref{tab:clumps_per_galaxy}), we noted that the models \textit{Zoobot-backbone} and \textit{{Zoobot-backbone-finetuned}} are closest to the distribution of the GZCS sample but tend to find slightly more normal clumps per galaxy. 

\begin{figure}
    \centering
    \subfloat[\centering Normal clump candidates.\label{fig:hist_redshift_clumps}]{{\includegraphics[width=0.8\columnwidth]{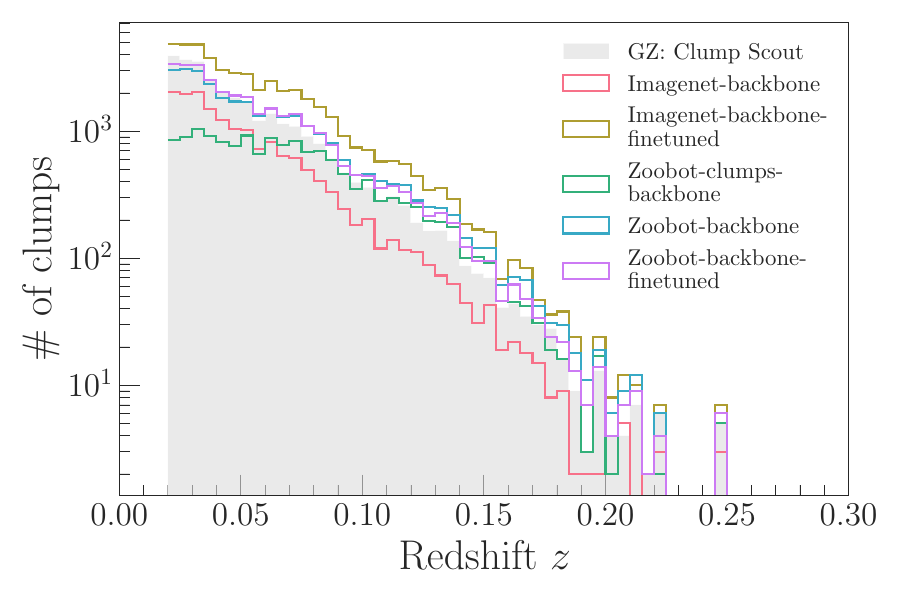} }}
    \\
    \subfloat[\centering Odd clump candidates.\label{fig:hist_redshift_odd_clumps}]{{\includegraphics[width=0.8\columnwidth]{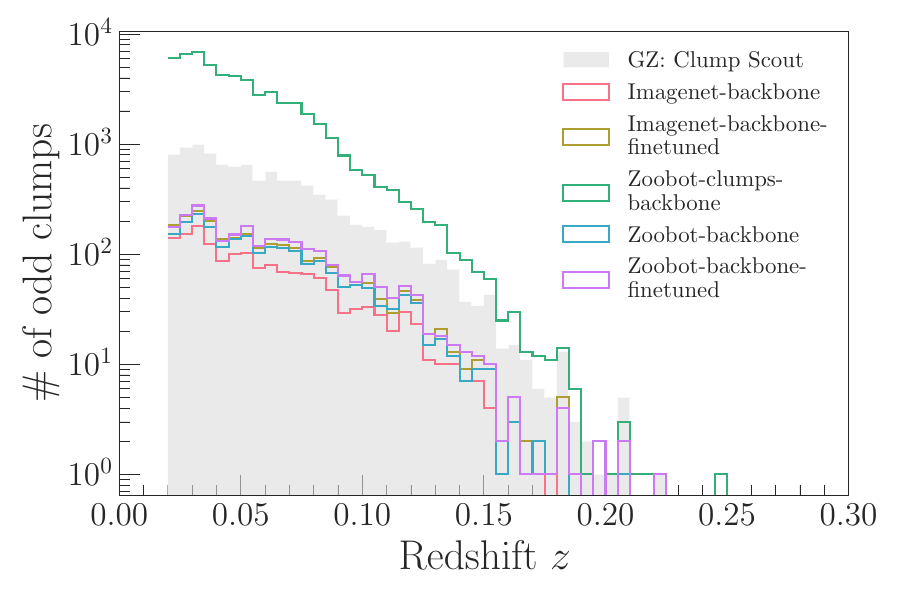} }}
    \caption[Histograms of detected clump candidates per redshift bin by the five different models.]{Histograms of detected clump candidates per redshift bin by the five different models, separately for normal and odd clumps. Detection score threshold is $\geq 0.3$. The underlying grey distributions are from GZCS volunteers' labels.}
\end{figure}

Considering the redshift of the host galaxies, the normal clump candidates predicted by \textit{Zoobot-backbone} and \textit{{Zoobot-backbone-finetuned}} are close to the GZCS-distribution of the normal clumps. Although these models do detect fewer clumps for host galaxies at redshift $0.02 - 0.04$, the other models differ substantially from the GZCS distribution (Figure \ref{fig:hist_redshift_clumps}). Specifically, the model \textit{Imagenet-backbone-finetuned} predicts far more clump candidates on the full redshift range of our sample of host galaxies, whereas the model \textit{Imagenet-backbone} detects fewer clumps compared to the number of clumps resulting from the GZCS project. We also observe that the number of predicted normal clumps from the model \textit{Zoobot-clumps-backbone} is far lower for redshifts $\lesssim 0.08$ but much higher for predicted odd clumps at all redshifts (Figure \ref{fig:hist_redshift_odd_clumps}). 

We also applied the same thresholds for the clump to host galaxy $u$-band flux ratio as we did for assessing the detection performance of the models (see Section \ref{sec:performance_flux_ratio}). The number of detected clumps with a $u$-band flux ratio of $\geq 3\%$ ($\geq 8\%$) reduces to $\sim 20-30\%$ ($\sim 6-11\%$). Again, the normal clump candidates predicted by the models \textit{Zoobot-backbone} and \textit{Zoobot-backbone-finetuned} are close to the GZCS-distribution of the normal clumps if either of the flux ratio thresholds are applied, but now the number of detected odd clumps is also similar to the number of odd clumps identified from the GZCS project (Table \ref{tab:clumps_per_galaxy}).

\begin{figure*}
    \centering
    \subfloat[\centering \textit{Imagenet-backbone}.\label{fig:clump_colour_normal1}]{{\includegraphics[width=0.33\textwidth]{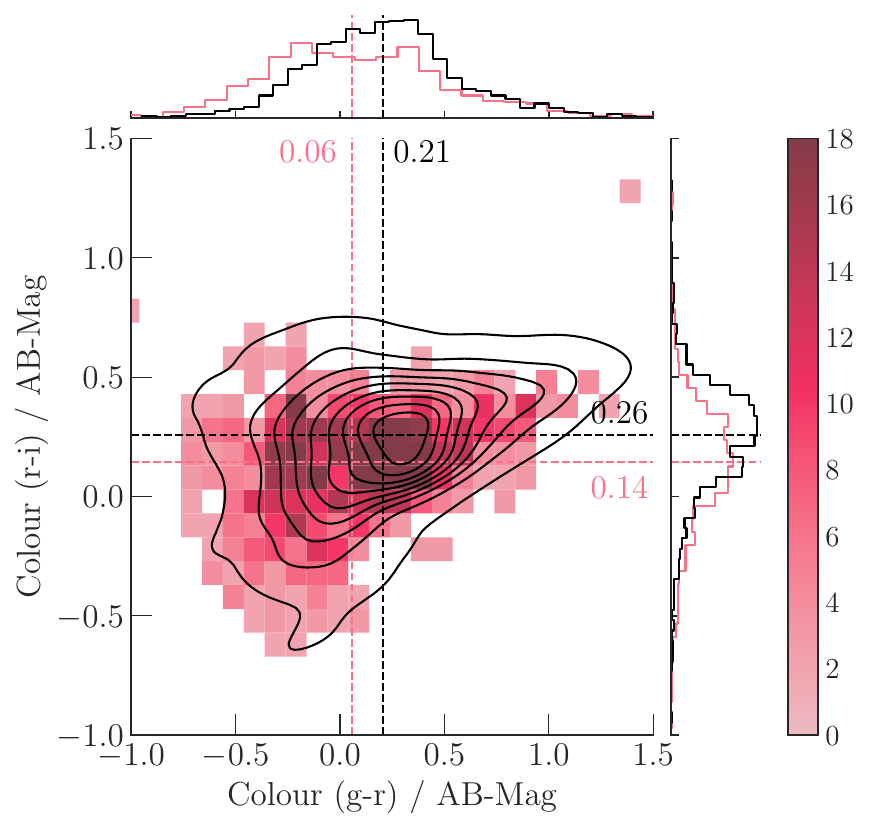} }}
    \subfloat[\centering \textit{Imagenet-backbone-finetuned}.\label{fig:clump_colour_normal2}]{{\includegraphics[width=0.33\textwidth]{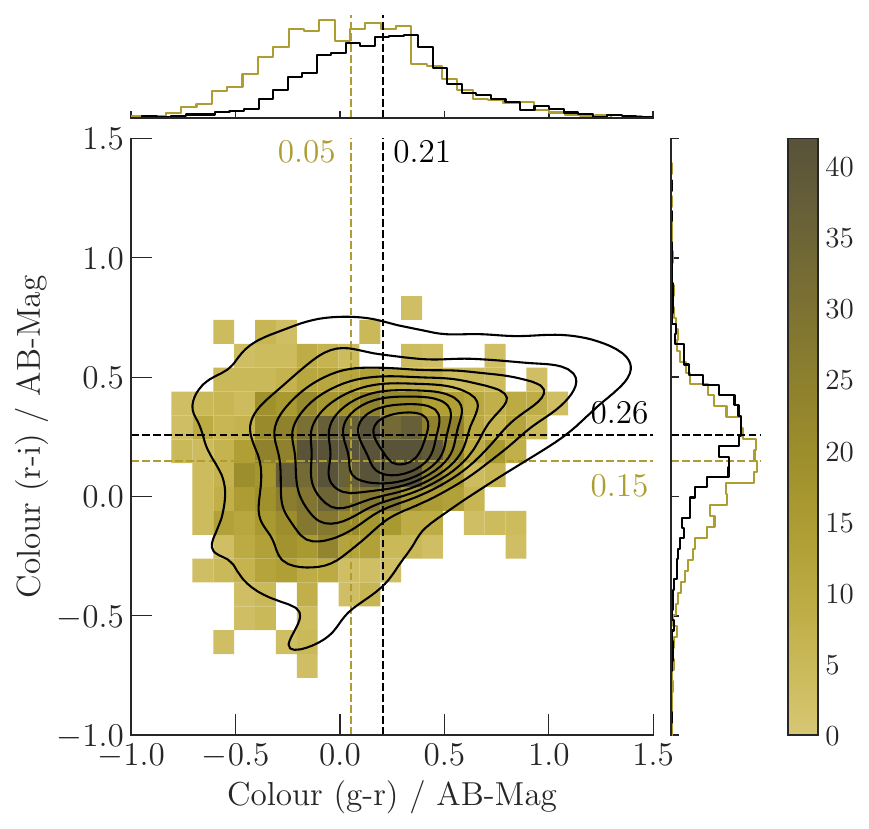} }}
    \subfloat[\centering \textit{Zoobot-clumps-backbone}.\label{fig:clump_colour_normal3}]{{\includegraphics[width=0.33\textwidth]{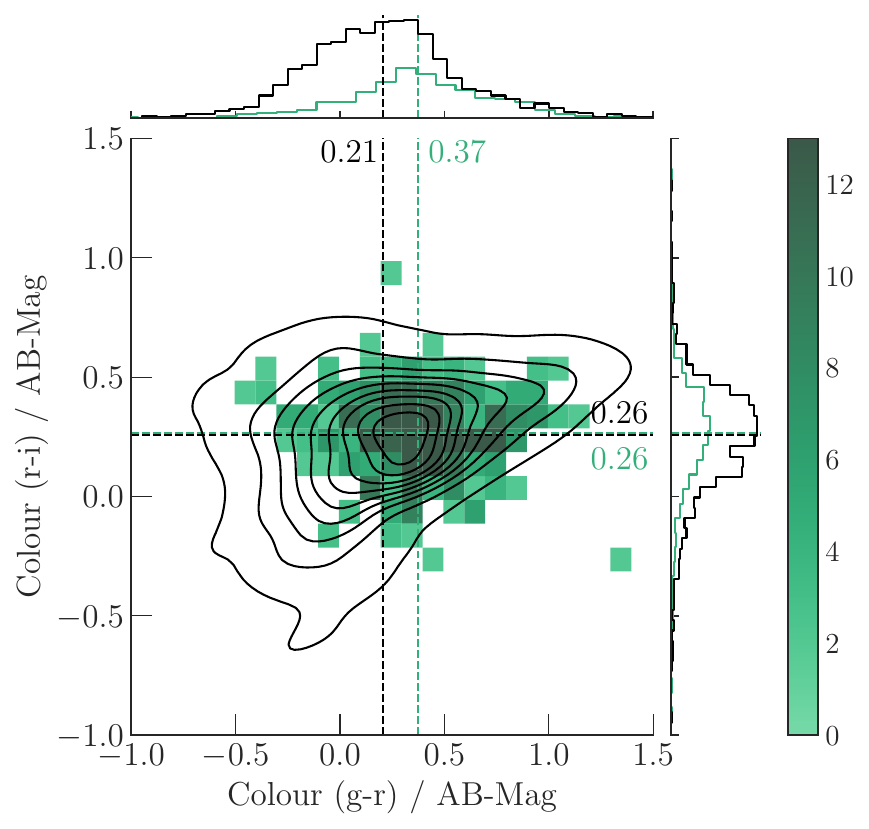} }}
    \\
    \subfloat[\centering \textit{Zoobot-backbone}.\label{fig:clump_colour_normal4}]{{\includegraphics[width=0.33\textwidth]{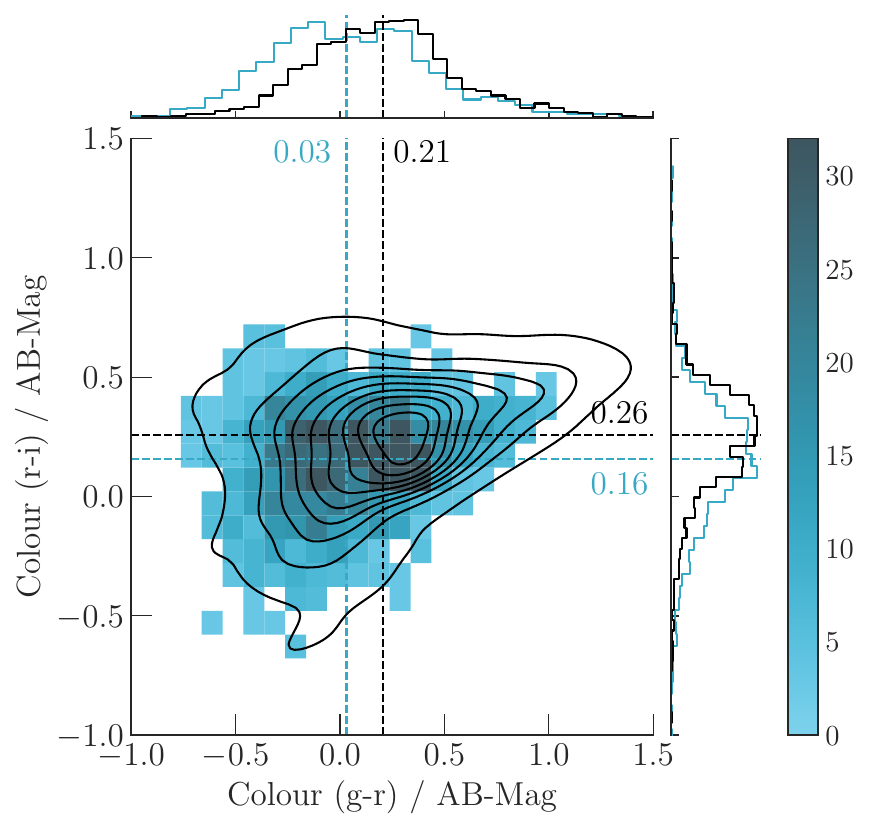} }}
    \subfloat[\centering \textit{Zoobot-backbone-finetuned}.\label{fig:clump_colour_normal5}]{{\includegraphics[width=0.33\textwidth]{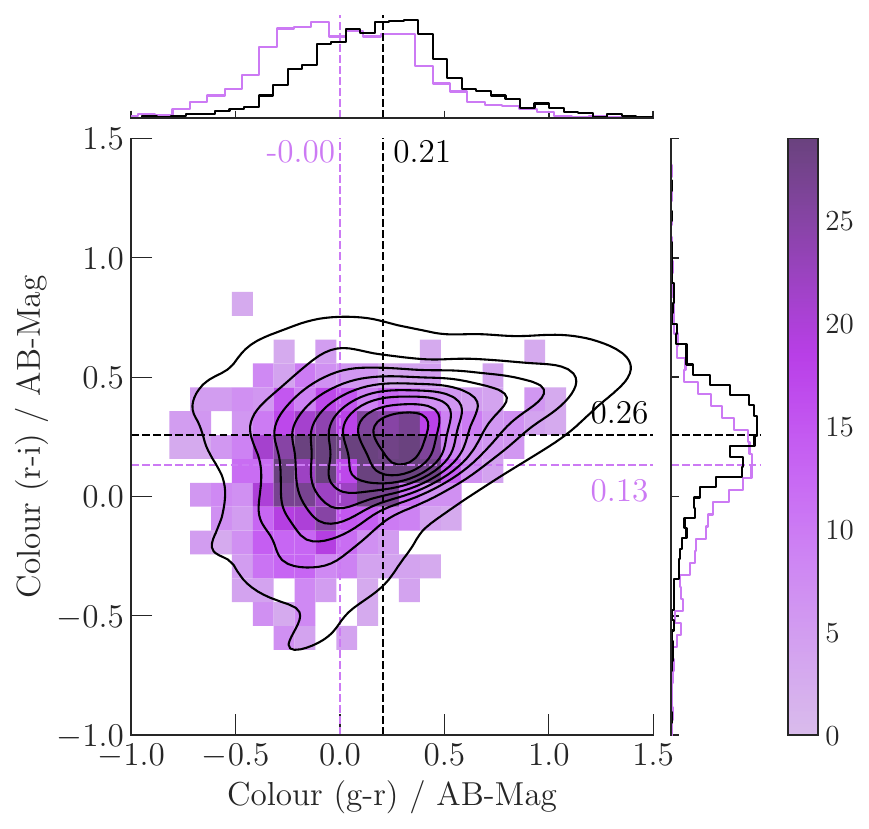} }}
    \caption[Colour-colour diagrams for normal clump candidates.]{Colour-colour diagrams for normal clump candidates with a $u$-band clump/galaxy flux ratio of $\geq 0.08$ for each of the five models. The colour bars indicate the specific counts of normal clumps for each $(g-r)/(r-i)$ bin (with bin size of 0.1). The small histograms on the top and right side of the plots are showing the distributions of $(g-r)$ and $(r-i)$ separately. The colour-colour distribution of the clumps annotated by the GZCS volunteers are overlaid with black contours and small histograms, also in black. Vertical and horizontal lines mark the median colour determined for the normal clumps, which are annotated with the corresponding values.}
    \label{fig:clump_colour_normal}
\end{figure*}

Plotted on a $(g-r)$/$(r-i)$ colour-colour diagram, the normal clump candidates predicted by most of the models (Figure \ref{fig:clump_colour_normal}) tend to be bluer than the clumps marked by the volunteers from GZCS. Only the normal clumps detected by \textit{Zoobot-clumps-backbone} are redder in comparison to the GZCS sample (Figure \ref{fig:clump_colour_normal3}).

\begin{figure*}
    \centering
    \subfloat[\centering \textit{Imagenet-backbone}.\label{fig:clump_colour_odd1}]{{\includegraphics[width=0.33\textwidth]{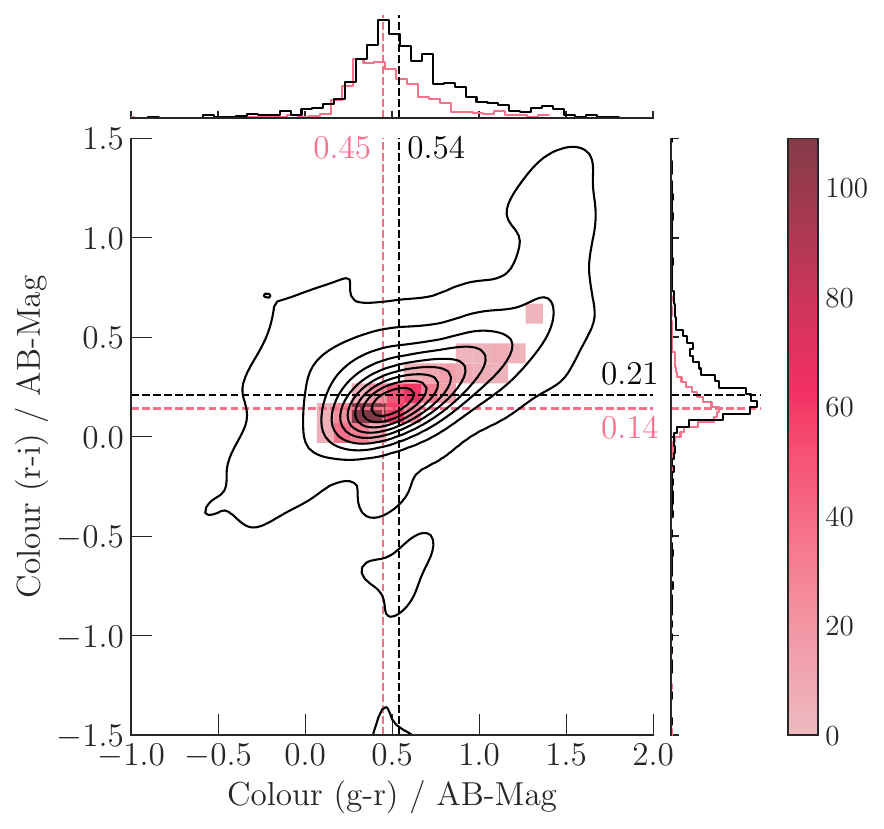} }}
    \subfloat[\centering \textit{Imagenet-backbone-finetuned}.\label{fig:clump_colour_odd2}]{{\includegraphics[width=0.33\textwidth]{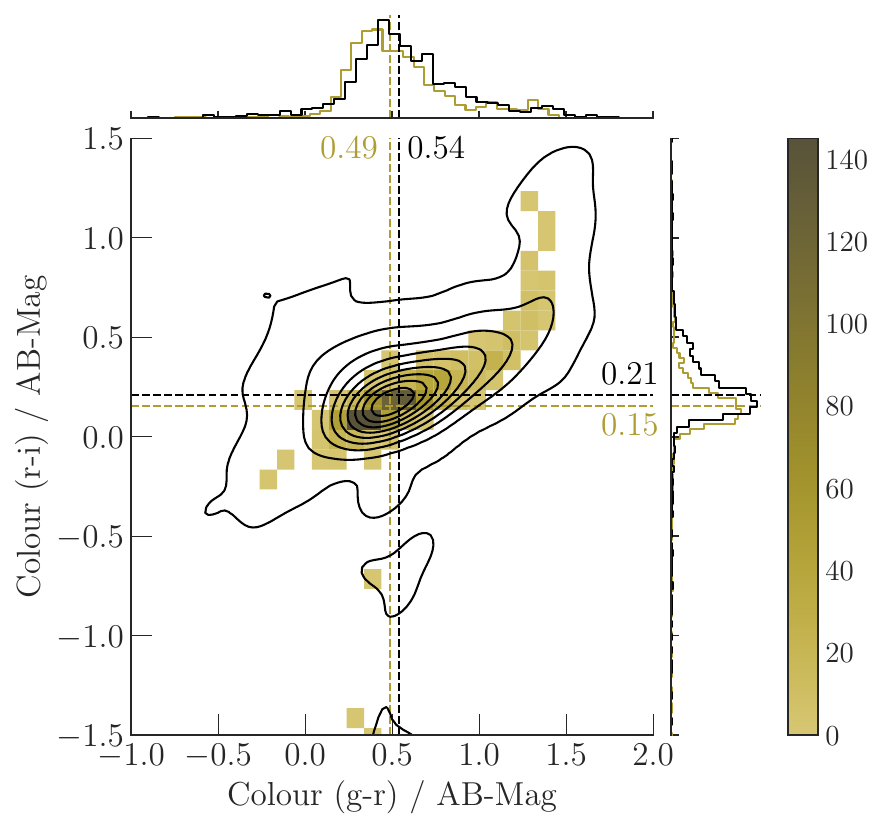} }}
    \subfloat[\centering \textit{Zoobot-clumps-backbone}.\label{fig:clump_colour_odd3}]{{\includegraphics[width=0.33\textwidth]{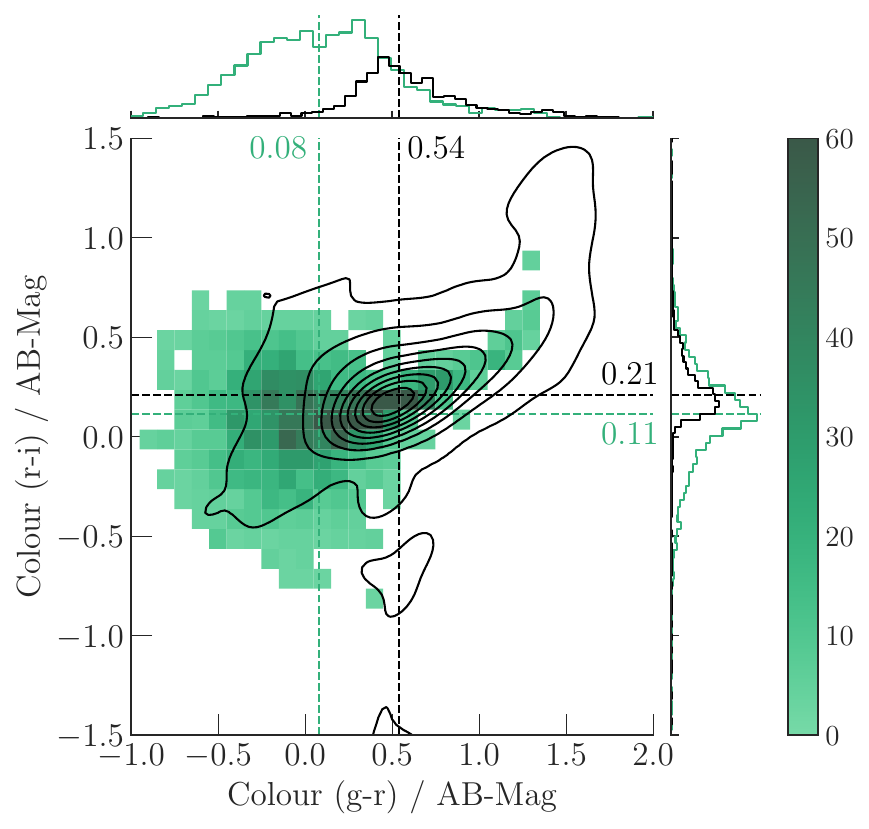} }}
    \\
    \subfloat[\centering \textit{Zoobot-backbone}.\label{fig:clump_colour_odd4}]{{\includegraphics[width=0.33\textwidth]{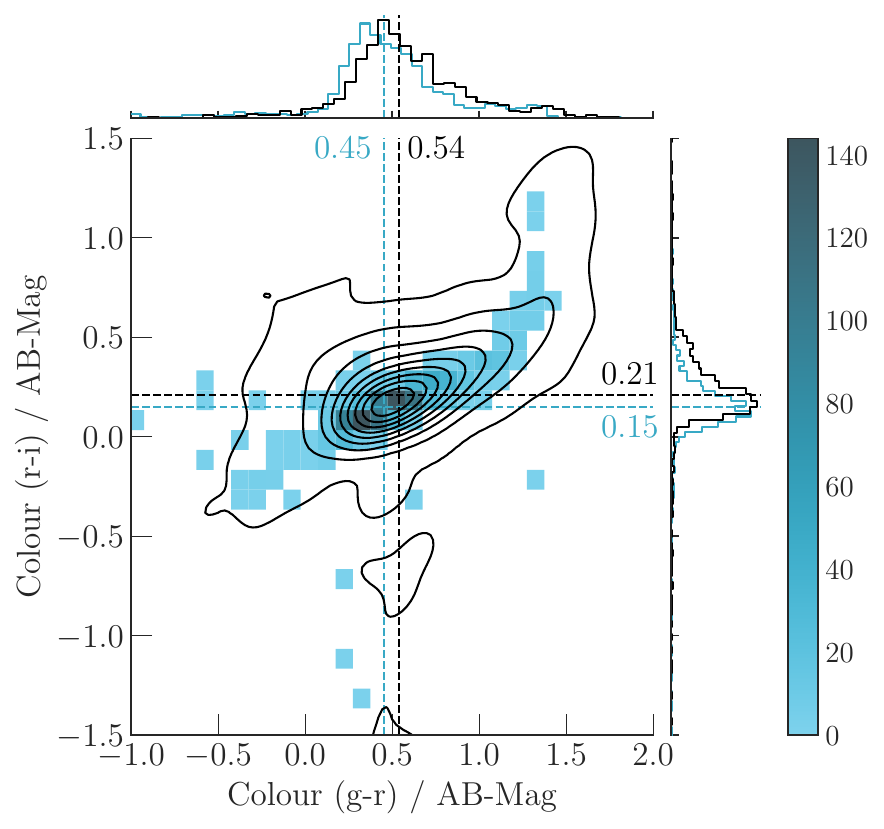} }}
    \subfloat[\centering \textit{Zoobot-backbone-finetuned}.\label{fig:clump_colour_odd5}]{{\includegraphics[width=0.33\textwidth]{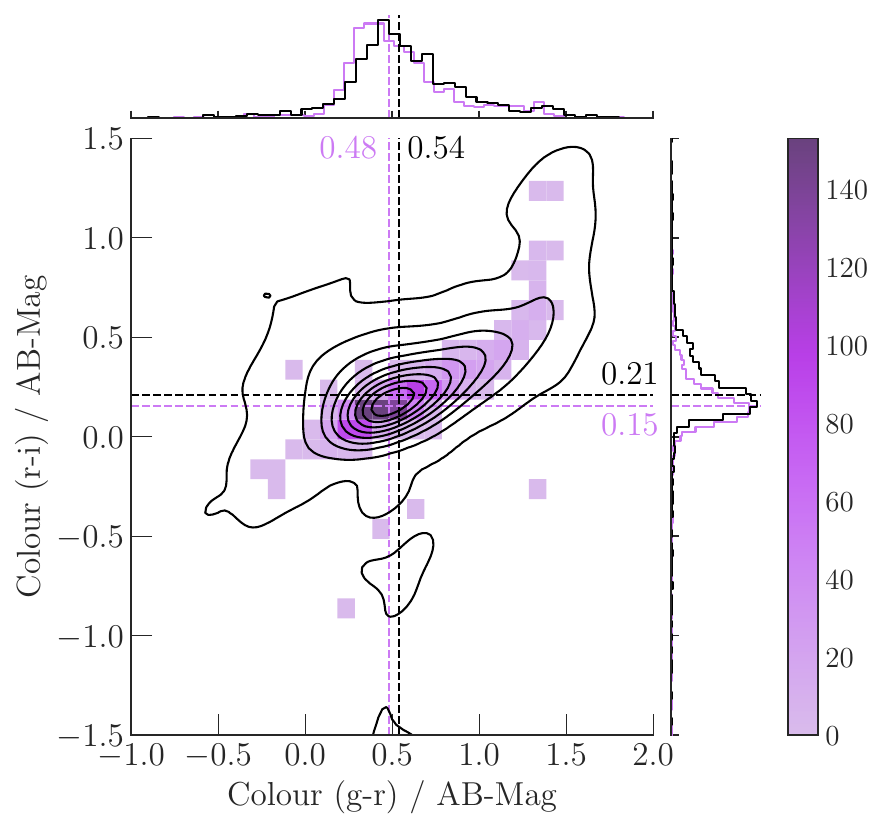} }}
    \caption[Colour-colour diagrams for odd clump candidates.]{Colour-colour diagrams for odd or unusual clump candidates with a $u$-band clump/galaxy flux ratio of $\geq 0.08$ for each of the five models. The colour bars indicate the specific counts of odd clumps for each $(g-r)/(r-i)$ bin (with bin size of 0.1). The small histograms on the top and right side of the plots are showing the distributions of $(g-r)$ and $(r-i)$ separately. The colour-colour distribution of the odd clumps annotated by the GZCS volunteers are overlaid with black contours and small histograms, also in black. Vertical and horizontal lines mark the median colour determined for the odd clumps, which are annotated with the corresponding values.}
    \label{fig:clump_colour_odd}
\end{figure*}

The predictions from this model also differ notably for odd clumps (Figure \ref{fig:clump_colour_odd}). \textit{Zoobot-clumps-backbone} not only detects far more odd clumps, but they also tend to be bluer and spread over a wider $(g-r)$ and $(r-i)$ range than the GZCS-distribution. This is in contrast to the odd clump candidate detections made by the other models which are closely resembling the $(g-r)$ and $(r-i)$ colour distribution from the GZCS sample and are following a tight locus of the colour-colour space, indicative of foreground stars. 

To see whether the host galaxy has an effect on our model predictions, we further compared the host galaxies' specific star-formation rates (sSFR), stellar masses and redshifts for which our models did detect normal clumps to the host galaxies for which the volunteers from GZCS have marked normal clumps. We obtained stellar masses $M_\star$ and sSFR for the host galaxies from the SDSS DR7 MPA-JHU value-added catalogue \citep{Kauffmann2003,Brinchmann2004}.
We caution the reader that these figures are only intended to compare the performance characteristics of our models with the performance characteristics of the GZCS volunteers. We make no claims about the true distribution of clumpy galaxies as a function of redshift, stellar mass or sSFR.

\begin{figure*}
    \centering
    \subfloat[\centering \textit{Imagenet-backbone}.\label{fig:clump_ssfr_z1}]{{\includegraphics[width=0.33\textwidth]{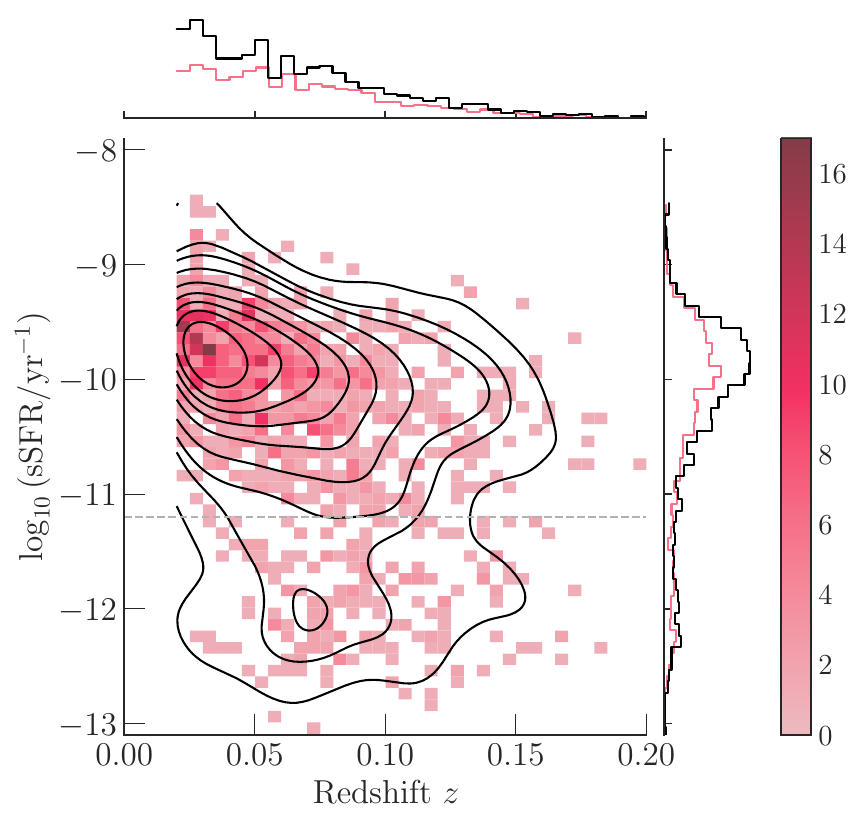} }}
    \subfloat[\centering \textit{Imagenet-backbone-finetuned}.\label{fig:clump_ssfr_z2}]{{\includegraphics[width=0.33\textwidth]{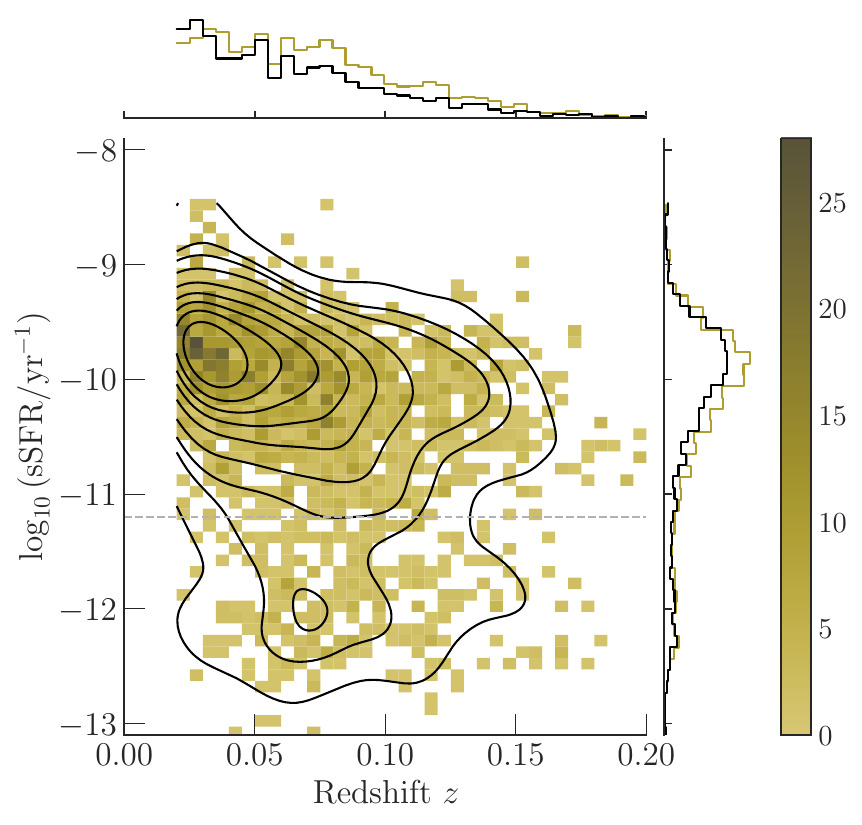} }}
    \subfloat[\centering \textit{Zoobot-clumps-backbone}.\label{fig:clump_ssfr_z3}]{{\includegraphics[width=0.33\textwidth]{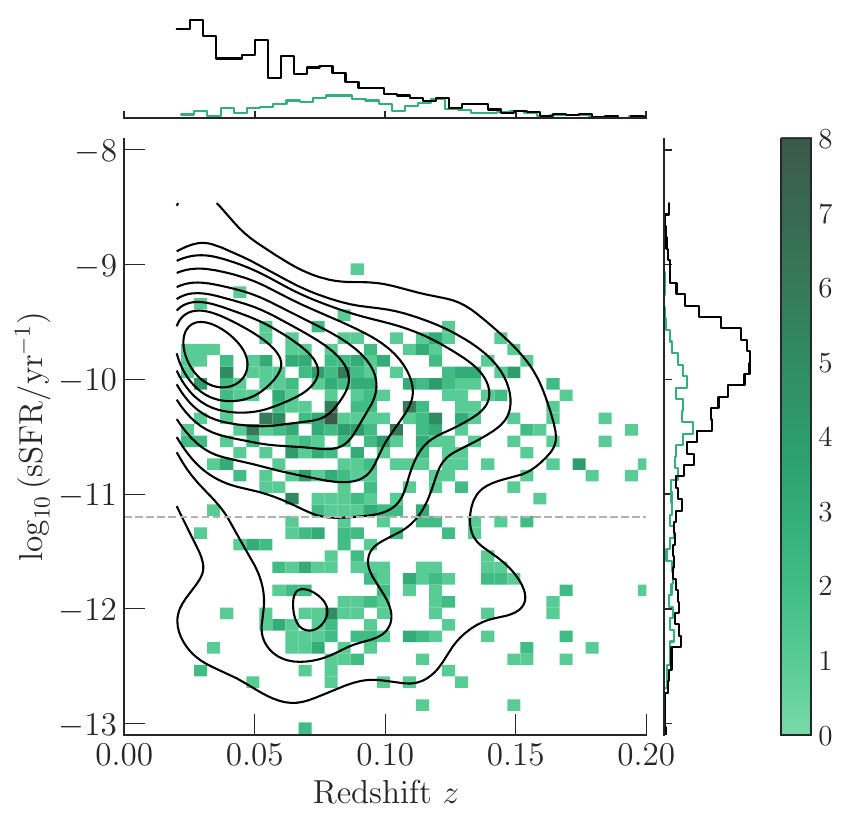} }}
    \\
    \subfloat[\centering \textit{Zoobot-backbone}.\label{fig:clump_ssfr_z4}]{{\includegraphics[width=0.33\textwidth]{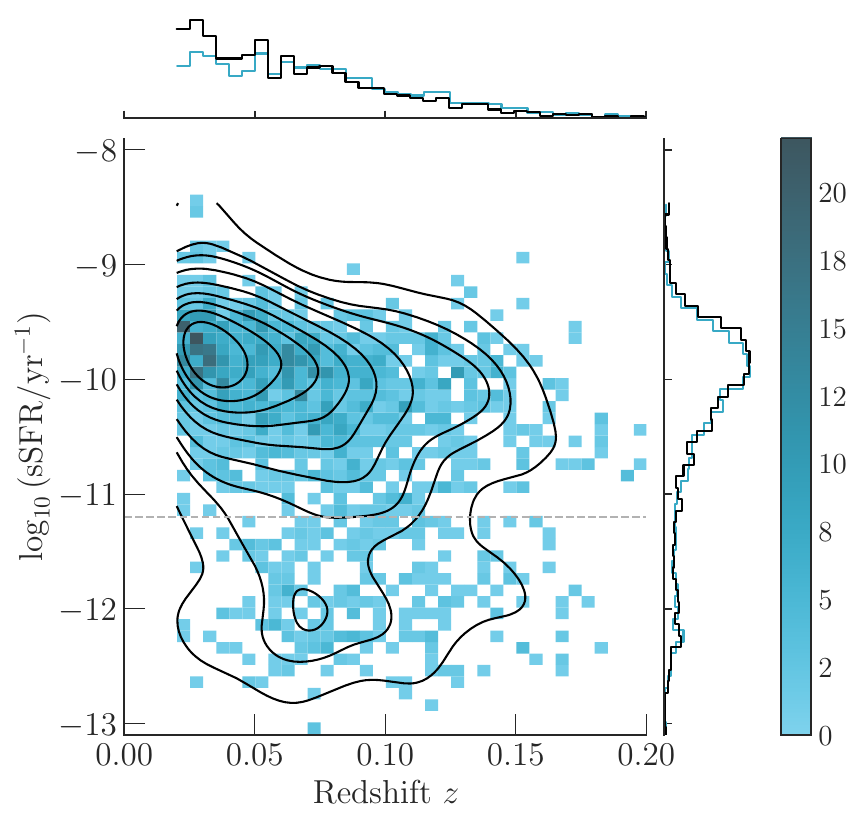} }}
    \subfloat[\centering \textit{Zoobot-backbone-finetuned}.\label{fig:clump_ssfr_z5}]{{\includegraphics[width=0.33\textwidth]{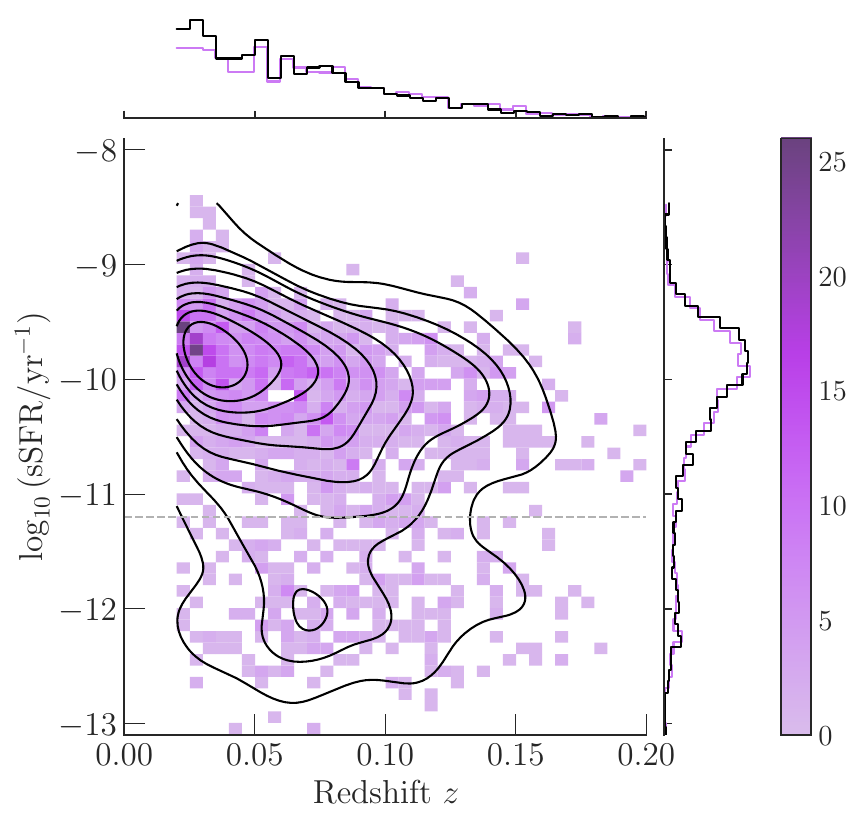} }}
    \caption[Specific star-formation rate vs. redshift of host galaxy for normal clump candidates.]{Specific star-formation rate (sSFR) vs. redshift $z$ of the host galaxy for normal clump candidates with a $u$-band clump/galaxy flux ratio of $\geq 0.08$. The colour bars indicate the specific counts of clumps for each $\log_{10}(\mathrm{sSFR})$ and $z$ bin (with bin sizes of 0.1 and 0.005, respectively). The small histograms on the top and right side of the plots are showing the distributions of $\log_{10}(\mathrm{sSFR})$ and $z$ separately. The distribution of the normal clumps annotated by the GZCS volunteers are overlaid using black contours and small histograms, also in black. The horizontal dashed line marks the separation between star-forming and quiescent galaxies at $\log_{10}(\mathrm{sSFR}) = -11.2$.}
    \label{fig:clump_ssfr_z}
\end{figure*}

Apart from \textit{Zoobot-clumps-backbone}, the distributions of the sSFR of the host galaxies containing predicted clumps are similar to those of the GZCS galaxies (Figure \ref{fig:clump_ssfr_z}). The majority of the normal clump candidates are found in star-forming ($\log_{10}(\mathrm{sSFR}) \gtrsim -11.2$) galaxies. At the higher end of the redshift range of our sample galaxies ($z > 0.05$), the distributions of the predicted clumps still resemble that of the GZCS-distribution for the models \textit{Zoobot-backbone} and \textit{Zoobot-backbone-finetuned} (Figures \ref{fig:clump_ssfr_z4} and \ref{fig:clump_ssfr_z5}). However, the model predictions from \textit{Imagenet-backbone} and \textit{Imagenet-backbone-finetuned} are notably different compared to the GZCS-galaxies for $\log_{10}(\mathrm{sSFR}) \gtrsim -11$ (Figures \ref{fig:clump_ssfr_z1} and \ref{fig:clump_ssfr_z2}).

\begin{figure*}
    \centering
    \subfloat[\centering \textit{Imagenet-backbone}.\label{fig:clump_mass_z1}]{{\includegraphics[width=0.33\textwidth]{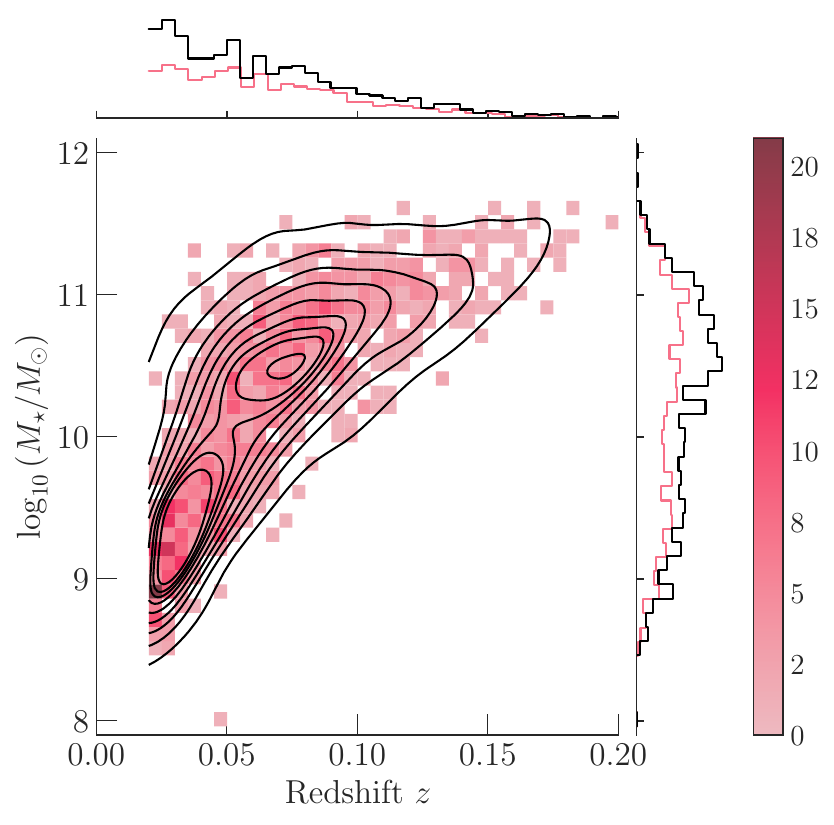} }}
    \subfloat[\centering \textit{Imagenet-backbone-finetuned}.\label{fig:clump_mass_z2}]{{\includegraphics[width=0.33\textwidth]{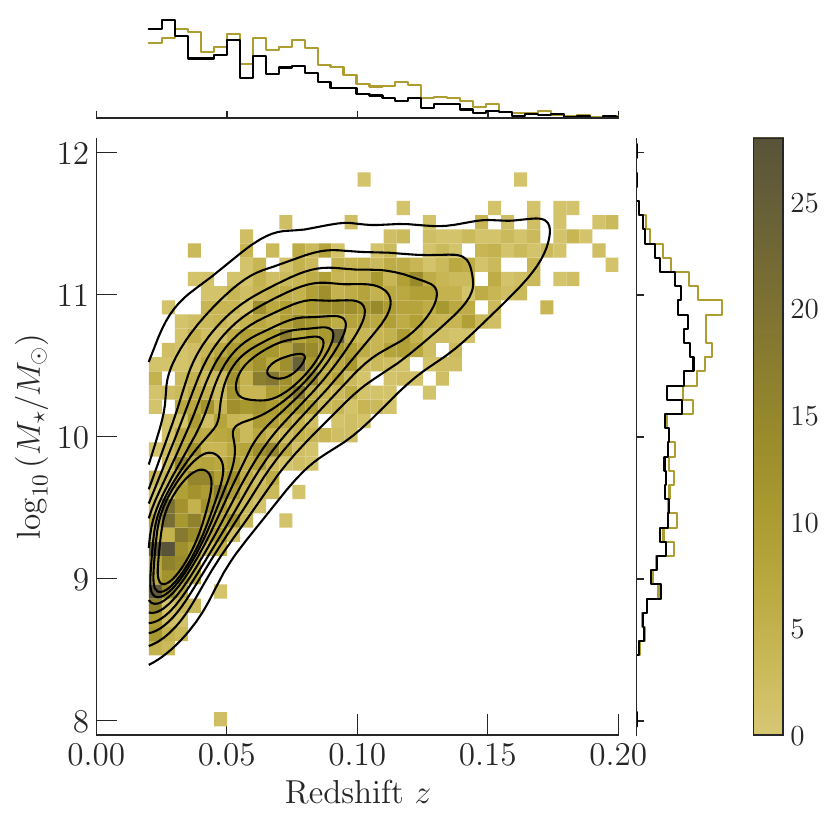} }}
    \subfloat[\centering \textit{Zoobot-clumps-backbone}.\label{fig:clump_mass_z3}]{{\includegraphics[width=0.33\textwidth]{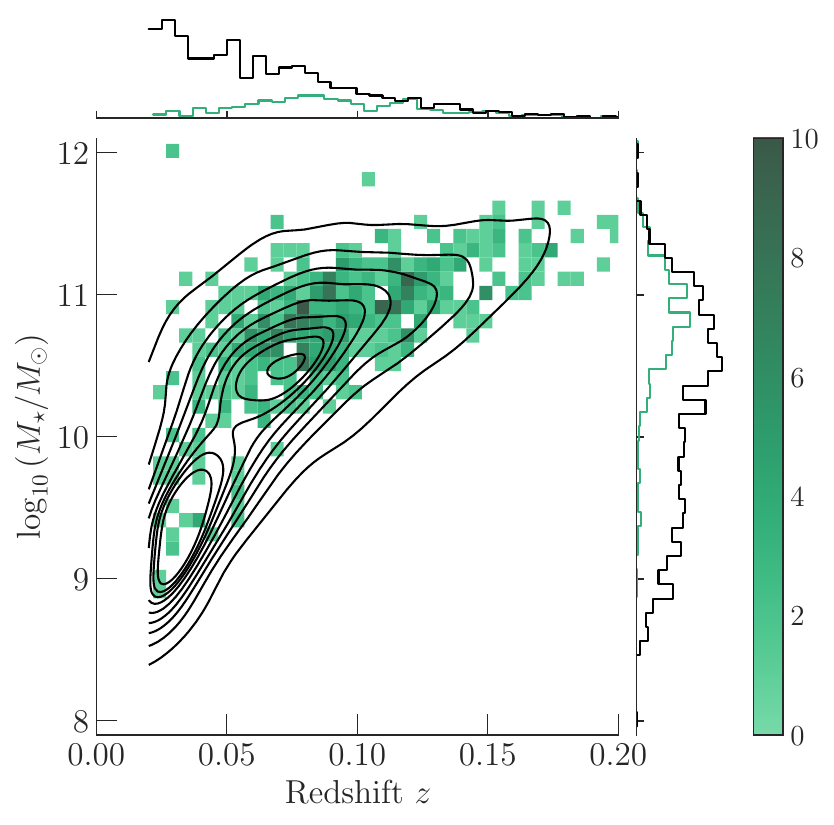} }}
    \\
    \subfloat[\centering \textit{Zoobot-backbone}.\label{fig:clump_mass_z4}]{{\includegraphics[width=0.33\textwidth]{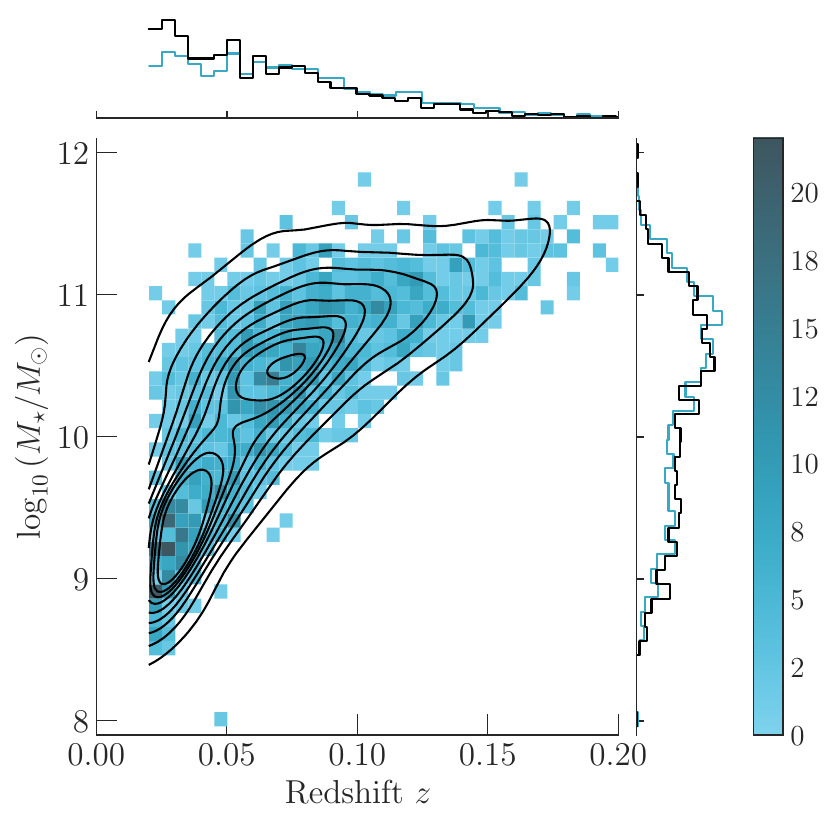} }}
    \subfloat[\centering \textit{Zoobot-backbone-finetuned}.\label{fig:clump_mass_z5}]{{\includegraphics[width=0.33\textwidth]{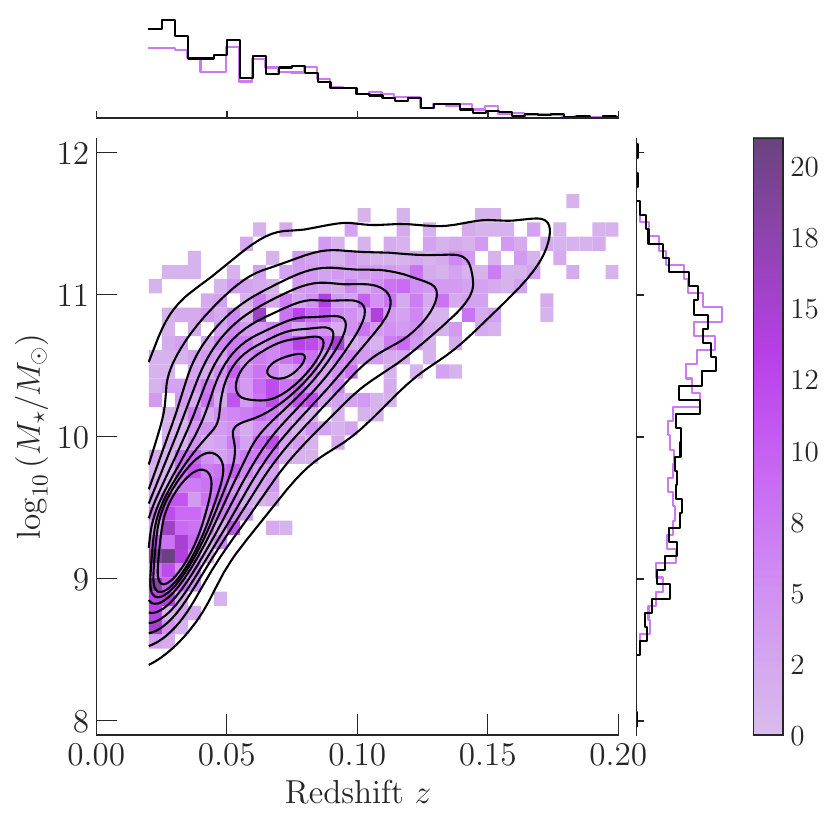} }}
    \caption[Stellar mass vs. redshift of host galaxy for normal clump candidates.]{Stellar mass $M_\star$ vs. redshift $z$ of the host galaxy for normal clump candidates with a $u$-band clump/galaxy flux ratio of $\geq 0.08$. The colour bars indicate the specific counts of clumps for each $\log_{10} (M_\star)$ and $z$ bin (with bin sizes of 0.1 and 0.005, respectively). The small histograms on the top and right side of the plots are showing the distributions of $\log_{10} (M_\star)$ and $z$ separately. The distribution of the normal clumps annotated by the GZCS volunteers are overlaid using black contours and small histograms, also in black.}
    \label{fig:clump_mass_z}
\end{figure*}

We also observed, that the model \textit{Imagenet-backbone} tends to predict fewer normal clumps in more massive galaxies ($\log_{10}(M_\star) \gtrsim 9.4$) for redshifts $z < 0.05$ (Figure \ref{fig:clump_mass_z1}), whereas \textit{Imagenet-backbone-finetuned} detects more normal clumps in comparison to the GZCS-distribution for galaxies with stellar masses of $\log_{10}(M_\star) \gtrsim 10.2$ (Figure \ref{fig:clump_mass_z2}). \textit{Zoobot-clumps-backbone}, on the other hand, produces normal clump candidates which are predominantly located in higher redshift galaxies with $\log_{10}(M_\star) \gtrsim 10.2$ (Figure \ref{fig:clump_mass_z3}). The \textit{Zoobot-backbone} and \textit{Zoobot-backbone-finetuned} models closely resemble the host galaxy distribution with clumps labelled by the GZCS volunteers (Figures \ref{fig:clump_mass_z4} and \ref{fig:clump_mass_z5}).

\subsubsection{Clump catalogue release and use}
We applied the model \textit{Zoobot-backbone} on the 53,613 galaxies from the original GZCS set (Table \ref{tab:sample_sizes}) and released a catalogue containing the detected clump candidates and their estimated properties along with this paper. The catalogue contains normal and odd clump candidates for a detection score $\geq 0.3$ together with measured fluxes and magnitudes for each of the $ugriz$-filter bands from SDSS. Table \ref{tab:catalogue} describes the columns in compact form.
\begin{table}
	\centering
	\caption[Description of the clump catalogue for the GZCS galaxies.]{Description of the clump catalogue for the GZCS galaxies.} 
        \label{tab:catalogue}
	\begin{tabular}{llll}
		\hline
		Columns & Property & Units & Source \\
		\hline
            1-2   & Object IDs & & SDSS \\
            3-8   & Clump candidate detection score, & & \\
                  & label and coordinates & & \\
            9-18  & Clump $ugriz$-fluxes & Jy & \\
                  & and errors & & \\
            19-33 & Clump $ugriz$-magnitudes & AB mag & \\
                  & and corrections & & \\
            34-37 & Clump colours & & \\
            38-40 & Est. clump/galaxy near-UV & & \\
                  & flux ratio ($u$-band) & & \\
            41-45 & Host galaxy coordinates, & & SDSS \\
                  & redshift and axis ratio & & \\
            46    & Host galaxy stellar mass (log) & $\log_{10}(M_\odot)$ & SDSS \\
            47    & Host galaxy sSFR (log) & $\log_{10}(\mathrm{yr}^{-1})$ & SDSS \\
            48-52 & Host galaxy $ugriz$-fluxes & Jy & SDSS \\
            53-67 & Host galaxy $ugriz$-magnitudes & AB mag & SDSS \\
                  & and corrections & & \\
		\hline
	\end{tabular}
\end{table}

When using the catalogue, we recommend excluding entries with the label `odd clumps' as these are very likely non-clump candidates (i.e. foreground stars, background galaxies or other point-like sources). Currently, our model framework does not include a classification score for the labels `normal clump' and `odd clump' that would allow for a more specific selection. Purity and completeness can be varied by filtering on the detection score, where a higher score threshold results in higher purity and vice versa (see Figure \ref{fig:recall_precision}).

We also caution the reader that we did not apply any survey completeness limits to this catalogue. Such completeness limits need to be applied for further scientific analysis of our sample. For an example, where galaxy stellar mass and redshift limits were applied, see \citet{Adams2022}.

\subsection{Clump candidate detection - HSC images}\label{sec:detections_hsc}
The Hyper Suprime-Cam Subaru Strategic Survey \citep[HSC SSP,][]{Aihara2017} partly overlaps with the SDSS footprint, but the wider aperture of the Subaru Telescope allows for much deeper imaging with the HSC and exposes many more morphological details. 

We did not train the models specifically for the HSC imaging data set or reproduce exactly the same image pre-processing as for the SDSS galaxy images (see Appendix \ref{sec:galaxy_images}). We note, however, that the \textit{Zoobot}-based feature extraction backbones had some existing `memory' learned from galaxy images with a higher spatial resolution than SDSS as such images have already been used in their pre-training as classifiers.

\begin{figure*}
    \centering
    \includegraphics[width=0.8\textwidth]{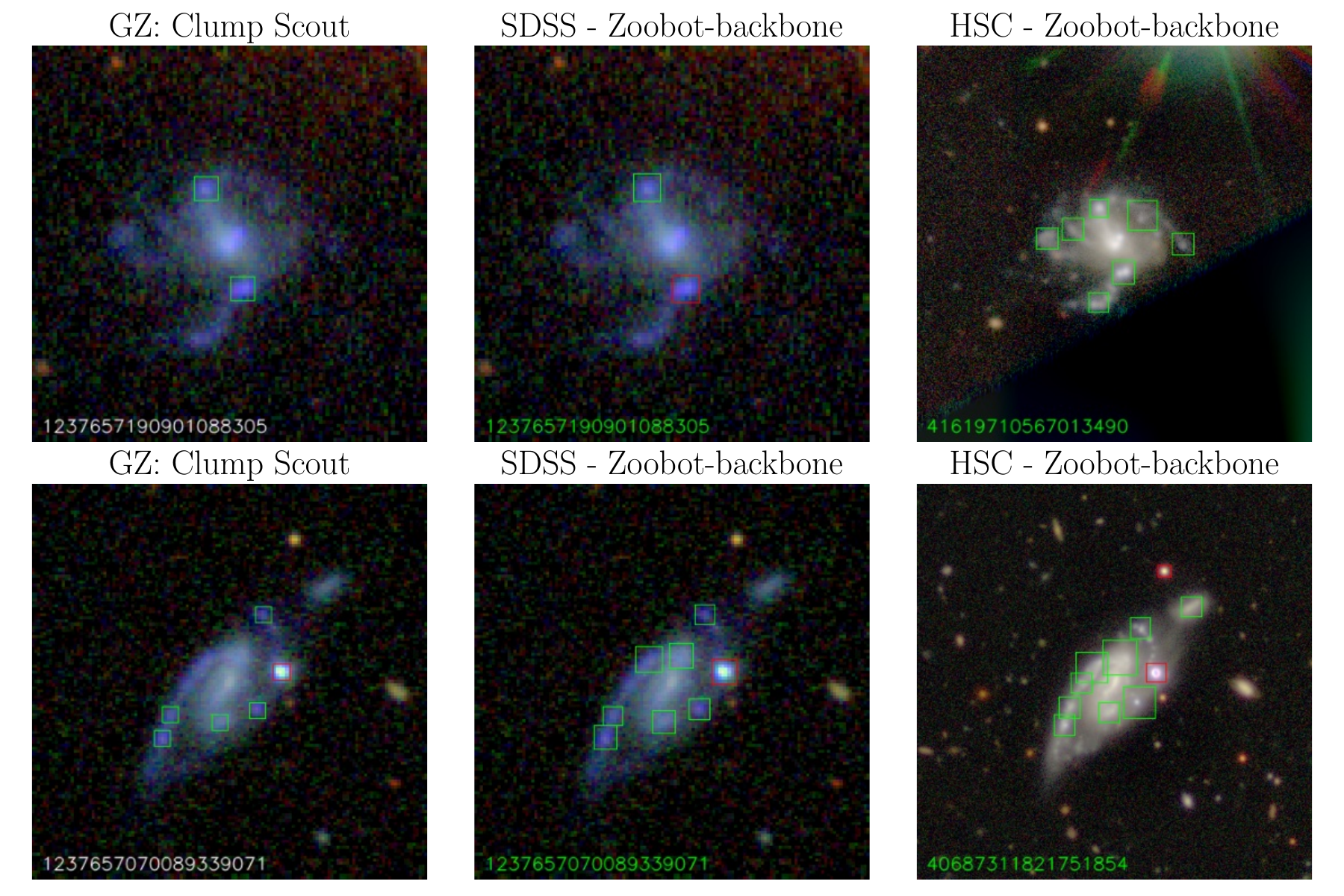}
    \caption[Clump candidates on HSC images.]{Clump candidates on two example HSC images. Column 1: SDSS image with volunteers' labels from GZCS, column 2: SDSS image with detections by \textit{Zoobot-backbone} and column 3: HSC image with detections by \textit{Zoobot-backbone}. Normal clumps are marked with green boxes, odd or unusual clumps with red boxes. The images are labelled with their SDSS-DR7 and HSC object numbers, respectively. Detection score threshold is $\geq 0.3$.}
    \label{fig:HSC_example}
\end{figure*}

A cross-match between the GZCS-catalogue and the HSC-catalogue found 2,424 objects surveyed by SDSS and the HSC SSP for the GZCS sample, which we used for a direct visual comparison of the clump candidates on images with different spatial resolution. Figure \ref{fig:HSC_example} shows such a comparison of two sample galaxies from both catalogues with the clump candidates overlaid. The higher resolution of HSC is immediately apparent and the \textit{Zoobot-backbone} is capable of detecting multiple clump candidates with varying sizes of the bounding boxes on galaxy images from this source. 

From visual inspection the \textit{Zoobot-backbone} model appears to produce the most reasonable detections. A comparison to the model \textit{Imagenet-backbone} can be seen from the additional examples in Figures \ref{fig:HSC_examples1}, \ref{fig:HSC_examples3} and \ref{fig:HSC_examples3} in the Appendix. \textit{Imagenet-backbone} appears to find far fewer clumps and some bounding boxes for detected objects are too large for clump-like objects.

\section{Discussion}\label{sec:discussion}
We have used a set of galaxy images annotated with clump markings from the Galaxy Zoo: Clump Scout citizen science project \citep{Adams2022} to train Faster R-CNN models for object detection which are capable of producing plausible predictions of clump locations within the host galaxies.

This supervised deep learning approach faces several challenges. The first challenge concerns the training (or ground-truth) data from which the models learn. Observational data has the advantage over simulated data as it does not require prior assumptions of the photometric and physical properties of the objects to be detected. On the other hand, even if the completeness of the sample has been corrected against expert labels and probabilistic algorithms applied to aggregate the volunteers' annotations into a consensus location and label \citep{Dickinson2022}, it is likely that the training set contains misidentifications and is missing genuine clump instances. 

Another challenge lies in how the imaging data is presented to the neural network. In this study, we have used three channels per image as input for our models (e.g. $gri$-bands converted into RGB for the SDSS and HSC images). We expect this approach to be superior to using only single-channel input data \citep[e.g.][]{HuertasCompany2020}, but this still needs to be validated.

In comparing the performance of \textit{Zoobot} as a feature extraction backbone against CNNs trained on `terrestrial' data sets we have also limited ourselves to these three channels. From the SDSS imaging data, two additional channels are available, namely the $u$- and $z$-band. To overcome these limitations we are considering using the feature extraction backbones in an ensemble configuration in future works. Possible configurations could consists of:
\begin{enumerate}
    \item two ResNet-based feature extraction backbones, where the sixth channel will be left unused, 
    \item an ensemble of five backbones, one backbone for each of the $ugriz$-bands.  
\end{enumerate}

Furthermore, fine-tuning a classification network before being used as a feature extraction backbone, like we tried with the \textit{Zoobot-clumps-backbone} model (see Appendix \ref{sec:zoobot_fine_tune}), leads to a worse detection performance. This is possibly caused by the feature extraction backbone `forgetting' previously learned features and now relying on different features for classifying clumpy galaxies. We expected the \textit{Zoobot-clumps-backbone} model to more closely resemble the GZCS population but instead, the model produces much bigger bounding boxes (Figure \ref{fig:sdss_example}), many more odd clump candidates (Figure \ref{fig:hist_redshift_odd_clumps}) and the normal clump candidates are mainly detected in more massive galaxies compared to the GZCS sample (Figure \ref{fig:clump_mass_z3}).

From an astrophysical point of view, we favour a more complete over a more pure set of detections. As \citet{Dickinson2022} pointed out, the volunteers' from GZCS tend to mark more faint features in galaxies as clumps than experts, especially when those features appear blue in colour. The authors also noted, that volunteers are more likely to mark a clump as `normal', despite being labelled by experts as `unusual' or `odd'. 

This disagreement makes the GZCS sample less suitable for acting as a benchmark to determine the purity of the model outputs. Instead, we argue that a higher completeness is better suited to scientific analyses. We find support for the robustness of our model completeness after we tested them on galaxy images with simulated clumps from \citet[][see Appendix \ref{sec:simulated_clumps}]{Adams2022}, where they were able to produce similar performance results. We therefore suggest that a final clump selection process needs to include a refinement of the model output by considering directly observable clump characteristics or derived photometric or physical properties (e.g. colour, flux and stellar mass estimates of the clumps). 

Taking into account the detection performances of the five FRCNN models as described in Section \ref{sec:performance} and the resulting clump population statistics in comparison to the GZCS data set, we selected the FRCNN model \textit{Zoobot-backbone} as the best and most robust detection model. We argue that:
\begin{enumerate}
    \item The detected clumps are closest to the GZCS population in terms of total number, clumps per galaxy and identified galaxies with at least one off-centre clump.
    \item Completeness and purity are among the highest, especially for the clump-galaxy flux-ratio of $\geq 3\%$.
    \item The model performs well after only being trained on relatively small data sets and is robust against over-fitting.
    \item It uses the unmodified \textit{Zoobot} CNN as the backbone feature extractor, for which no problem-specific fine-tuning is necessary.
    \item This FRCNN model will likely benefit from the continued training of \textit{Zoobot} for galaxy classifications on more and more imaging data sets.
\end{enumerate}

Utilising the `feature knowledge' \textit{Zoobot} has already acquired can help to facilitate the extension of deep learning-based object detection to a wide range of morphological features observed in galaxies, e.g. bars, rings or spiral arms. Furthermore, extending Faster R-CNN from object detection to object instance segmentation will add a valuable alternative to `traditional' segmentation methods using contrast detection techniques with SExtractor \citep{Bertin1996}, for example. A framework like Mask R-CNN \citep{He2017}, which uses a similar architecture to Faster R-CNN, will likely benefit from a pre-trained feature extraction backbone like \textit{Zoobot}.

The ability of our models not only to detect clumps but also to classify the detections into normal and odd clumps can greatly facilitate the removal of foreground star contamination in massive sets of galaxy images. We present in this paper first promising results but will continue to train and evaluate our models with new training data which will be specifically labelled for this purpose.

\section{Conclusions}\label{sec:conclusion}
This paper has presented a deep learning model for detecting giant star-forming clumps in low-redshift galaxies. We developed object detection models using the Faster R-CNN framework and trained the models on real observations from the Galaxy Zoo: Clump Scout project. These 18,772 low-redshift galaxy images, taken from the Sloan Digital Sky Survey, were annotated with clump markings by non-expert volunteers and aggregated to 39,745 potential clump locations using a probabilistic aggregation algorithm.

We tested five Faster R-CNN models with different feature extraction backbones, all of which are based on the ResNet50 architecture but initialised with either terrestrial or domain-specific pre-trained weights in different training modes. We trained each model on 20 different training sets with varying sample sizes, ranging from 500 to 18,772 galaxy images with marked clump locations. For all 100 training runs, we evaluated the detection performance using the standard COCO metrics and also determined completeness and purity within the GSFC-specific context after applying necessary post-processing steps.

The key results are summarised in the following points:
\begin{enumerate}
    \item Deep learning-based object detection models have been trained on large samples of real observational data instead of simulated data. This paper has shown that Faster R-CNN can be successfully applied for detecting clumps in galaxy images. The models we present in this paper are capable of producing clump candidate detections with a completeness and purity of $\geq 0.8$ on SDSS imaging data.
    \item The same models can be used without additional training (`out-of-the-box') on imaging data from the HSC SSP, which have increased spatial resolution compared to the data, that has been used for training the models. 
    \item Using \textit{Zoobot} as a feature extraction backbone for the FRCNN model has shown the effectiveness of transfer learning within an astrophysical context. The models using the unmodified \textit{Zoobot} feature extractor are robust against over-fitting and produce the best results for detecting giant star-forming clumps.
    \item This paper has also shown how domain adaptation made it possible to apply FRCNN models to problem sets which are generally too small for a `training from scratch' approach. The final model, \textit{Zoobot-backbone}, achieved a high detection performance while only being trained on $\sim 5,000$ samples or $39\%$ of the full training set. To achieve good prediction results, the effort of creating labelled data sets can be reduced as training the model does not require large data sets and will be less computationally expensive.
    \item The model output can be made suitable for further scientific analysis after necessary post-processing steps and corrections for sample completeness have been applied.
\end{enumerate}

\section*{Acknowledgements}
We thank Miguel Aragon and the second anonymous referee for their useful and constructive comments that led to improvements in this manuscript.

JP acknowledges funding from the Science and Technology Facilities Council (STFC) Grant Code ST/X508640/1. MW acknowledges funding from the Science and Technology Facilities Council (STFC) Grant Code ST/R505006/1. DA, LFF and KBM acknowledge partial funding from the US National Science Foundation Award IIS 2006894 and NASA Award 80NSSC20M0057.

This research made use of the open-source Python scientific computing ecosystem, including NumPy \citep{Harris2020}, Matplotlib \citep{Hunter2007}, seaborn \citep{Waskom2021} and Pandas \citep{McKinney2010}. This research made use of Astropy, a community-developed core Python package for Astronomy \citep{astropy2022} and the associated Python libraries APLpy \citep{Robitaille2019} and photutils Python package \citep{Bradley2023}.

For the DL model development the Python frameworks Tensorflow Object Detection API \citep{Huang2017} and Pytorch/Torchvision \citep{Paszke2019} were used.

This publication uses data generated via the Zooniverse.org platform, development of which is funded by generous support, including a Global Impact Award from Google, and by a grant from the Alfred P. Sloan Foundation.

The authors acknowledge the Minnesota Supercomputing Institute (MSI, \url{https://www.msi.umn.edu/}) at The University of Minnesota for providing high-performance computing (HPC) resources that have contributed to the research results reported within this paper.

\section*{Data Availability}
The data underlying this paper were used in \citet{Adams2022} and can be obtained as a machine-readable table by downloading the associated article data from \url{https://doi.org/10.3847/1538-4357/ac6512}. The final models and code are made publicly available at: \url{https://github.com/ou-astrophysics/Faster-R-CNN-for-Galaxy-Zoo-Clump-Scout}. A detailed catalogue of Giant Star-forming Clumps, detected for the full set of Galaxy Zoo: Clump Scout galaxies observed by SDSS, can be downloaded from \url{https://doi.org/10.5281/zenodo.8228890}.



\bibliographystyle{mnras}
\bibliography{frcnn} 




\appendix
\section{Creating the galaxy images} \label{sec:galaxy_images}
\subsection{SDSS galaxy images} \label{sec:galaxy_images_sdss}
The SDSS galaxy images were created as cutouts from the SDSS DR15 Legacy survey data. To ensure a comparable visual size of the target galaxies, we scaled the cutouts to six times the 90-percent $r$-band Petrosian radius while keeping a scale of $0.396$ arcsec per pixel to match the native SDSS resolution. We then created RGB-composite images from the three single $g$, $r$ and $i$ band FITS, where the $g$, $r$ and $i$ map to red, green and blue channels using `Lupton'-scaling \citep{Lupton2004}:
\begin{eqnarray}\label{eq:lupton}
  I^{\prime }_{x} = \frac{1}{Q}\mathrm{asinh}\left[Q\cdot \frac{\left(\frac{I_{x}}{\beta _{x}}-m\right)}{\alpha }\right],
\end{eqnarray}
where $I_x$ is the input pixel intensity in band $x$ and $I^{\prime }_{x}$ is the scaled pixel intensity. Table \ref{tab:lupton_params} lists the parameters used for the GZCS-cutouts.
\begin{table}
	\centering
	\caption[Parameters Lupton-scaling.]{Parameters Lupton-scaling for $g$, $r$ and $i$ bands.} 
        \label{tab:lupton_params}
	\begin{tabular}{lrrr} 
		\hline
		Parameter & Band $g$ & Band $r$ & Band $i$\\
		\hline
		$Q$ & 7 & 7 & 7 \\
		Stretch $\alpha$ & 0.2 & 0.2 & 0.2 \\
		Minimum $m$ & 0 & 0 & 0 \\
		Channel scales $\beta$ & 0.7 & 1.17 & 1.818 \\
		\hline
	\end{tabular}
\end{table}

In a final step, we resized the RGB-composite images to $400 \times 400$ pixels so that the visual sizes of each central galaxy is similar but with varying resolutions.



The SDSS DR15 PhotoPrimary table \citep{Aguado2019} lists a redshift range of $0.02 \leq z \leq 0.25$ with a median value of $z_{\text{median}} = 0.05$ for the clumpy galaxies from GZCS. The $r$-band PSF-FWHM from SDSS for the individual observations varies from $0.60'' \leq \text{PSF}_{\text{FWHM}, r} \leq 2.08''$ with a median value of $\text{PSF}_{\text{median}} = 1.12''$. For these redshift ranges cosmological angular size - redshift relations can be simplified giving a median size of objects seen at an angle of one PSF-FWHM of $1.14$ kpc. This corresponds to the initial assumption that clumps with $\sim 1$ kpc physical sizes are unresolved with SDSS. We note, however, that clumps marked by the volunteers in galaxies at the higher end our our sample redshift range are unresolved up to a size of $\sim 5$ kpc and might not necessarily represent genuine GSFCs.

\subsection{HSC galaxy images}
We obtained the HSC galaxy cutouts using the command-line tools to access the PDR3 Data Access Service from HSC SSP. We applied a field of view (FOV) of $60\arcsec$ which matches the median FOV from the GZCS cutouts. The $g$, $r$ and $i$ filter bands were used to create a RGB-composite image using an $\textrm{asinh}$ stretch to the images:
\begin{equation}
    I^{\prime }_{x} = \frac{ \textrm{asinh}(e^{10}\,I_{x}) / \textrm{asinh}(e^{10}) + 0.05 }{0.72},
\end{equation}
where $I_x$ is the input pixel intensity in band $x$ and $I^{\prime }_{x}$ is the scaled pixel intensity. Finally, the RGB-composites were resized to $400 \times 400$ pixels.

\section{Fine-tuning \textit{Zoobot} for classifying clumpy galaxies}
\label{sec:zoobot_fine_tune}
We used the set of the GZCS imaging data after it has been reduced by the aggregation algorithm from \citet{Dickinson2022} to develop a classifcation CNN specifically for separating clumpy and non-clumpy galaxies (\textit{Zoobot Clumps}). The 45,643 images were split into a training set (36,514 or $80\%$), validation set (4,564 or $10\%$) and a test set (4,565 or $10\%$) with the same distribution of clumpy and non-clumpy galaxies as the full data set (Table \ref{tab:dataset_classifier}). A galaxy containing at least one off-centre clump, either a `normal' or `odd' clump (see Section \ref{sec:gzcs}), after the consensus label aggregation from \citet{Dickinson2022}, was given the class `clumpy' and `w/o clumps' otherwise.
\begin{table}
	\centering
	\caption[Train, validation and test data sets for the \textit{Zoobot Clumps} classification model.]{Train, validation and test data sets for the \textit{Zoobot Clumps} classification model.} 
        \label{tab:dataset_classifier}
	\begin{tabular}{lrrrrr}
		\hline
		Label & ID & Train & Validation & Test & Total \\
		\hline
		w/o clumps & 0  & 20,503    & 2,541     & 2,612     & 25,656    \\
                       &    & (56.15\%) & (55.67\%) & (57.22\%) & (56.21\%) \\
            clumpy     & 1  & 16,011    & 2,023     & 1,953     & 19,987    \\
                       &    & (43.85\%) & (44.33\%) & (42.78\%) & (43.79\%) \\
            total      & -- & 36,514    & 4,564     & 4,565     & 45,643    \\
                       &    & (100.00\%)& (100.00\%)& (100.00\%)& (100.00\%)\\
		\hline
	\end{tabular}
\end{table}

During training, the model was presented not only with the original set of input galaxy images but also with variations of it. This image augmentation helps to improve the generalisation ability of the model and increases the subset of the learning data. The variations were mostly randomly applied and consisted of the following techniques:
\begin{enumerate}
    \item random resizing, keeping the aspect ratio between 0.9 and 1.1,
    \item random cropping to $224 \times 224$ pixels within a 10\% margin of the original image,
    \item random horizontal flip,
    \item random rotation of $90^{\circ}$ and
    \item normalisation of the pixel values in each channel.
\end{enumerate}

We train two \textit{Zoobot Clumps} versions, based on the ResNet50 \citep{He2016} and EfficientNetB0 \citep{Tan2019} architecture, over 100 epochs without extensive hyper-parameter tuning. Learning rate and optimiser settings were varied around values gained from expert knowledge (H. Dickinson and M. Walmsley, priv. comm.), but resulting model performance did not change significantly. We chose the \textit{Adam} optimiser \citep{Kingma2014} over the standard stochastic gradient descent (SGD) optimiser as it introduces an adaptive learning rate and increases computation speed. We show the parameters used for developing  \textit{Zoobot Clumps} in Table \ref{tab:params_classifier}. 
\begin{table}
	\centering
	\caption[Parameters used for training the \textit{Zoobot Clumps} classification model.]{Parameters used for training the \textit{Zoobot Clumps} classification model.} 
        \label{tab:params_classifier}
	\begin{tabular}{lll}
		\hline
		   & ResNet50 & EfficientNetB0 \\
		\hline
		Python framework & PyTorch & PyTorch, Tensorflow \\
		Infrastructure & NVIDIA A100  & NVIDIA A100   \\
                           & SXM4-40GB GPU& SXM4-40GB GPU \\
		Batch size & 32 & 32 \\
		Optimiser & Adam & Adam \\
		Initial learning rate & $10^{-4}$ & $10^{-4}$ \\
		Epochs & 100 & 100 \\ 
		Accuracy & $0.8291 \pm 0.0109$ & $0.8484 \pm 0.0104$ \\
                     & (at epoch=62) & (at epoch=74) \\
		\hline
	\end{tabular}
\end{table}

The models with the best classification performance achieved an accuracy of $0.8291 \pm 0.0109$ and $0.8484 \pm 0.0104$ (within a $2\sigma$ confidence interval) based on the ResNet50 and EfficientNetB0 architecture, respectively.

\section{Training run details}\label{sec:training_runs}
For the several training runs, we randomly assigned the remaining 18,772 galaxy images, after we have applied all exclusions (see Table \ref{tab:sample_sizes}), into $20$ run-groups where the group size increased exponentially from 500 to 18,772. Each run-group was further split into a training (70\%), validation (20\%) and test set (10\%). The split was also done randomly but using a stratification based on the ratio of odd (or unusual) to normal clumps in each galaxy to maintain a comparable distribution of both clump classes in all groups. For the whole data set the ratio $(\text{odd clumps} / \text{normal clumps}) = 0.293 \pm 0.004$ and the ratio of $(\text{clumps} / \text{galaxy}) = 2.12 \pm 0.009$. Table \ref{tab:sizes_run_groups} lists the sample sizes of the various run-groups and sets.

Note, that only galaxies with at least one off-centre clump were used for training. True negative samples are obtained from areas in the image where no instances of the objects to be detected are located. Additional images containing no instances of the objects are usually not required as the number of negative and positive anchor boxes need to be balanced for the RPN as otherwise a bias towards negative samples will occur \citep{ren2015}.
\begin{table}
	\centering
	\caption[Sample sizes for each set per run-group.]{Number of galaxy images and annotated clumps (in brackets) for each set per run-group.} 
        \label{tab:sizes_run_groups} 
	\begin{tabular}{crrrr}
		\hline
		  Run- & Training & Validation & Test & Total \\
            group & & & & \\ 
		\hline
            1  & 349    & 100   & 51    & 500    \\ 
               &(752)   &(223)  &(104)  &(1,079) \\ \hline
            2  & 424    & 120   & 61    & 605    \\ 
               &(893)   &(244)  &(128)  &(1,265) \\ \hline
            3  & 512    & 146   & 74    & 732    \\ 
               &(1,056) &(292)  &(151)  &(1,499) \\ \hline
            4  & 620    & 177   & 89    & 886    \\ 
               &(1,298) &(366)  &(194)  &(1,858) \\ \hline
            5  & 751    & 214   & 108   & 1,073  \\ 
               &(1,671) &(456)  &(224)  &(2,351) \\ \hline
            6  & 908    & 260   & 130   & 1,298  \\ 
               &(1,923) &(557)  &(257)  &(2,737) \\ \hline
            7  & 1,099  & 314   & 158   & 1,571  \\ 
               &(2,361) &(685)  &(316)  &(3,362) \\ \hline
            8  & 1,330  & 380   & 191   & 1,901  \\ 
               &(2,788) &(822)  &(429)  &(4,039) \\ \hline
            9  & 1,610  & 460   & 231   & 2,301  \\ 
               &(3,426) &(953)  &(482)  &(4,861) \\ \hline
            10 & 1,949  & 557   & 279   & 2,785  \\ 
               &(4,145) &(1,190)&(589)  &(5,924) \\ \hline
            11 & 2,358  & 674   & 338   & 3,370  \\ 
               &(5,027) &(1,426)&(690)  &(7,143) \\ \hline
            12 & 2,855  & 816   & 408   & 4,079  \\ 
               &(6,082) &(1,733)&(849)  &(8,664) \\ \hline
            13 & 3,455  & 987   & 494   & 4,936  \\ 
               &(7,323) &(2,076)&(1,095)&(10,494)\\ \hline
            14 & 4,181  & 1,195 & 598   & 5,974  \\ 
               &(8,808) &(2,526)&(1,289)&(12,623)\\ \hline
            15 & 5,061  & 1,443 & 724   & 7,230  \\ 
               &(10,712)&(3,025)&(1,510)&(15,247)\\ \hline
            16 & 6,124  & 1,750 & 876   & 8,750  \\ 
               &(12,782)&(3,730)&(1,759)&(18,271)\\ \hline
            17 & 7,412  & 2,118 & 1,060 & 10,590 \\ 
               &(15,851)&(4,429)&(2,272)&(22,552)\\ \hline
            18 & 8,971  & 2,563 & 1,282 & 12,816 \\ 
               &(18,986)&(5,415)&(2,680)&(27,081)\\ \hline
            19 & 10,857 & 3,102 & 1,552 & 15,511 \\ 
               &(22,959)&(6,633)&(3,259)&(32,851)\\ \hline
            20 & 13,140 & 3,754 & 1,878 & 18,772 \\
               &(27,825)&(7,941)&(3,979)&(39,745)\\
		\hline
	\end{tabular}
\end{table}

\section{COCO metrics}\label{sec:coco}
For object detection models performance is typically evaluated by average precision metrics. A common set of metrics is used for the COCO Object Detection Challenge \citep[COCO metrics,][]{Lin2014}. 

Given the prediction score (objectness) $c_i$ as the probability whether an anchor box $i$ contains an object or not, a threshold value $c_0 \in [0, 1]$ is set so that if $c_i \geq c_0$, box $i$ is defined to contain an object and not, if $c_i < c_0$. We calculated the Intersection over Union (IoU, see Section \ref{sec:post_processing}) for each bounding box containing objects with respect to the ground-truth, in this case the annotations from the GZCS-volunteers. If the IoU is $\geq 0.5$, for example, the detection is a True Positive (TP), otherwise a False Positive (FP). On the other hand, if a bounding box has high enough overlap with a ground-truth object (IoU is $\geq 0.5$) but the objectness score is below the threshold, so $c_i < c_0$, then this is called a False Negative (FN). The IoU-threshold can be varied depending on the specific task. 

Precision $p$ can then be defined as 
\begin{equation}
    p(c_0) = \frac{\text{\#TP}(c_0)}{\text{\#TP}(c_0) + \text{\#FP}(c_0)}
\end{equation}
and recall $r$ as
\begin{equation}
    r(c_0) = \frac{\text{\#TP}(c_0)}{\text{\#TP}(c_0) + \text{\#FN}(c_0)},
\end{equation}
where $\text{\#TP}(c_0)$, $\text{\#FP}(c_0)$ and $\text{\#FN}(c_0)$ are the number of true positives, false positives and false negatives, respectively, depending on the prediction score threshold $c_0$.

Precision and recall are equivalent to purity and completeness, respectively, which are more commonly used in an astrophysical context. Completeness is the number of objects in a data set that are detected over the number that exists. Purity is the number of true detections over the number of all detections. High values for both metrics show that a detector is returning accurate results (`high precision') as well as returning a majority of all true results (`high completeness').

We observed a trade-off between precision and recall which is typically for object detection problems. High precision can be achieved with low score thresholds $c_0$ which results in a lower recall and vice-versa (see Figure \ref{fig:recall_precision} for an example where purity and completeness are used instead of precision and recall). Plotting precision against recall for a discrete set of score thresholds, e.g. $c_n \in [0.0, 0.1, \dots , 1.0]$, the area under this curve is used as a model comparison metric, called average precision or AP:
\begin{equation}
    \text{AP} = \sum_n (r_n - r_{n-1}) p_n, 
\end{equation}
where $r_n$ and $p_n$ are the corresponding values for recall and precision specific to the score threshold $c_n$. In other words, the AP summarizes such a $r-p$ plot as the weighted mean of precision achieved at each threshold, using the step increase in recall from the previous threshold as the weight. A value close to $1$ represents both high recall and high precision.

For multi-class detection problems the mean average precision $\overline{\text{AP}}$ is defined by:
\begin{equation}
    \overline{\text{AP}} = \frac{1}{k} \sum_{i=1}^{k} \text{AP}_i , 
\end{equation}
for the $k>1$ classes.

Instead of iterating through discrete values of the detection score $c_n$, different thresholds for the IoU are used for calculating the average recall (AR). From the recall-IoU curve, where $\text{IoU} \in [0.5, 1.0]$, the area under the curve, multiplied by two, is used as the value for AR:
\begin{equation}
    \text{AR} = 2 \int_{0.5}^{1.0} r(\text{IoU})\, \text{d}(\text{IoU}).
\end{equation}
This is then averaged over all classes $k$ to define the mean average recall $\overline{\text{AR}}$:
\begin{equation}
    \overline{\text{AR}} = \frac{1}{k} \sum_{i=1}^{k} \text{AR}_i . 
\end{equation}

Mean average precision $\overline{\text{AP}}$ and mean average recall $\overline{\text{AR}}$ can be combined into a single performance metric. This metric is called F1-score $f_1$ and is defined as:
\begin{equation}
    f_1 = 2 \times \frac{\overline{\text{AP}}\,\overline{\text{AR}}}{\overline{\text{AP}} + \overline{\text{AR}}}.
\end{equation}
The F1-score can be used to compare different models, especially if they vary in precision and recall and if both metrics are equally important for evaluating detection performance. 

\section{Clump photometry}\label{sec:photometry}
\begin{figure}
    \centering
    \includegraphics[width=0.7\columnwidth]{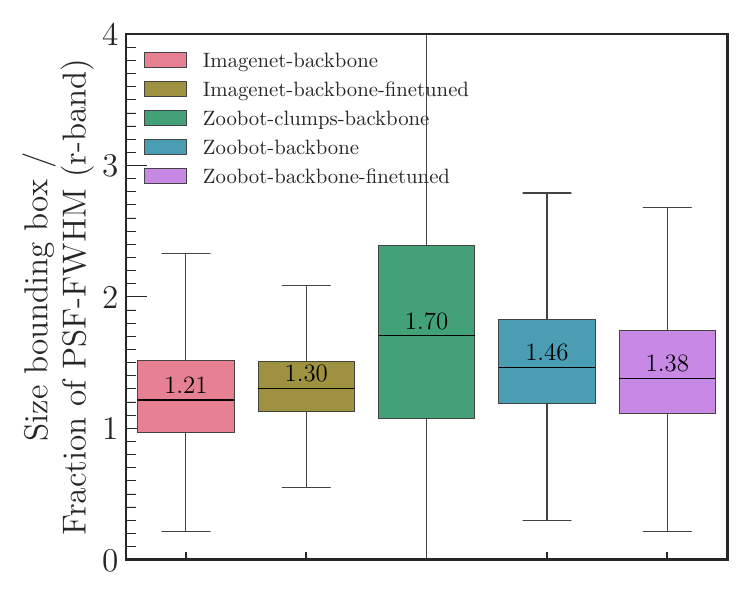}
    \caption[Distribution of bounding box sizes.]{Distribution of bounding box sizes for all five models as fractions of the $r$-band PSF-FWHM. The size of a bounding box is described by the radius of a circle from the centre of the box to the closest side of the box. The annotations show the median values of the sizes.}
    \label{fig:bbox_size_psf}
\end{figure}

The flux of each clump was measured from each of the $ugriz$-bands FITS using the \emph{photutils} Python package \citep{Bradley2023}. Figure \ref{fig:bbox_size_psf} shows the distribution of the sizes of the bounding boxes described by the radius of a circle from the centre of the box to the closest side of each box. Allowing for some margin, we chose an aperture with a radius of $1.125 \times$ the band-specific PSF-FWHM centred on the clump midpoint. We acknowledge, that this fixed aperture is not suitable for the wider range of bounding box sizes output by the model \textit{Zoobot-clumps-backbone}. As this model tends to produce bigger bounding boxes for similar object detections (see Section \ref{sec:detections_gzcs}, Figures \ref{fig:sdss_examples1}, \ref{fig:sdss_examples2} and \ref{fig:sdss_examples3} in the Appendix), we kept the same aperture size for all models for better comparison.

Annuli spanning three to five PSF-FWHM where used to compute the median background flux around the aperture. We multiplied this per-pixel value with the aperture area and then subtracted the result from the flux measured in the clump aperture.

\begin{figure}
    \centering
    \subfloat[\centering RGB image with clumps inside (green) and outside (red) the galaxy's spatial extent marked.]{{\includegraphics[width=0.49\columnwidth]{figures/galaxy_mask3.png} }}
    \subfloat[\centering $r$-band FITS with background and clump mask.]{{\includegraphics[width=0.49\columnwidth]{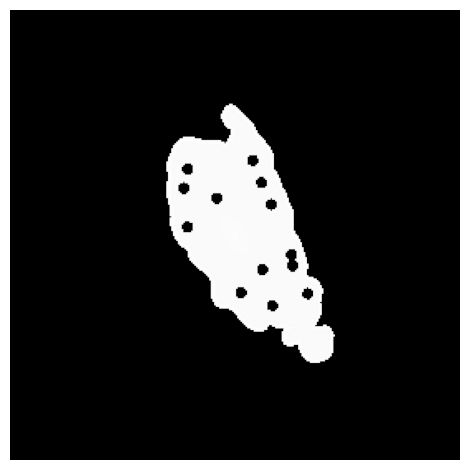} }}
    \caption[Masking background and other clumps.]{For clump photometry measurements the host galaxy background and the other clumps are masked for correct background subtraction.}
    \label{fig:galaxy_mask_photometry}
\end{figure}

Here, background refers to the diffuse light of the host galaxy in which the clumps are embedded. There are several factors that can impact the background flux estimate for each clump. For clumps close to the rims of the galaxy the background estimate will be affected by the area outside the galaxy extent and not only resulting from the diffuse light of the host galaxy. Also, adjacent clumps might fall into the area of the annulus and will obscure the background estimate. To mitigate these effects, we took photometry measurements after masking the area outside, and all other identified clumps within, the host galaxy (Figure \ref{fig:galaxy_mask_photometry}). 

We also corrected the background-subtracted fluxes for the flux loss due to the small aperture sizes as it was assumed that the clumps are unresolved at this scale. Using a Gaussian profile for the PSFs, the fluxes were multiplied by a factor of $\sim 1.03$ for the aperture correction.

Flux values as observed by SDSS are reported in nanomaggies \citep{Stoughton2002} and were converted into Jansky using the factor $3.631\times 10^{-6}$ Jy/nMgy. Also, we converted the flux values $f$ into AB magnitudes $m_{\text{AB}}$ \citep{Oke1983} using:
\begin{equation}
    m_{\text{AB}} = 22.5 - 2.5 \log_{10}(f).
\end{equation}

Further corrections to the AB magnitudes were applied for galactic extinction. For each clump location the reddening $E(B-V)$ is retrieved from the \citet{Schlegel1998} dust maps and converted into an extinction $A_\lambda$ applied to each $ugriz$ AB magnitude using the tabulated factors from \citet{Schlafly2011}. We determined the clump colours $(u-g)$, $(g-r)$, $(r-i)$ and $(i-z)$ as differences between the background-subtracted clump magnitudes.

\section{Galaxy images with simulated clumps}\label{sec:simulated_clumps}
Besides the real observational data, galaxy images containing simulated clumps were made available from \citet{Adams2022}. This data set consists of 84,565 clumps placed in 26,736 galaxies with comparable characteristics to the main GZCS sample. The simulated clumps were placed randomly within the central galaxy's spatial extent but with a higher probability for inner-galactic area ($p=0.75$) compared to the outer, low surface brightness area ($p=0.25$).

Luminosity, mass and spectral properties of the artificial clumps were carefully modelled to span a wide property range allowing for a robust completeness and purity assessment of the object detection models. The detailed process for generating these images with simulated clumps can be found in \citet{Adams2022}.

\begin{figure}
    \centering
    \includegraphics[width=1\columnwidth]{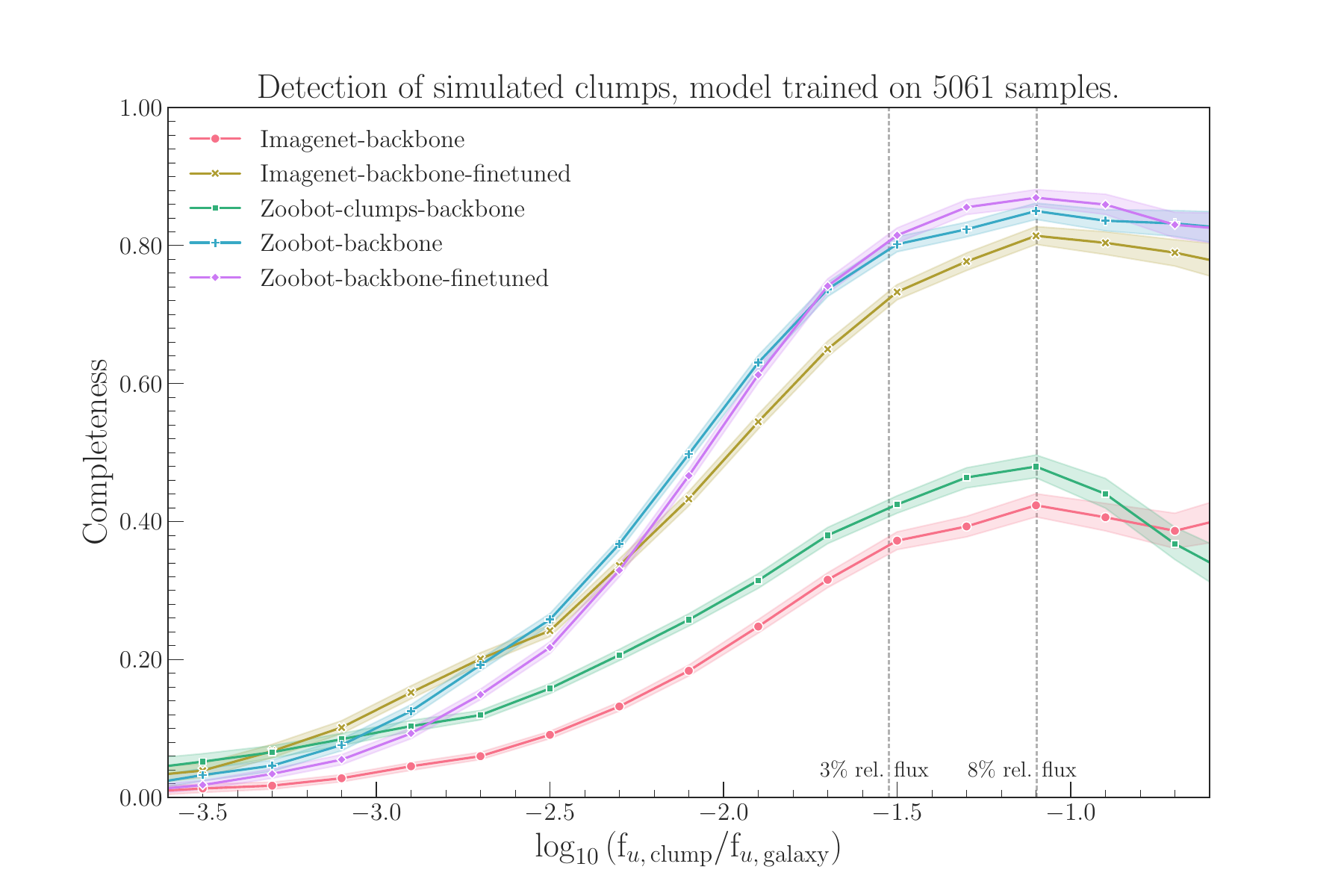}
    \caption[Model completeness per relative clump flux (simulated clumps).]{Model completeness per relative clump flux for training run 15 with a training sample size of 5,061 at a score threshold of $\geq 0.3$  (simulated clumps). Shaded areas showing the $95\%$ confidence interval.}
    \label{fig:completeness_sim_run_15}
\end{figure}

\begin{figure}
    \centering
    \includegraphics[width=1.0\columnwidth]{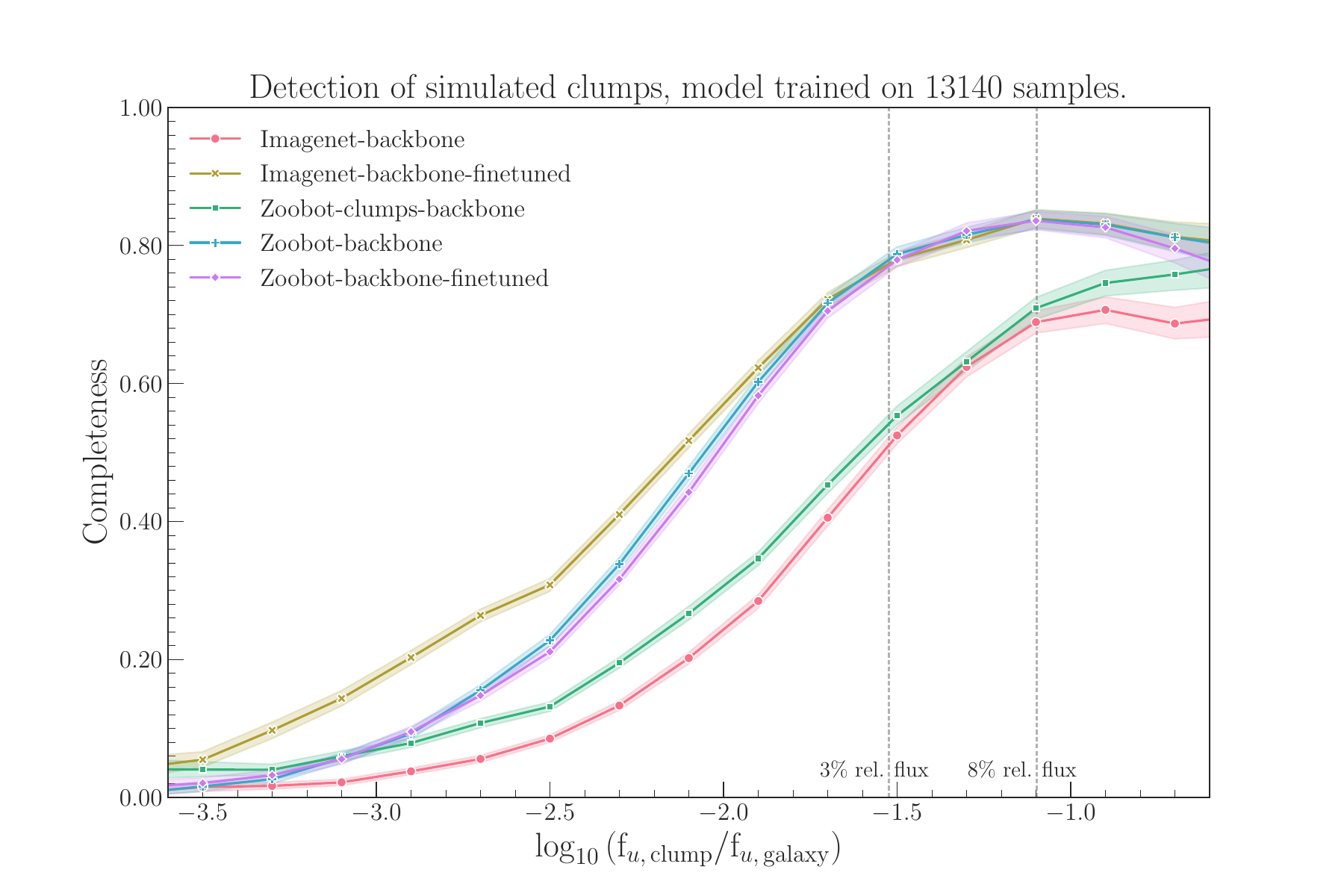}
    \caption[Model completeness per relative clump flux for training run 20 (simulated clumps).]{Model completeness per relative clump flux for training run 20 at a score threshold of $\geq 0.3$ (simulated clumps). Shaded areas showing the $95\%$ confidence interval.}
    \label{fig:completeness_sim_run_20}
\end{figure}

We evaluated our models on the set of 26,736 galaxies with simulated clumps and measured lower overall completeness and purity compared to the levels reached by the models after we applied them to the real clumps. The simulated clumpy galaxies created by \citet{Adams2022} contain many more faint clumps which were used for probing the volunteers' recovery capability and completeness during the GZCS project. As faint clumps have rarely been labelled by the volunteers in the data sets with real observations used to train the FRCNN models, we expected detection performance to be lower for faint clumps and, hence, in overall detection completeness. This can be seen from Figure \ref{fig:completeness_sim_run_15} and \ref{fig:completeness_sim_run_20}, where completeness drops for a clump-galaxy flux ratio below $3\%$ for all models.

With focus on the clump-specific flux ratio ranges the ranking of the models in terms of completeness performance is similar to what we observed from the detections on the real imaging data. Both FRCNN models using a version of the unmodified \textit{Zoobot} as their feature extraction backbone are reaching the highest completeness levels regardless of how many samples were used for training. \textit{Imagenet-backbone-finetuned} is capable of reaching similar completeness levels but only after intensive training. The completeness curve for this model is rising after being trained on 5,061 samples (Figure \ref{fig:completeness_sim_run_15}) to 13,140 samples in the training set (Figure \ref{fig:completeness_sim_run_20}). Again, \textit{Zoobot-clumps-backbone} and \textit{Imagenet-backbone} are much worse compared to the other models.

\section{Additional visual examples of detected clumps in SDSS and HSC galaxy images}\label{sec:additional_examples}
The following images show nine different examples of galaxies annotated by the volunteers from GZCS, where we compare the detections from all five FRCNN models with the volunteers' markings. All models were trained on the full training set (13,140 galaxies, Table \ref{tab:sample_sizes}) and for all detections we applied a detection score threshold of $\geq 0.3$.

\begin{figure*}
    \centering
    \includegraphics[width=1.\textwidth]{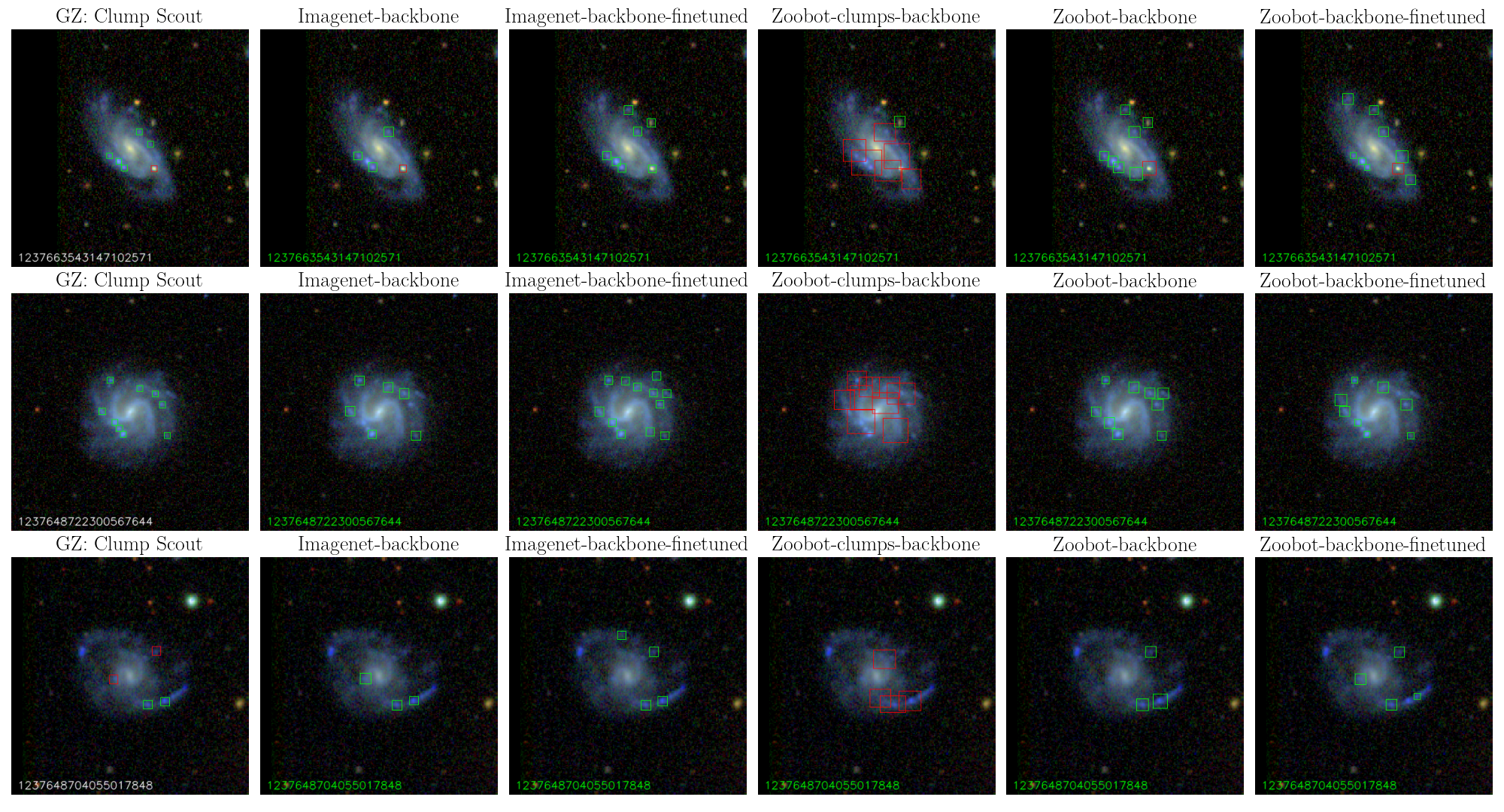}
    \caption[Comparison of detections on SDSS images from all models.]{Comparison of detections on SDSS images from all models. Column 1: SDSS image with volunteers' labels from GZCS, column 2: with object detections by \textit{Imagenet-backbone}, column 3: with object detections by \textit{Imagenet-backbone-finetuned}, column 4: with detections by \textit{Zoobot-clumps-backbone}, column 5: with detections by \textit{Zoobot-backbone} and column 6: with object detections by \textit{Zoobot-backbone-finetuned}. Normal clumps are marked with green boxes, odd or unusual clumps with red boxes. The images are labelled with their SDSS-DR7 object numbers. Detection score threshold is $\geq 0.3$.}
    \label{fig:sdss_examples1}
\end{figure*}

\begin{figure*}
    \centering
    \includegraphics[width=1.\textwidth]{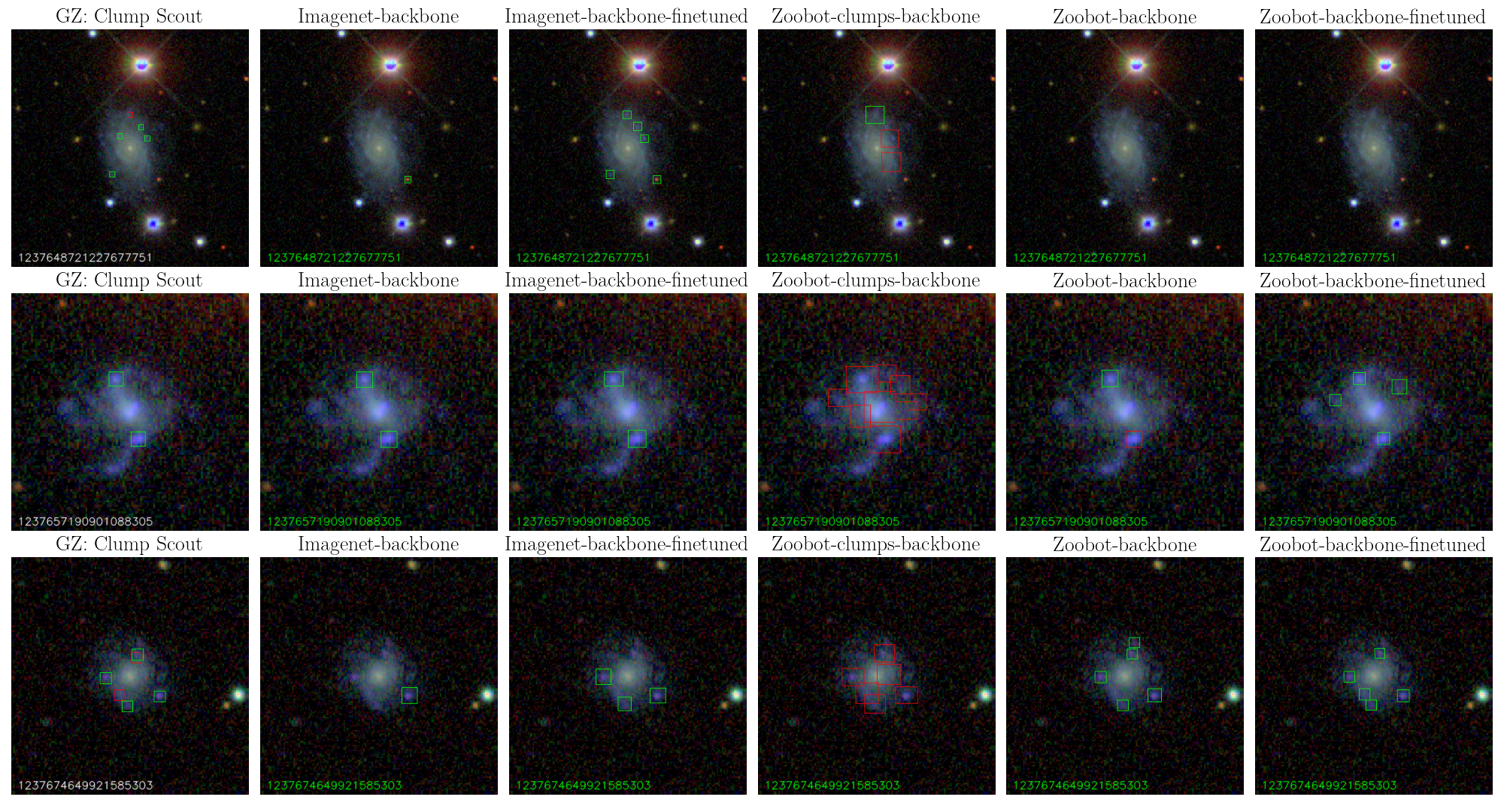}
    \caption[Comparison of detections on SDSS images from all models (2).]{Comparison of detections on SDSS images from all models. Images and detections as described in the previous figure.}
    \label{fig:sdss_examples2}
\end{figure*}

\begin{figure*}
    \centering
    \includegraphics[width=1.\textwidth]{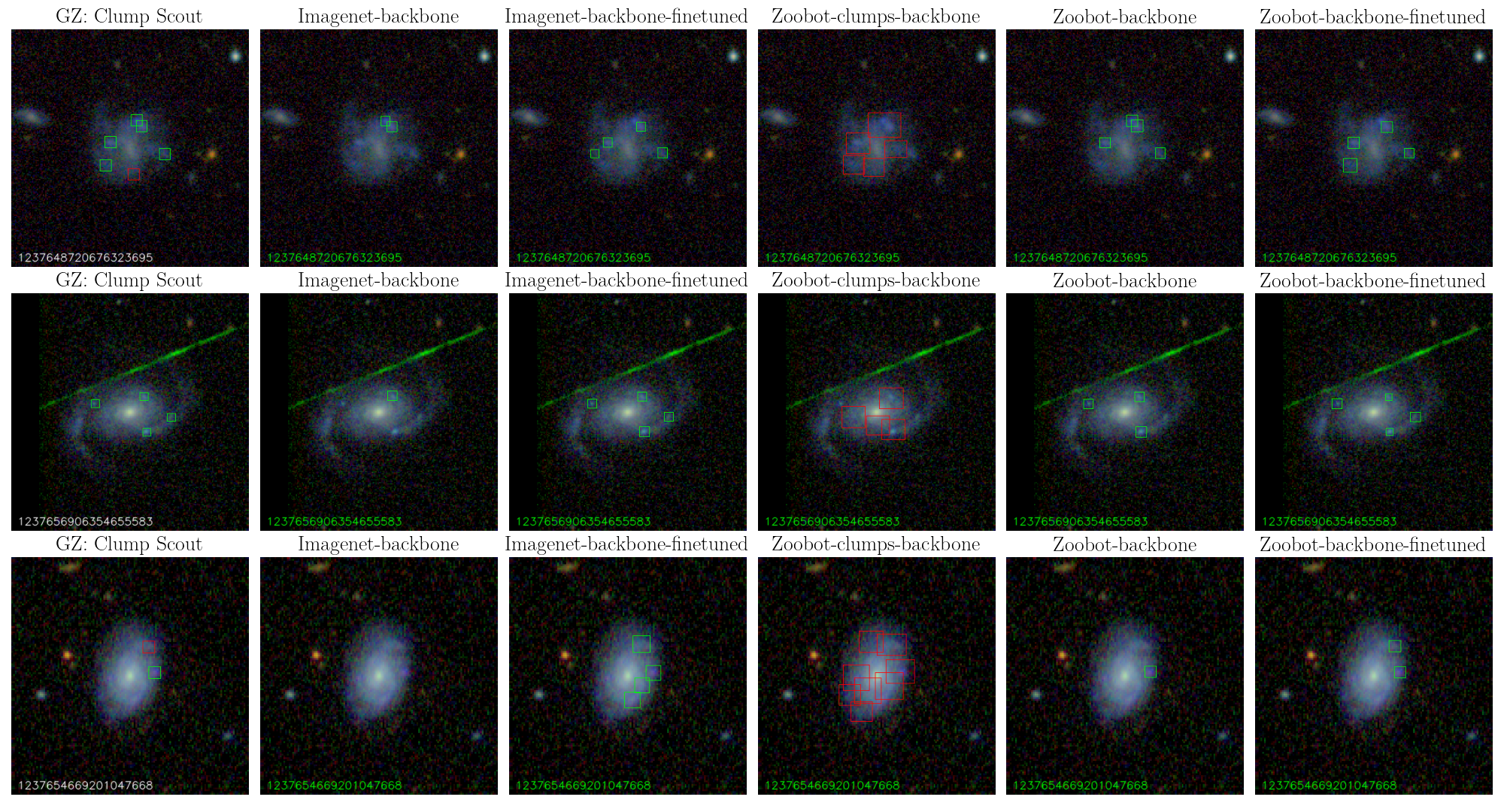}
    \caption[Comparison of detections on SDSS images from all models (3).]{Comparison of detections on SDSS images from all models. Images and detections as described in the previous figure.}
    \label{fig:sdss_examples3}
\end{figure*}

Similar to Figure \ref{fig:sdss_example}, we compare the clump candidates for SDSS galaxies in Figures \ref{fig:sdss_examples1}, \ref{fig:sdss_examples2} and \ref{fig:sdss_examples3}. The galaxy images were chosen to cover many of the challenges the models can face during the object detection inference process, e.g. prominent foreground stars and image artifacts (Figure \ref{fig:sdss_examples3}, for example).

\begin{figure*}
    \centering
    \includegraphics[width=1.\textwidth]{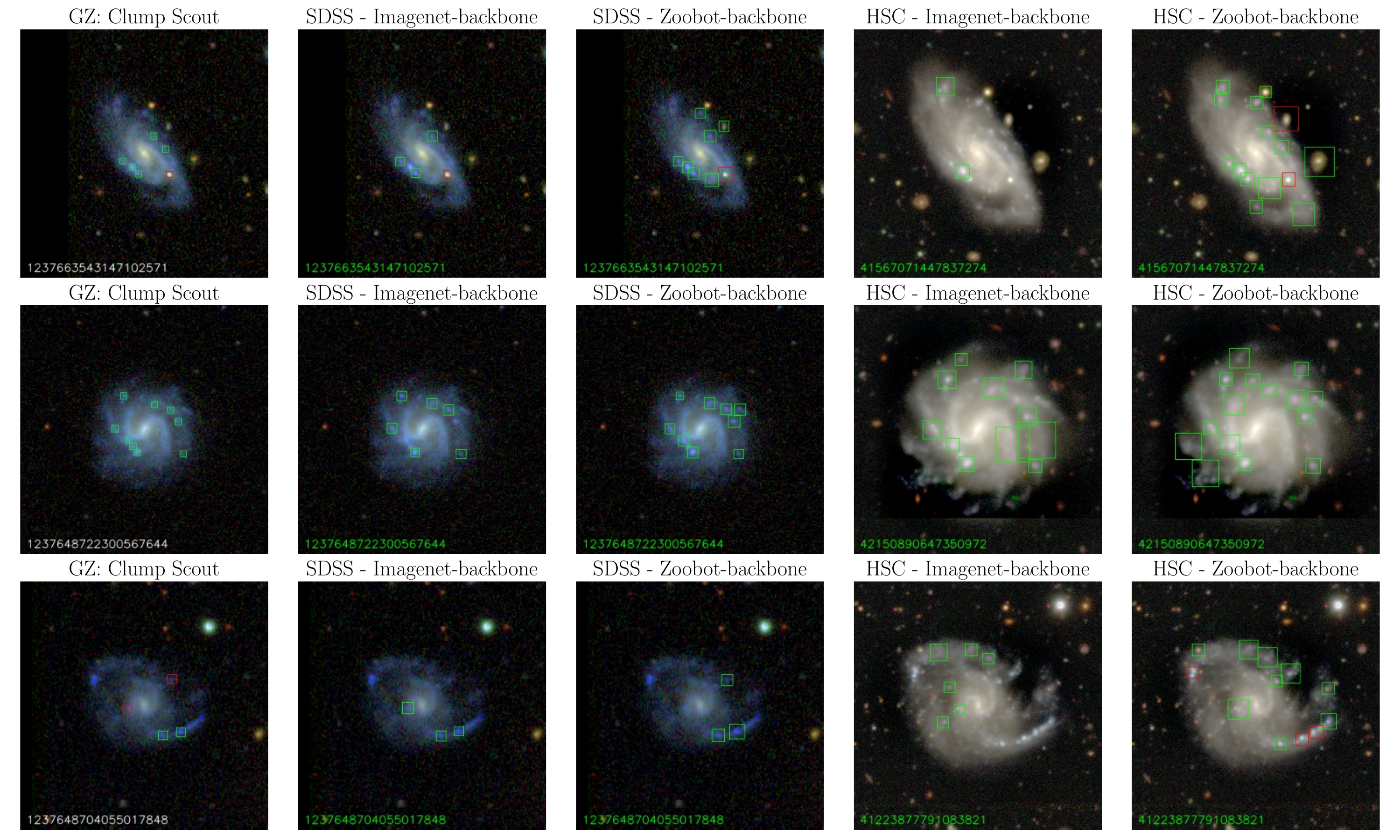}
    \caption[Comparison of detections on SDSS and HSC images.]{Comparison of detections on SDSS and HSC images. Column 1: SDSS image with volunteers' labels from GZCS, column 2: SDSS image with object detections by \textit{Imagenet-backbone}, column 3: SDSS image with detections by \textit{Zoobot-backbone}, column 4: HSC image with object detections by \textit{Imagenet-backbone} and column 5: HSC image with detections by \textit{Zoobot-backbone}. Normal clumps are marked with green boxes, odd or unusual clumps with red boxes. The images are labelled with their SDSS-DR7 and HSC object numbers, respectively. Detection score threshold is $\geq 0.3$.}
    \label{fig:HSC_examples1}
\end{figure*}

\begin{figure*}
    \centering
    \includegraphics[width=1.\textwidth]{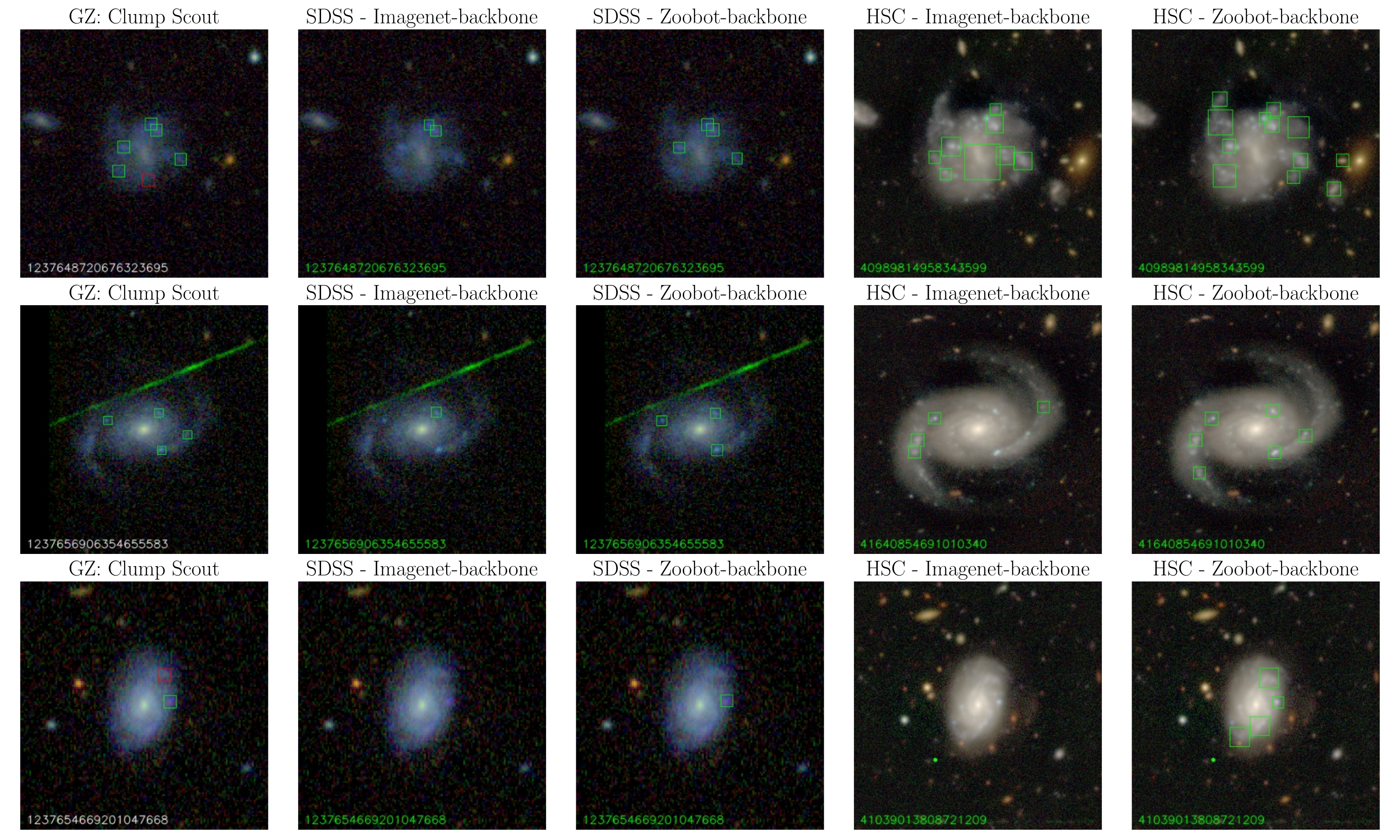}
    \caption[Comparison of detections on SDSS and HSC images (3).]{Comparison of detections on SDSS and HSC images. Images and detections as described in the previous figure.}
    \label{fig:HSC_examples3}
\end{figure*}

In addition, Figures \ref{fig:HSC_examples1}, \ref{fig:HSC_examples3} and \ref{fig:HSC_examples3} compare the clump markings and detections on the same nine galaxies with detections made on HSC galaxies, for which we found a cross-match with our GZCS sample. We compare only the detection results from the models \textit{Imagenet-backbone} and \textit{Zoobot-backbone} for clarity and to highlight the differences between a terrestrial and a domain-specific/astrophysical feature extraction backbone.

\begin{figure*}
    \centering
    \includegraphics[width=1.\textwidth]{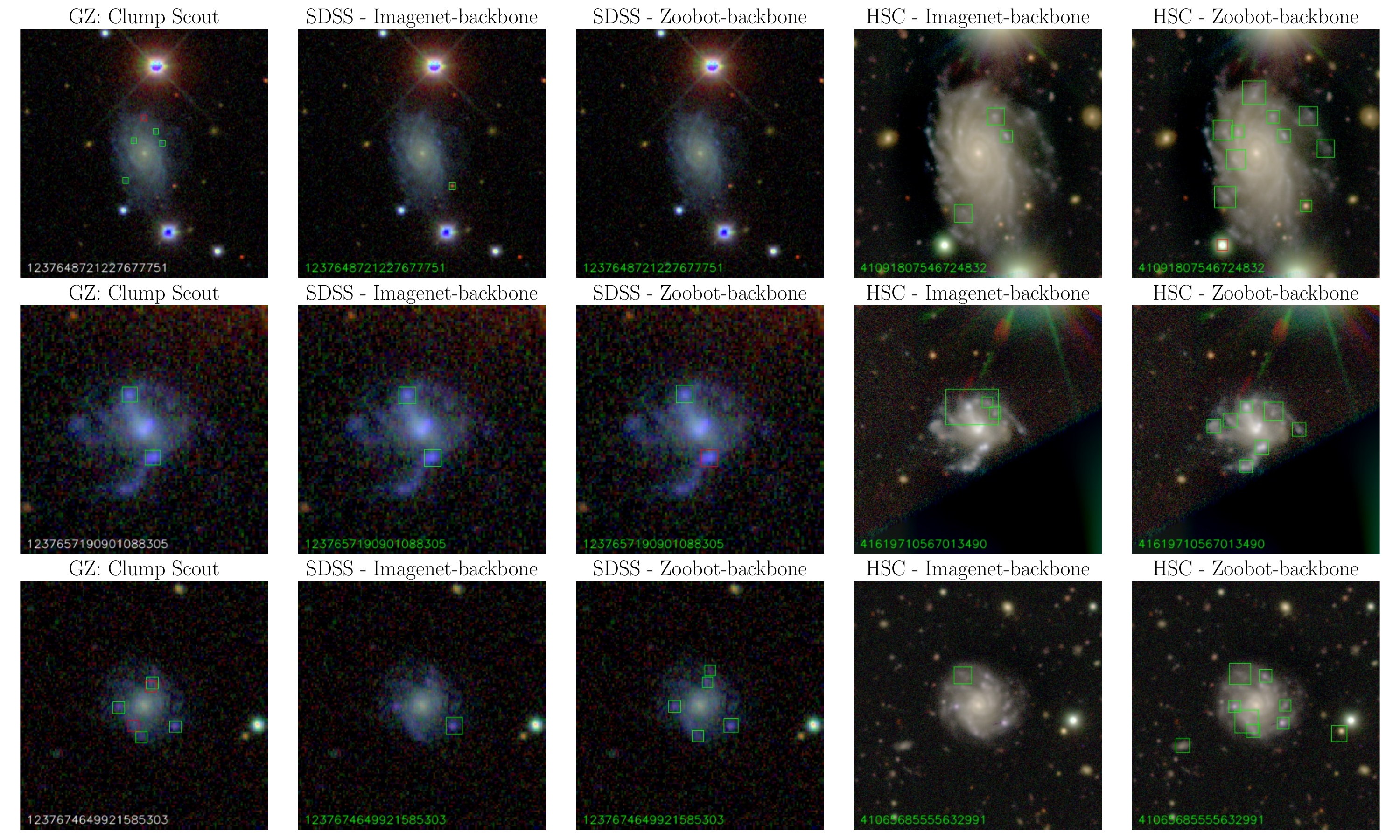}
    \caption[Comparison of detections on SDSS and HSC images (2).]{Comparison of detections on SDSS and HSC images. Images and detections as described in the previous figure.}
    \label{fig:HSC_examples2}
\end{figure*}

\bsp	
\label{lastpage}
\end{document}